%% file: manuscript.tex
\documentclass[aoas]{imsart}

\RequirePackage{amsthm,amsmath,amsfonts,amssymb}
\RequirePackage[authoryear]{natbib}
\RequirePackage[colorlinks,citecolor=blue,urlcolor=blue]{hyperref}
\RequirePackage{graphicx}
\usepackage{scribe}
\usepackage{algorithm}
\usepackage{xr-hyper}
\makeatletter
\newcommand*{\addFileDependency}[1]{
  \typeout{(#1)}
  \@addtofilelist{#1}
  \IfFileExists{#1}{}{\typeout{No file #1.}}
}
\makeatother

\startlocaldefs
\theoremstyle{plain}

\newtheorem{theorem}{Theorem}[section]

\theoremstyle{remark}

\endlocaldefs

\begin{document}

\begin{frontmatter}
\title{Modified treatment policy effect estimation with weighted energy distance}
\runtitle{Modified treatment policy effect estimation with weighted energy distance}

\begin{aug}
\author[A]{\fnms{Ziren Jiang}},
\author[B]{\fnms{Jared D. Huling}}

\address[A]{Division of Biostatistics and Health Data Science, University of Minnesota,\\ Minneapolis, MN, U.S.A.}
\address[B]{Division of Biostatistics and Health Data Science, University of Minnesota,\\ Minneapolis, MN, U.S.A. huling@umn.edu}

\end{aug}

\begin{abstract}
The causal effects of continuous treatments are often characterized through the average dose response function, which is challenging to estimate from observational data due to confounding and positivity violations. Modified treatment policies (MTPs) are an alternative approach that aim to assess the effect of a modification to observed treatment values and work under relaxed assumptions. Estimators for MTPs generally focus on estimating the conditional density of treatment given covariates and using it to construct weights. However, weighting using conditional density models has well-documented challenges. Further, MTPs with larger treatment modifications have stronger confounding and no tools exist to help choose an appropriate modification magnitude. This paper investigates the role of weights for MTPs showing that to control confounding, weights should balance the weighted data to an unobserved hypothetical target population that can be characterized with observed data. Leveraging this insight, we present a versatile set of tools to enhance estimation for MTPs. We introduce a distance that measures imbalance of covariate distributions under the MTP and use it to develop new weighting methods and tools to aid in the estimation of MTPs. Using our methods we study the effect of mechanical power of ventilation on in-hospital mortality.
\end{abstract}


\end{frontmatter}

\input{section/section_01.tex}

\input{section/section_02.tex}

\input{section/section_03.tex}

\input{section/section_04.tex}

\input{section/section_05.tex}

\input{section/section_06.tex}

\input{section/section_07.tex}

\input{section/section_08.tex}

\input{section/section_09.tex}

\bibliographystyle{imsart-nameyear}
\bibliography{Bibliography}

\makeatletter\@input{xx_supp.tex}\makeatother
\end{document}


\begin{frontmatter}
\title{Supplementary materials for ``Modified treatment policy effect estimation with weighted energy distance''}
\runtitle{Modified treatment policy effect estimation with weighted energy distance}

\begin{aug}



\author[A]{\fnms{Ziren Jiang}},
\author[B]{\fnms{Jared D. Huling}}

\address[A]{Division of Biostatistics and Health Data Science, University of Minnesota,\\ Minneapolis, MN, U.S.A.}
\address[B]{Division of Biostatistics and Health Data Science, University of Minnesota,\\ Minneapolis, MN, U.S.A. huling@umn.edu}
\end{aug}

\end{frontmatter}

\section{Notation}
This section defines key notation that will be used throughout the Supplementary Material.
\begin{itemize}
    \item $Y \in \mathbb{R}$: the observed response value.
    \item $A \in \mathbb{R} $: the observed treatment value.
    \item $\X \in \mathbb{R}^p$: the vector of pre-treatment covariates.
    \item $\{\X_i,A_i,Y_i\}_{i=1}^n$: the collection of data from an observational study with sample size $n$. 
    \item $(\mathcal{X},\mathcal{A})$: the joint support of the random variables $\X$ and $A$.
    \item $q(\x,a)$: a policy that modifies the treatment value according to the input of patient characteristics $\x$ and observed treatment value $a$. The function $q(\x,a)$ can be a modified treatment policy (MTP) if it satisfies the condition that $q(\x,a)\in \mathcal{A}(\x,a)$ for any $(\x, a) \in (\mathcal{X},\mathcal{A})$.
    \item $\mathcal{A}_{\x}\equiv \{a:(\x,a)\in (\mathcal{X},\mathcal{A})\}$: the support of $A$ given $\X=x$.
    \item $\mu^q\equiv \mathbb{E}[Y^{q(\X,A)}]$: mean potential outcome under the policy $q$.
    \item $\{I_{j,\x}\}_{j=1}^{J(\x)}$: the partition of $\mathcal{A}_{\x}$ for a given $\x$ on which the policy $q(\x,a), a\in I_{j,\x}$ has a differentiable inverse function $h_j(\x,\cdot)$ for $\forall j=1,...,J(\x)$. 
    \item $I_{j,\x}(a)$: indicator function such that $I_{j,\x}(a)=1$ if $a\in I_{j,\x}$ and $I_{j,\x}(a)=0$ otherwise.
    \item $\mu(\x,a)\equiv E(Y|\X=\x,A=a)$: conditional mean outcome function. 
    \item $F_{\X,A}^{q}$: CDF of the random variable $(\X,q(\X,A))$ where $q(\X,A)$ be the treatment value under the policy $q$. Here, we refer to $F_{\X,A}^{q}$ as the MTP-shifted CDF. 
    \item $F_{n,\X,A}^{q}\equiv \frac{1}{n}\sum_{i=1}^n\sum_{j=1}^{J(\X_i)}I_{j,\X_i}(A_i)I(\X_i\leq\x,q_j(\X_i,A_i)\leq a)$: the empirical CDF of the random variable $(\X,q(\X,A))$ with the data $\{\X_i,A_i\}_{i=1}^n$. $F_{n,\X,A}^{q}$ be the empirical estimator of the target CDF $F_{\X,A}^{q}$ which we refer to as the target empirical CDF. 
    \item $F_{n,\w,\X,A}\equiv \frac{1}{n}\sum_{i=1}^n w_i I(\X_i\leq\x,A_i\leq a)$: the weighted empirical CDF of the random variable $(\X,A)$ with the data $\{\X_i,A_i\}_{i=1}^n$ corresponding to the weights $\w=\{w_i\}_{i=1}^n$. 
    \item $\mathcal{E}(F_{n,\w,\X,A}, F^{q}_{n,\X,A})$: weighted energy distance of two empirical CDFs $F_{n,\w,\X,A}$ and $F^{q}_{n,\X,A}$. See Section 3 for the specific formula for calculation.
    \item $\w^e_{n}$: (unpenalized) energy balancing weights for covariate balancing. See Section 4 of the Supplementary Material for the specific definition. 
    \item $\w^{p}_{n}$: penalized energy balancing weights for covariate balancing. See Section 5.1 for the definition.
    \item $\mu^q_{\w}\equiv\sum_{i=1}^n w_iY_i$: Estimator of the mean potential outcome $\mu^q$ with the balancing weights $\w$.
    \item $\mu^q_{AG}$: augmented estimator of the mean potential outcome $\mu^q$. See Section 6 for the definition.

\end{itemize}

\section{Additional introduction of weighted energy distance}
\subsection{Definition of energy distance} \label{suppsec2.1}
The energy distance \citep{szekely2013energy} is a measure of distance between two multivariate distributions. The energy distance between the $d$-dimensional independent random variables $\boldsymbol{G}$ and $\boldsymbol{H}$ where $\boldsymbol{U}, \boldsymbol{U}' \overset{\mathrm{iid}}{\sim} \boldsymbol{G}$ and $\boldsymbol{V},\boldsymbol{V}' \overset{\mathrm{iid}}{\sim} \boldsymbol{H}$ is defined as 
\begin{equation}
    \mathcal{E}(\boldsymbol{G},\boldsymbol{H})=2\mathbb{E}(|\boldsymbol{U}-\boldsymbol{V}|_d)-\mathbb{E}(|\boldsymbol{U}-\boldsymbol{U}'|_d)-\mathbb{E}(|\boldsymbol{V}-\boldsymbol{V}'|_d),
\end{equation}
where $\mathbb{E}(|\boldsymbol{U}|_d)<\infty$, $\mathbb{E}(|\boldsymbol{V}|_d)<\infty$ so that the expectations are well-defined.

With the energy distance between two random vectors defined, the energy distance between two empirical CDFs can be similarly defined. Let $\G_n$ be the empirical CDF of sample $\{\boldsymbol{U}_i\}_{i=1}^n$ where $\boldsymbol{U}_i \overset{\mathrm{iid}}{\sim} \boldsymbol{G}$ and $\boldsymbol{H}_m$ be the empirical CDF of sample $\{\boldsymbol{V}_i\}_{i=1}^m$ where $\boldsymbol{V}_i \overset{\mathrm{iid}}{\sim} \boldsymbol{H}$. The energy distance between $\boldsymbol{G}_n$ and $\boldsymbol{H}_n$ is defined as
\begin{equation}
    \begin{split}
        \mathcal{E}(\G_n, \boldsymbol{H}_m)&=\frac{2}{nm}\sum_{i=1}^n\sum_{j=1}^m||\boldsymbol{U}_i-\boldsymbol{V}_j||_2-\frac{1}{n^2}\sum_{i=1}^n\sum_{j=1}^n||\boldsymbol{U}_i-\boldsymbol{U}_j||_2\\
        &-\frac{1}{m^2}\sum_{i=1}^m\sum_{j=1}^m||\boldsymbol{V}_i-\boldsymbol{V}_j||_2.
    \end{split}
\end{equation}

Note that, the definition of the energy distance ensures that it is rotation invariant in $d$-dimensional space. Here we provide some of the other key properties of the energy distance described in \citet{szekely2013energy}. The first property, made rigorous in the following Lemma, is that the energy distance $\mathcal{E} (\mathbf{G}, \mathbf{H})$ is non-negative and equals zero if and only if $\boldsymbol{U}$ and $\boldsymbol{V}$ are identically distributed.

\begin{lemma}[Proposition 1 from \citet{szekely2013energy}]
    If the $d$-dimensional random variables $\boldsymbol{U}\sim\boldsymbol{G}$ and $\boldsymbol{V}\sim\boldsymbol{H}$ are independent with $\mathbb{E}(|\boldsymbol{U}|_d)+\mathbb{E}(|\boldsymbol{V}|_d)<\infty$, and $\varphi_G$, $\varphi_H$ denote their respective characteristic functions, then the energy distance can be expressed as
    \begin{equation}
        \mathcal{E} (\boldsymbol{G},\boldsymbol{H}) =\frac{1}{c_d}\int_{R^d}\frac{|\varphi_G(t)-\varphi_H(t)|^2}{|t|^{d+1}_d}dt,
    \end{equation}
    where 
    \begin{equation}
        c_d=\frac{\pi^{(d+1)/2}}{\Gamma(\frac{d+1}{2})}
    \end{equation}
    and $\Gamma(\cdot)$ is the complete gamma function. Thus $\mathcal{E} (\boldsymbol{G},\boldsymbol{H}) \geq 0$ with equality  if and only if $\boldsymbol{G}$ and $\boldsymbol{H}$ are identical.
\end{lemma}

\subsection{Alternative metric: maximum mean discrepancy (MMD)} \label{suppsec2.2}
Here, we introduce the maximum mean discrepancy as an alternative metric that can be used to construct the optimal weights. Maximum mean discrepancy (MMD), originally proposed in the context of machine learning, is defined as the distance between embeddings of distributions to reproducing kernel Hilbert spaces (RKHS). With a different choice of kernel function, MMD provides a flexible definition of the distance between two distributions. In fact, the energy distance may be interpreted as a special case of MMD with a particular kernel function \citep{sejdinovic2013equivalence}. As such, it is an attractive alternative to our use of the energy distance to characterize distributional imbalance, however we do not prove the analytical properties of the balancing weights with MMD as the metric. One can simply switch the energy distance to the MMD in our methods. The MMD between two empirical CDFs $\G_n$ and $\boldsymbol{H}_m$ is defined as 
\begin{equation}
    \begin{split}
        \gamma^2_{ker}&=\frac{1}{n^2}\sum_{i=1}^n\sum_{j=1}^nk(\U_i,\U_j)+\frac{1}{m^2}\sum_{i=1}^m\sum_{j=1}^mk(\V_i,\V_j)-\frac{2}{nm}\sum_{i=1}^n\sum_{j=1}^m k(\U_i,\V_j),
    \end{split}
\end{equation}
where $k(\cdot, \cdot)$ is a valid kernel function. A common choice is the Gaussian kernel
\begin{equation}
    k_v(\x,\y)=\exp \left(-\frac{||\x-\y||^2}{2v^2}\right),
\end{equation}
where $v$ is a key tuning parameter that must be specified. In the simulation experiments of this paper, we use this kernel with the median heuristic $v=\textnormal{median}\{||\U_i-\V_j||^2|1<i<m, 1<j<n\}$ \citep{garreau2017large} as the choice of the tuning parameter. However, the use of the median heuristic is not fully justified and may be suboptimal in many scenarios. 

\section{Generalization to the stochastic interventions}\label{suppsec:SI}

Stochastic interventions \citep{munoz2012population} are a type of alternative causal parameter for estimating the effects of either standard or continuous interventions. Instead of fixing the shifted intervention $q(\X,A)$ to one single value, stochastic interventions view the shifted intervention as a random draw from a user-specified shifted intervention distribution. In \citet{munoz2012population}, the authors define the shifted intervention distribution as $\textnormal{Pr}(A=a|\X=\x)=f_{A|\X}(a-\delta(\x)|\X=\x)$ where $f_{A|\X}(a|\X=\x)$ is the conditional distribution of the treatment, and $\delta(\x)$ is a known function that characterize the stochastic intervention. Stochastic interventions are a general framework that can incorporate a variety of estimands, such as the incremental propensity score intervention proposed by \citet{kennedy2019nonparametric}. Thus, extensions of our framework to stochastic interventions greatly increases the generality of our proposed methods.

Letting $Y^\delta$ denote the counterfactual outcome under the stochastic intervention $\delta$, the \citet{munoz2012population} give the following identification result:
\begin{equation}
    \begin{split}
        \mathbb{E}(Y^q)=&\int_{a\in\mathcal{A}}\int_{\x\in\mathcal{X}} \mu(\x,a)f_{A|\X}(a-\delta(\x)|\X=\x)f_{\X}(\a)d\x da\\
        =& \int_{A\in\mathcal{A}}\int_{\X\in\mathcal{X}} \mu(\x,a)d F_{\X,A}(\x,a+\delta(\x)).
    \end{split}
\end{equation}

It is straightforward to prove that $F_{\X,A}(\x,a+\delta(\x))$ is the CDF of random variable $(\X,Q(\X,A)) = (\X,A-\delta(\X))$ where $Q(\X,A)=A-\delta(\X)$. Therefore, our result of the estimation of the causal effect of MTP can be easily generalized to stochastic interventions with the particular shifted distribution based on the function $q(\x,a)=a-\delta(\x)$. 

In \citep{diaz2021nonparametric}, the authors considered a more general rule for stochastic intervention effect in the longitudinal studies and showed its connection to modified treatment policies. For simplicity, we only consider the scenario with one time point. For any user-given function ${d}(\x,a)$, Theorem 1 of \citep{diaz2021nonparametric} gives the following identification result:
\begin{equation}
    \mathbb{E}(Y^{{d}})=\mathbb{E}(\mu(\X,{d}(\X,A)))=\int_{A\in\mathcal{A}}\int_{\X\in\mathcal{X}} \mu(\x,{d}(\x,a)) dF_{\X,A}(\x,a).
\end{equation}
We further require that ${d}(\x,a)$ (as a function of $a$) to be at least piece-wise invertible (assumption A3 in section \ref{sec:2} of the manuscript). With this assumption, we will have exactly the same identification result as in (\ref{eq:identification}) with $q={d}$.

With the identification results for stochastic interventions, we can straightforwardly extend the error decomposition for MTPs to stochastic interventions. Assume we have the weighted estimator $\hat{\mu}^{q}_{\w}=\frac{1}{n}\sum_{i=1}^{n} w_i Y_i$, and denote $\mu^q=\mathbb{E}(Y^q)$ then we have
\begin{align}
       \hat{\mu}^{q}_{\w}-\mu^q={} & \underbrace{\int_{(\mathcal{X},\mathcal{A})}\mu(\boldsymbol{x},a)d(F_{n,\w,\X,A}-F_{n,\X,A}^{q})(\boldsymbol{x},a)}_{\text{error due to confounding bias}} \\
        &+\underbrace{\int_{(\mathcal{X},\mathcal{A})}\mu(\boldsymbol{x},a)d(F_{n,\X,A}^{q}-F_{\X,A}^{q})}_{\text{sampling error}} +\frac{1}{n}\sum_{i=1}^n w_i\epsilon_i, 
\end{align}
which is identical to the error decomposition of MTP with a prespecified, deterministic function $q(\x,a)$.

\section{Additional illustration of using weighted energy distance for MTPs}
\subsection{Simulation results for the method of choosing a feasible MTP scale}\label{sec:suppapplication1}
Here, we provide the simulation results to demonstrate the control of type I error for our method of choosing a feasible MTP scale. We use the same data-generating mechanism as the first simulation scenario in the manuscript Section \ref{sec:7} with sample size $n=100, 200, 400, 800$, and number of covariates $p=10, 20, 40, 80$ with continuous treatment. The detailed data-generating mechanism is described in the Supplementary Material Section \ref{sec:suo_sim}. Under the null hypothesis, there is no difference between the observed population and the shifted population, i.e., $q(\x,a)=a$. For the alternative hypothesis, we adopt $q(\x,a)=a-10$ as our shift function. Two permutation strategies are under evaluation in this simulation study, the first strategy is the one we present in the manuscript section \ref{sec:4.1} where we randomly partition the dataset $\{\X_i,A_i\}_{i=1}^n$ into two subsamples $\{\X_{I_k},A_{I_k}\}_{k=1}^{n/2}$ and $\{\X_{J_k},A_{J_k}\}_{k=1}^{n/2}$. The permutation is conducted on the combined sample $\{\Tilde{w}_i, \Tilde{\X}_i, \Tilde{A}_i\}_{i=1}^n = (\{w_{I_k}, \X_{I_k},A_{I_k}\}_{k=1}^{n/2}, \{w_{J_k}, \X_{J_k},q(\X_{J_k}, A_{J_k})\}_{k=1}^{n/2})$ of size $n$. The second strategy is to use the entire dataset in the permutation without any partition, i.e., we use the sample $\{\Tilde{w}_i, \Tilde{\X}_i, \Tilde{A}_i\}_{i=1}^2n = (\{w_{i}, \X_{i},A_{i}\}_{i=1}^{n}, \{w_{j}, \X_{j},q(\X_{j}, A_{j})\}_{j=1}^{n})$ of size $2n$ for the permutation. We explore the second strategy to assess the challenges presented by the samples being correlated.

We run 2000 experiments for each simulation condition. For each simulation experiment, the p-value for the test is calculated as the proportion of times that $T_b>T$ among the $b=1,...,B$ permutations, where $T_b$ is the weighted energy distance of the two permuted samples for the $b$th permutation and $T$ is the weighted energy distance of the weighted observed sample and the MTP-shifted sample (see the main manuscript Section \ref{sec:4.1} for the definition). Under the null hypothesis of no population shift, we examine the control of type I error rate for our test by evaluating whether the p-values are uniformly distributed among $[0,1]$. Under the alternative hypothesis of existing population imbalance, we compare the power of the two tests with different permutation strategies. The power is estimated by the proportion of p-values smaller than $0.05$. 

\begin{figure}[ht]
    \centering
     \includegraphics[width=\textwidth]{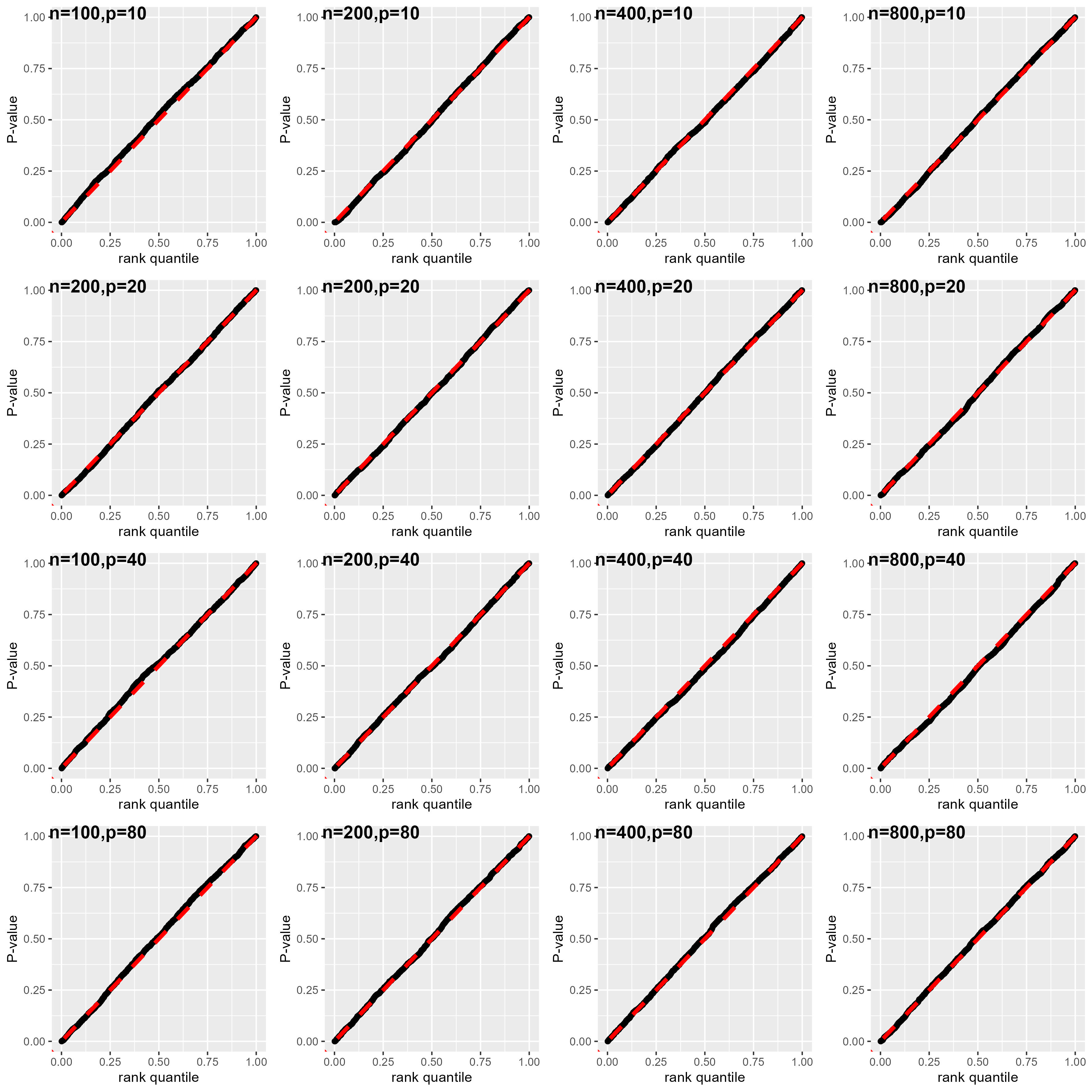}
    \caption{Plot of ranked p-value with the corresponding rank quantile for the first permutation strategy. The x-axis is the quantile of the ranked p-value and the y-axis is the p-value. The p-values are demonstrated to be uniformly distributed between $[0,1]$ under the null hypothesis.}
    \label{fig:application1_3}
\end{figure}

\begin{figure}[ht]
    \centering
     \includegraphics[width=\textwidth]{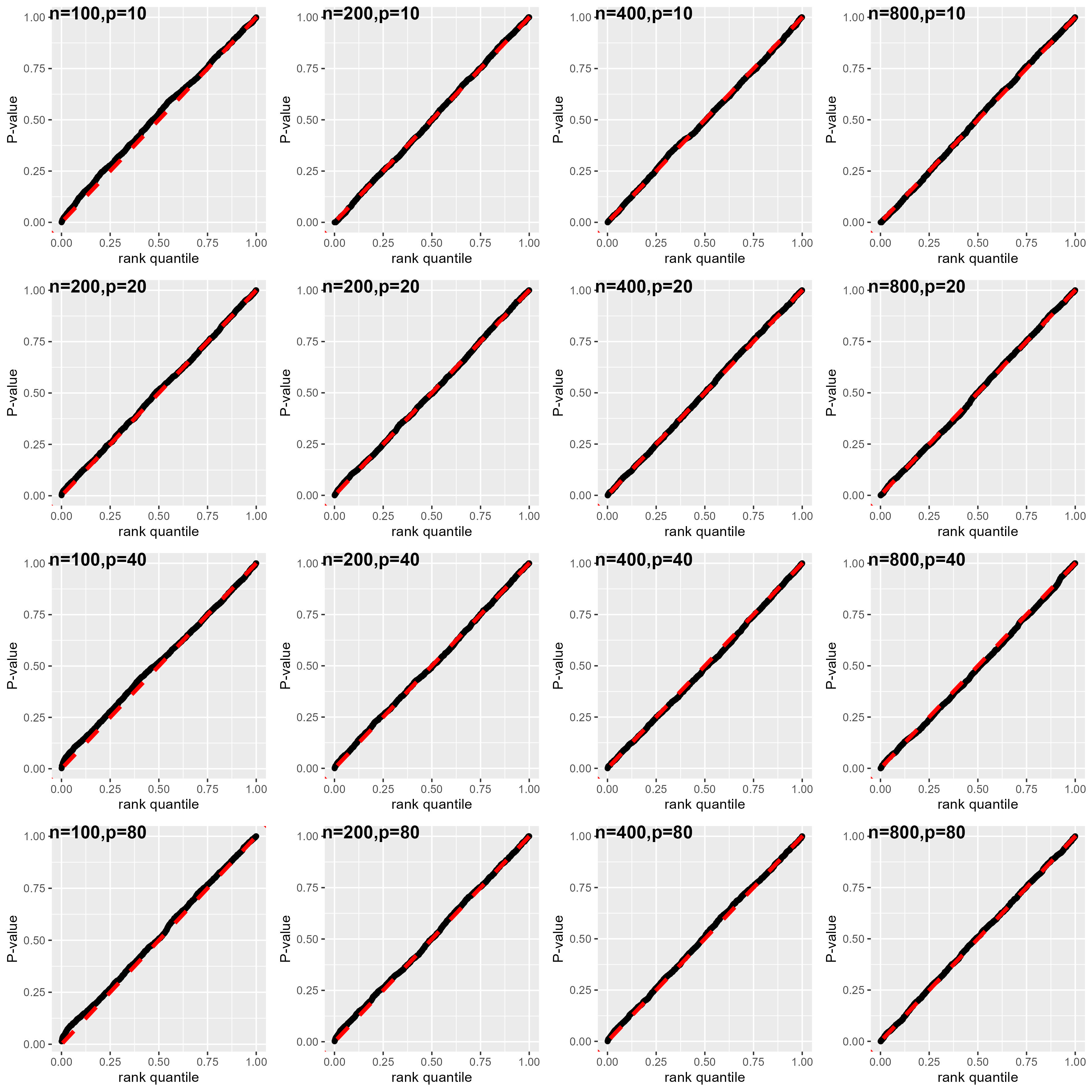}
    \caption{Plot of ranked p-value with the corresponding rank quantile for the first permutation strategy. The x-axis is the quantile of the ranked p-value and the y-axis is the p-value. The p-values are demonstrated to be uniformly distributed between $[0,1]$ under the null hypothesis.}
    \label{fig:application1_4}
\end{figure}

Figure \ref{fig:application1_3} and \ref{fig:application1_4} display the plot of the ordered p-value and its rank for all simulation conditions. For both strategies, the p-values are demonstrated to be uniformly distributed between $[0,1]$ under the null hypothesis. Therefore, both strategies demonstrate their ability to control the type I error. Table \ref{Table:appl1} displays the proportion of p-values less than 0.05 under the null and alternative hypothesis for the two permutation strategies. Under the null hypothesis, both strategies 1 and 2 control the type I error around the nominal level of 0.05. Under the alternative hypothesis of existing population imbalance, the first method has the larger power in most scenarios, especially when the power is not close to 1. From the simulation results, although the second strategy includes a larger sample size for the permutation test, the correlation among the two samples $\{w_{i}, \X_{i},A_{i}\}_{i=1}^{n}$ and $\{w_{j}, \X_{j},q(\X_{j}, A_{j})\}_{j=1}^{n}$ may actually decrease the power. Therefore, we adopt the first strategy which is formally presented in the main manuscript.

\begin{table}[]
\centering
\caption{Empirical frequency of the p-value less than 0.05 under the null and alternative hypothesis for the two permutation strategies. Strategy 1 uses the uncorrelated sample while Strategy 2 uses the correlated sample. Each row represents a simulation condition, where $n$ is the sample size and $p$ is the number of covariates. }\label{Table:appl1}
\resizebox{0.7\textwidth}{!}{
\begin{tabular}{cccccc}
 &\multicolumn{2}{c}{Under null}&&\multicolumn{2}{c}{Under alternative}\\
 \cmidrule{2-3}\cmidrule{5-6}
Condition & Strategy 1 & Strategy 2 && Strategy 1 & Strategy 2 \\
n=100                  p=10 & 0.055      & 0.055      && 0.943      & 0.935      \\
n=200                  p=10 & 0.059      & 0.054      && 0.983      & 0.986      \\
n=400                  p=10 & 0.045      & 0.044      && 0.999      & 1          \\
n=800                  p=10 & 0.048      & 0.049      && 1          & 1          \\[6pt]
n=100                  p=20 & 0.055      & 0.055      && 0.84       & 0.805      \\
n=200                  p=20 & 0.055      & 0.052      && 0.936      & 0.932      \\
n=400                  p=20 & 0.05       & 0.048      && 0.981      & 0.984      \\
n=800                  p=20 & 0.051      & 0.049      && 0.999      & 1          \\[6pt]
n=100                  p=40 & 0.044      & 0.043      && 0.773      & 0.709      \\
n=200                  p=40 & 0.055      & 0.055      && 0.843      & 0.822      \\
n=400                  p=40 & 0.046      & 0.046      && 0.903      & 0.904      \\
n=800                  p=40 & 0.053      & 0.052      && 0.97       & 0.973      \\[6pt]
n=100                  p=80 & 0.062      & 0.057      && 0.736      & 0.574      \\
n=200                  p=80 & 0.05       & 0.05       && 0.783      & 0.697      \\
n=400                  p=80 & 0.047      & 0.048      && 0.816      & 0.774      \\
n=800                  p=80 & 0.045      & 0.045      && 0.828      & 0.811     
\end{tabular}
}

\end{table}

\subsection{Evaluation of arbitrary weights for a given MTP and dataset}\label{sec:additionalillu}
\begin{figure}[ht]
    \centering
   \includegraphics[width=0.8\textwidth]{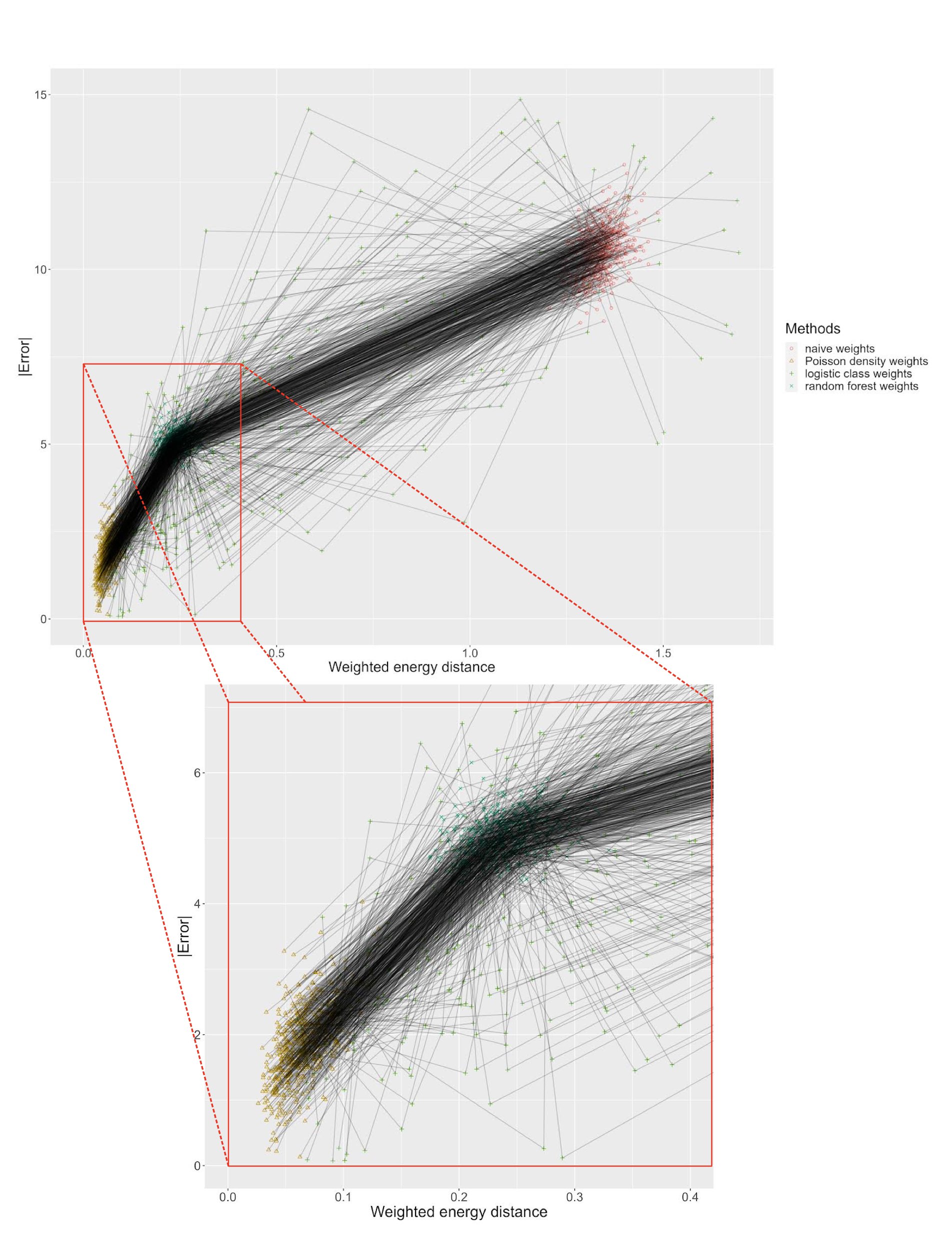}
    \caption{Scatter plot illustrating the association between the weighted energy distance and the absolute estimation error. Data is derived from a simulation study with continuous intervention and parameters $n=800, p=80$. Results from the same simulation replication are connected with lines. Methods with a larger weighted energy distance generally correspond to a greater absolute error.}
    \label{fig:application2_1}
\end{figure}

This subsection provides additional illustration of the correlation between the weighted energy distance after covariate balancing and the corresponding estimation error. This demonstrates the effectiveness our method of choosing balancing methods with the smallest weighted energy distance. Figures \ref{fig:application2_1} depicts the correlation between the weighted energy distance and the estimation error for four distinct weights: naive weights (which equally weight all sample points), random forest weights, logistic class weights, and Poisson density weights (which match the true data generating mechanism). The figure displays simulation results for $n=800$ and $p=80$ under the simulation study \#1 described in the simulation section. Each point on the graph represents the energy distance (x-axis) and the estimation error (y-axis) of one weight, and a line corresponds to a particular simulation dataset. Within each simulation scenario, there is a noticeable trend between the weighted energy distance and the absolute value of the estimation error, suggesting that the method with the smallest energy distance tends to yield the smallest estimation error for a given dataset. 

\begin{figure}[ht]
    \centering
     \includegraphics[width=\textwidth]{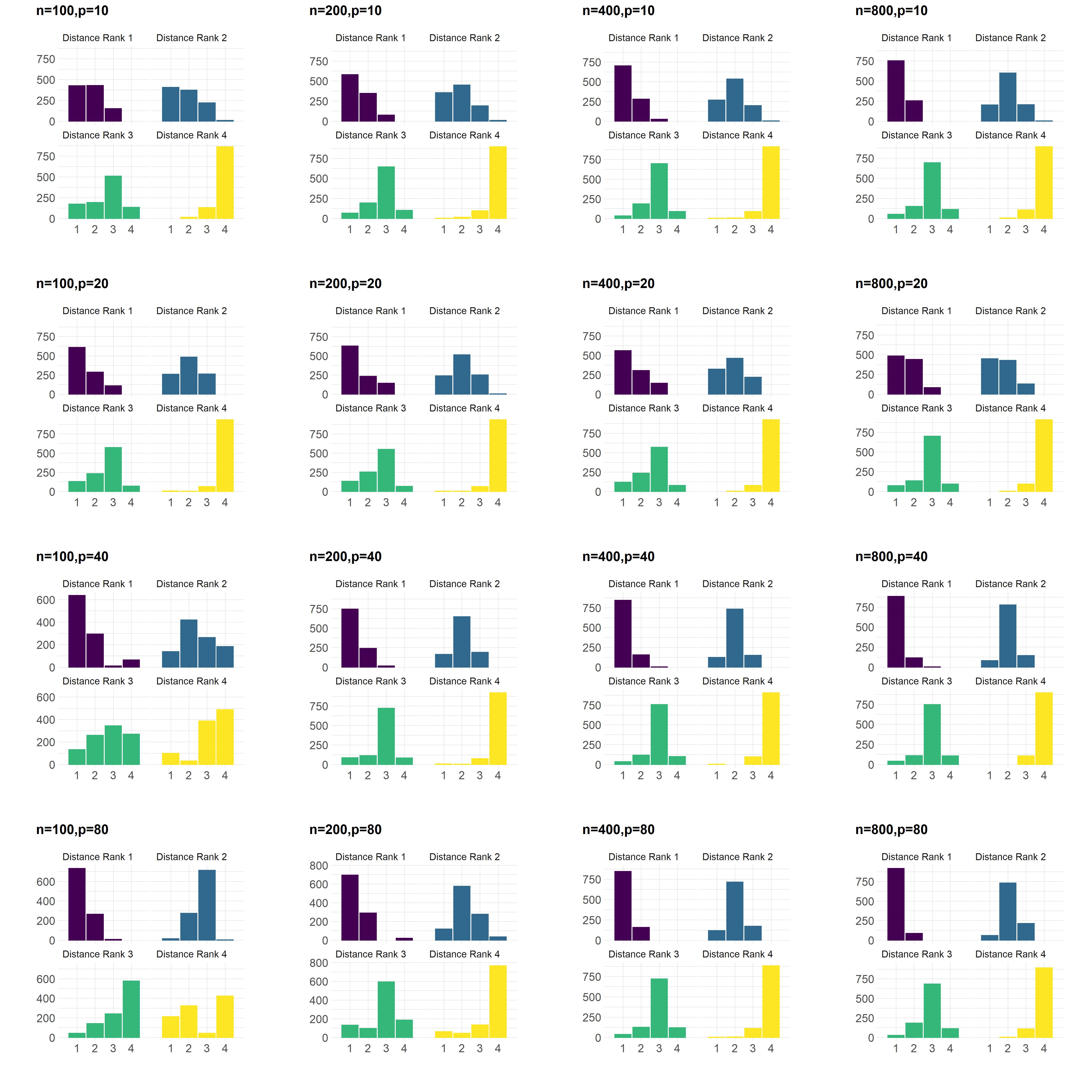}
    \caption{Collection of histograms illustrates the relationship between the rank of estimation error and the rank of weighted energy distance under simulation conditions with discrete interventions. More details on the simulation can be found in suplementary Section \ref{sec:suo_sim}. For all simulation conditions, methods with the lowest weighted energy distance generally demonstrate the best performance. This outcome further validates our approach of selecting different balancing methods based on their weighted energy distance for a given MTP and dataset.}
    \label{fig:application2_3}
\end{figure}

We further demonstrate the performance of this method using all 16 simulation conditions (with $n=100$ to $n=800$ and $p=10$ to $p=80$) of the simulation study \#1 used in our simulation section for discrete treatments (which enables the use of Poisson density). Figure \ref{fig:application2_3} summarizes the relationship between the energy distance rank and the error rank. We can see that for most of the simulation conditions, the chosen method with the smallest weighted energy distance has the best performance in terms of estimation error. From the figure, we can also see that our method for choosing different balancing methods generally performs better for larger sample sizes and more complex scenarios (with more covariates). However, the energy distance still serves as a good criterion to determine the best weighting approach for a given MTP and dataset.

\clearpage

\clearpage

\clearpage

\section{Energy balancing estimator}\label{sec:5.1}
\subsection{An unpenalized energy balancing estimator}
In this section, we provide an unpenalized version of our energy balancing estimator (which we refer as energy balancing estimator). As stated in the paper, the existence of extreme weights is a key hurdle for a weighted estimator to achieve root-$n$ consistency. Therefore, the following unpenalized energy balancing estimator requires additional restrictions for extreme weights in order to be root-n consistent. The energy balancing weights $\w^{e}_n$ are defined as the weights $\w^{e}_n=\{w^e_{i,n}\}_{i=1}^n$ such that
\begin{equation}\label{equ:UnpenaObj}
    \begin{split}
        \w^{e}_n\in &\underset{\w=\{w_1,...,w_n\}}{\arg\min}\mathcal{E}(F_{n,\w,\X,A},F_{n,\X,A}^{q})\\
        &\textrm{subject to} \sum_{i=1}^n w_i=n \textrm{ ,and } w_i\geq 0 \textrm{ for all } i.
    \end{split}
\end{equation}
The corresponding energy balancing estimator of the mean potential outcome under the MTP is 
\begin{equation}
    \hat{\mu}^{q}_{\w^{e}_n}=\sum_{i=1}^n w_{i,n}^eY_i.
\end{equation}

Like the penalized energy balancing weights, the purpose of the energy balancing weights is to minimize the energy distance between the weighted empirical CDF and the MTP-shifted empirical CDF. As shown in (\ref{the:3.1}) and Theorem \ref{the:3.2} in the main text, the energy distance is a measure of the discrepancy between two distributions. Thus, by using the energy balancing weights, the weighted empirical distribution and the shifted empirical distribution (under MTP) are balanced. Consequently, we can expect that the corresponding energy balancing estimate will in general have a small bias due to the connection between the distributional balance and bias in estimating MTP effects. Analogous to the penalized energy balancing weights, we prove the following properties of the energy balancing weights. Theorem \ref{the:5.1} proves that by minimizing the weighted energy distance, the energy balancing weights makes the weighted empirical CDF converges to the true CDF under the MTP.

\begin{theorem}\label{othe:5.1}
Assume $\mathbb{E}(||\X||_2|A)<\infty$, $\mathbb{E}(||\X||_2)<\infty$, and assumption A0-A3 in the main text Section \ref{sec:2} hold. Let $\w_n^e$ be the energy balancing weights. Then, we have
\begin{equation}
    \lim_{n\to\infty} F_{n,\w_n^e, \X,A}(\x,a)=F_{\X,A}^{q}(\x,a)
\end{equation}
almost surely for every continuity point $(\x,a)\in ( \mathcal{X},\mathcal{A})$. Furthermore,
\begin{equation}
    \lim_{n\to\infty}\mathcal{E}(F_{n,\w_n^e, \X,A},F_{n,\X,A}^{q})=0
\end{equation}
holds almost surely.
\end{theorem}
Following the above result, we directly have the asymptotic unbiasness of the energy balancing estimator $\hat{\mu}^{q}_{\w^{e}_n}$ by applying the Portmanteau Theorem.

\begin{theorem}\label{othe:5.2}
Suppose the conditions of Theorem \ref{othe:5.1} hold, and the conditional mean (potential) outcome $\mu(\x,a)$ is bounded and continuous on $(\mathcal{X},\mathcal{A})$. Then, $\hat{\mu}^{q}_{\w^{e}_n}$ is an asymptotically unbiased estimator of $\mu^q$.
\end{theorem}

To ensure root-$n$ consistency of the energy balancing estimator one can impose a restriction to avoid these extreme weights. One approach is to introduce an additional penalization term for the sum of squared weights, resulting in penalized energy balancing weights (see the main paper for more detail). Alternatively, one could also assume or impose in the optimization problem that the energy balancing weights meet the condition $w^e_{i,n}\leq C n^{1/3}$ for all $i=1,...,n$, where $C$ is a real number. This assumption of no extreme weights is essential for the following theorem demonstrating the root-$n$ consistency of the energy balancing estimator. Although we assume this holds, it is straighforward to impose this condition in the optimization problem as all theoretical results can straightforwardly be proved with such a modified optimization problem as in \citet{huling2023independence}. 

\begin{theorem}\label{othe:5.3}
Assume the conditions in Theorem \ref{othe:5.1}. Let $\mathcal{H}$ be the native space induced by the radial kernel $\Phi(.)=-||.||_2$ on $\mathcal{X}$. Suppose the following mild conditions hold:
\begin{itemize}
    \item \textnormal{\textbf{C-1:}} $\mu(\cdot,\cdot)\in\mathcal{H}$
    \item \textnormal{\textbf{C-2:}} Var $[\mu(\X,A)]<\infty$ 
    \item \textnormal{\textbf{C-3:}} Var $[Y|\X = \x,A=a]$ is bounded over $(\x,\a)\in(\X,\A)$.
    \item \textnormal{\textbf{C-4:}} $\mathbb{E}[g^2(\W,\W',\W'',\W''')]\leq\infty$ where $\W,\W',\W'',\W'''\stackrel{i.i.d.}{\sim}F_{\X,A}^{q}$ is a vector with $\W=(\X,A)$,  $h(\w)=\frac{f^q_{\X,A}(\x,a)}{f_{\X,A}(\x,a)}$, and the kernel function $g(.)$ is defined as:
\begin{equation}
    g(\w,\w',\w'',\w''')=h(\w)||\w-\w''||_2+h(\w')||\w'-\w'''||_2-h(\w)h(\w')||\w-\w'||_2-||\w'''-\w''||_2
\end{equation}
    \item \textnormal{\textbf{C-5:}} The energy balancing weights $\w^e_n$ satisfies $w^e_{i,n}\leq C n^{1/3}$ for some constant $C>0$ independent of $n$ for $i=1,\dots,n$. 
\end{itemize}

Then, the proposed EBW estimator $\hat{\mu}^{q}_{\w^{e}_n}$ is root-n consistent, i.e.:
\begin{equation}
    \sqrt{\mathbb{E}_{\X,\A,\Y}[(\hat{\mu}^{q}_{\w^{e}_n}-\mu^{q})^2]}=\mathcal{O}(n^{-1/2}).
\end{equation}
\end{theorem}
We briefly comment on the consequence of the assumptions. \textbf{C-1} restricts the conditional mean function $\mu(\x,a)$ to belong to the native space $\mathcal{H}$. As demonstrated in \citet{huling2020energy}, this condition can be viewed as a smoothness assumption on the conditional mean function $\mu(\x,a)$. \textbf{C-2} restricts the conditional mean function to have finite variance and \textbf{C-3} requires the conditional variance to be bounded. These two conditions are both mild and fairly weak in practice. \textbf{C-4} requires the kernel $g(\cdot)$ to have a finite second moment and \textbf{C-5} assumes that there are no extreme weights for the estimation. Similar to \citet{huling2023independence}, condition \textbf{C-5} can be directly imposed in the optimization problem and does not impact the asymptotic properties of the weights. 

\subsection{Implementation detail for minimizing the weighted energy distance}\label{suppSec: codingdetail}
In this section, we provide the implementation details of our algorithm for solving the unpenalized and penalized energy balancing weights. First, consider the optimization problem \ref{equ:UnpenaObj} for unpenalized energy balancing weights:
\begin{equation*}
    \begin{split}
        \w^{e}_n\in &\underset{\w=\{w_1,...,w_n\}}{\arg\min}\mathcal{E}(F_{n,\w,\X,A},F_{n,\X,A}^{q})\\
        &\textrm{subject to} \sum_{i=1}^n w_i=n \textrm{ ,and } w_i\geq 0 \textrm{ for all } i.
    \end{split}
\end{equation*}
where 
\begin{align*}
\mathcal{E}(&F_{n,\w,\X,A},F_{n,\X,A}^{q}) = {}  \frac{1}{n^2}\bigg\{2\times\sum_{i=1}^n\sum_{k=1}^n\sum_{j=1}^{J(\X_k)}I_{j,\X_k}(A_k)w_i||(\X_i,A_i)-(\X_k,q_j(\X_k,A_k))||_2 \nonumber\\
        &-\sum_{i=1}^n\sum_{k=1}^n w_i w_k ||(\X_i,A_i)-(\X_k,A_k)||_2 \nonumber\\
        &-\sum_{i=1}^n\sum_{j=1}^{J(\X_i)}\sum_{k=1}^n\sum_{j'=1}^{J(\X_k)} I_{j,\X_i}(A_i) I_{j',\X_k}(A_k)||(\X_i,q_j(\X_i,A_i))-(\X_k,q_{j'}(\X_k,A_k))||_2 \bigg\}\\
        &= \frac{1}{n^2}\bigg\{  2\times\sum_{i=1}^n\sum_{k=1}^n\ w_i||(\X_i,A_i)-(\X_k,q(\X_k,A_k))||_2 \nonumber \\
        &-\sum_{i=1}^n\sum_{k=1}^n w_i w_k ||(\X_i,A_i)-(\X_k,A_k)||_2 \nonumber -\sum_{i=1}^n\sum_{k=1}^n  ||(\X_i,q(\X_i,A_i))-(\X_k,q(\X_k,A_k))||_2    \bigg\}\\
        &= \w^T\mathbb{Q}\w + 2 \w^T \mathbf{b} - \sum_{i=1}^n\sum_{k=1}^n  ||(\X_i,q(\X_i,A_i))-(\X_k,q(\X_k,A_k))||_2
\end{align*}
where $q(\x,a)=\sum_{j=1}^{J(\x)}I_{j,\x}(a) q_j(\x,a)$, $\w=\{w_1,...,w_n\}$, $\mathbb{Q}$ be a matrix with $||(\X_i,A_i)-(\X_j,A_j)||_2$ the element of the $i$-th row and $k$-th column, and $\mathbf{b}$ be a vector with the $i$-th element being $\sum_{k=1}^n||(\X_i,A_i)-(\X_k,q(\X_k,A_k))||_2$. 

Therefore, the optimization problem becomes:
\begin{equation*}
    \begin{split}
        \w^{e}_n\in &\underset{\w=\{w_1,...,w_n\}}{\arg\min}\w^T\mathbb{Q}\w+ 2 \w^T \mathbf{b}\\
        &\textrm{subject to} \sum_{i=1}^n w_i=n \textrm{ ,and } w_i\geq 0 \textrm{ for all } i.
    \end{split}
\end{equation*}
which is a linear optimization problem with linear restrictions that can be handled efficiently by the existing algorithms. For example, we use the R package OSQP \cite{osqp} to solve the energy balancing weights. 

Similarly, for the penalized energy balancing weights, the optimization problem can be transformed as:
\begin{equation*}
    \begin{split}
        \w^{e}_n\in &\underset{\w=\{w_1,...,w_n\}}{\arg\min}\w^T(\mathbb{Q}+\lambda\mathbb{I})\w+ 2 \w^T \mathbf{b}\\
        &\textrm{subject to} \sum_{i=1}^n w_i=n \textrm{ ,and } w_i\geq 0 \textrm{ for all } i.
    \end{split}
\end{equation*}
where $\lambda$ be a user-specified penalty parameter and $\mathbb{I}$ is the identity matrix.

\section{Supplementary simulation study materials}\label{sec:suo_sim}
\subsection{Data-generating mechanism}\label{sec:datagene}
This subsection contains full details of the data-generating mechanisms used in the main text. They are described as follows:
\begin{itemize}
    \item The covariates $(X_{i1},...,X_{ip})$ for subject $i=1,...,n$ are generated follows one of the three distributions. Binomial distribution with successful probability $0.7$ or a uniform distribution with $\min=0,\max=1$ or a uniform distribution with $\min=-1,\max=1$.
    \item For each subject $i=1,...,n$, the treatment mean $\lambda_i$ is generated through a cubic function of covariates $(X_{i1},...,X_{ip})$. 
    \begin{equation*}
        \lambda_i=X_{i1}^2-X_{i2}^2+X_{i1}*X_{i2}+X_{i4}+X_{i5}+X_{i7}+X_{i8}+X_{i9}+\sum_{j=11}^p \frac{(X_{ij}+X_{ij}^2+X_{ij}^3)}{\sqrt{p-4}}
    \end{equation*}
    \item If the treatment type is discrete, generate $A_i\sim \text {Poisson}(\lambda_i)$. If the treatment type is continuous, generate $A_i\sim 2\lambda_i\cdot \text{Beta}(2,2)$.
    \item For simulation \#1, given the covariates $(X_{i1},...,X_{ip})$ and treatment $A_i$, generate the outcome $Y_i\sim N(\mu_i,1)$ where the mean $\mu_i$ follows,
    \begin{equation*}
        \begin{split}
            &\mu_i=\bigg(-1-\frac{1}{2}X_{i3}+\frac{1}{2}X_{i4}+\frac{1}{2}X_{i5}+\sum_{j \text{ is even},j=12}^p \frac{1}{2(p-5)^2}X_{ij}\bigg)+\bigg(\left(\frac{A_i-20}{2}\right)^2-12\bigg)\\
            &\times \bigg(1-\frac{1}{2}X_{i1}^2-\frac{1}{2}X_{i2}^2+X_{i1}X_{i2}+\frac{1}{2}(X_{i7}+X_{i8}+X_{i10})-\sum_{j \text{ is odd},j=11}^p \frac{1}{2(p-5)^2}X_{ij}\bigg)^2
        \end{split}
    \end{equation*}
    \item For simulation \#2 with the more complex data-generating mechanism, given the covariates $(X_{i1},...,X_{ip})$ and treatment $A_i$, generate the outcome $Y_i\sim N(\mu_i,1)$ where the mean $\mu_i$ follows,
    \begin{equation*}
        \begin{split}
            \mu_i= &\bigg(\frac{3}{(p-5)^{1/3}}\bigg)\bigg(-1-\frac{1}{2}X_{i3}+\frac{1}{2}X_{i4}+\frac{1}{2}X_{i5}+\sum_{j \text{ is even},j=12}^p (\frac{1}{2}X_{ij}+X_{ij}^2\\
            &+X_{ij}(X_{i(j-10)}^2+2X_{i(j-9)}))\bigg)+
            \bigg((\frac{A_i-20}{2})^2-6\bigg)\bigg(\frac{3}{(p-5)^{1/3}}\bigg)\\
            &\times \bigg(1-0.5X_{i1}^2-0.5X_{i2}^2+X_{i1}X_{i2}+0.5(X_{i7}+X_{i8}+X_{i10})\\
            &\quad\quad -\sum_{j \text{ is odd},j=11}^p (-\frac{1}{2}X_{ij}+0.3X_{ij}^2+0.3X_{ij}(X_{i(j-8)}^2+2X_{i(j-7)}))\bigg)
        \end{split}
    \end{equation*}

    \item The modified treatment policy is $q(\x,a)$ designed to  depend on the original treatment $a$, 
\begin{equation*}
     q(\x,a) =
    \begin{cases}
      a+5 & \text{if $0<a<10$}\\
      a+3 & \text{if $10<a<20$}\\
      a+1 & \text{if $20<a$}.
    \end{cases}  
\end{equation*}
\end{itemize}

\subsection{Additional simulation results}\label{sec:simresult}
In this subsection, we present the simulation results of MSE in the simulation study \#1 and the additional simulation results for the simulation study \# 2. 

\begin{figure}[ht]
    \centering
     \includegraphics[width=\textwidth]
     {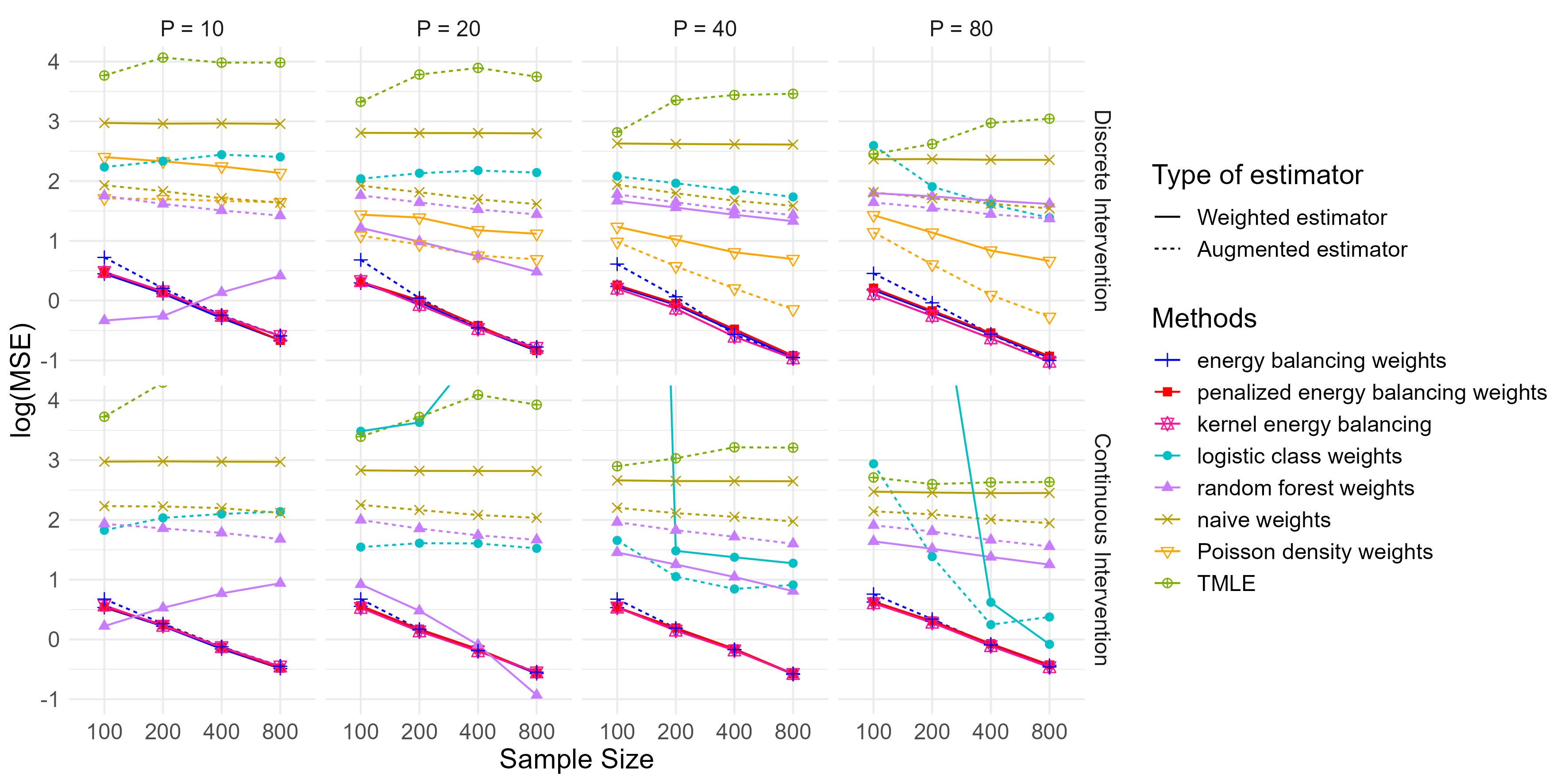}
    \caption{Simulation results for the logarithm of the mean squared error (MSE) across different sample sizes, type of intervention values, and the dimensionality of covariates. The data-generating mechanism is moderately complex (simulation \# 1). Balancing methods are displayed in different colors and shapes of points. Weighted estimators are displayed in solid lines and the augmented estimator are in dashed lines.}
    \label{fig:simulation1_3}
\end{figure}

\begin{figure}[ht]
    \centering
     \includegraphics[width=\textwidth]{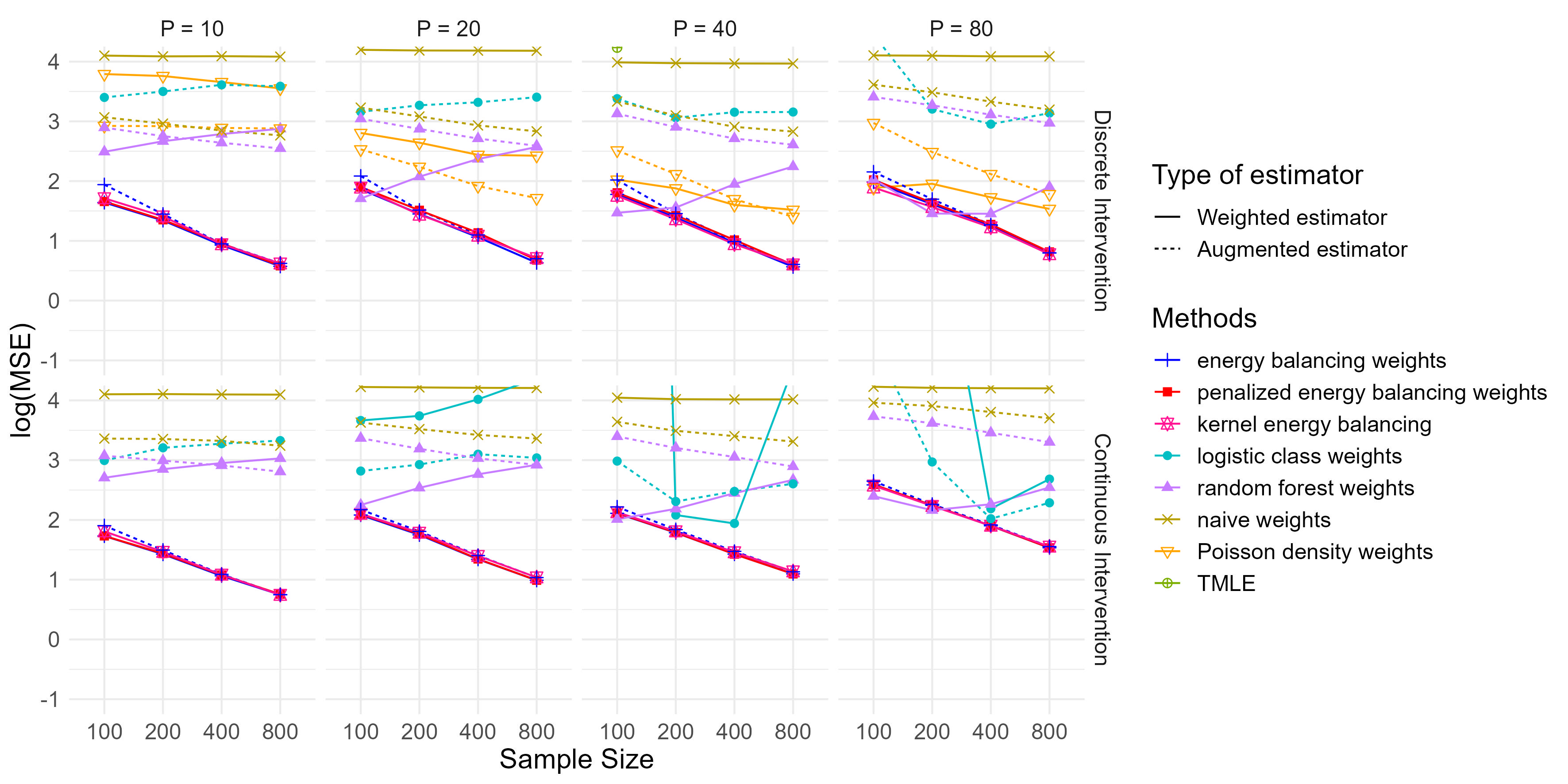}
    \caption{Simulation results for the logarithm of the mean squared error (MSE) across different sample sizes, type of intervention values, and the dimensionality of covariates. The data-generating mechanism is extremely complex (simulation \# 2). Balancing methods are displayed in different colors and shapes of points. Weighted estimators are displayed in solid lines and the augmented estimator are in dashed lines.}
    \label{fig:simulation2_3}
\end{figure}

The MSE results under simulation setting \#1 are presented in Figure \ref{fig:simulation1_3} and the MSE results under simulation setting \# 2 are presented in Figure \ref{fig:simulation2_3}.
Regarding the MSE, all three of our methods exhibit very similar performance, with the most significant benefits observed for small sample sizes. This reflects the finite sample advantage of our methods. In comparison, the random forest weights, Poisson density weights, and logistic class weights have substantially higher biases, which do not improve significantly with larger sample sizes. Although the random forest weights sometimes have the smallest bias with small sample sizes, they appear to become worse with the increase of the sample sizes, indicating considerable model misspecification in this condition.

\begin{figure}[ht]
    \centering
     \includegraphics[width=\textwidth]
     {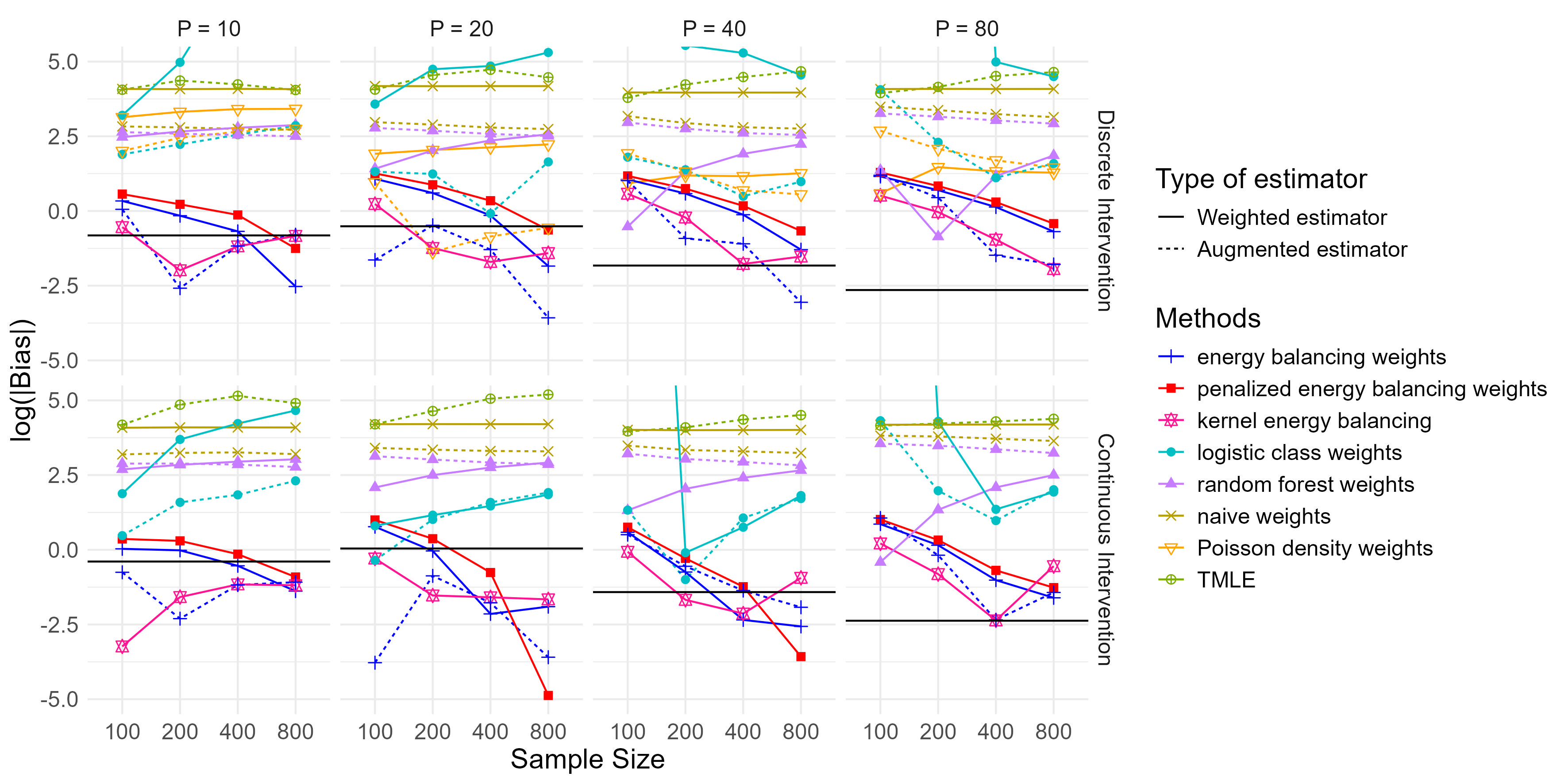}
    \caption{Simulation results for the logarithm of the absolute value of bias across different sample sizes, type of intervention values, and the dimensionality of covariates. The data-generating mechanism is extremely complex (simulation \# 2). Balancing methods are displayed in different colors and shapes of points. Weighted estimators are displayed in solid lines and the augmented estimator are in dashed lines.}
    \label{fig:simulation2_1}
\end{figure}

\begin{figure}[ht]
    \centering
     \includegraphics[width=\textwidth]
    {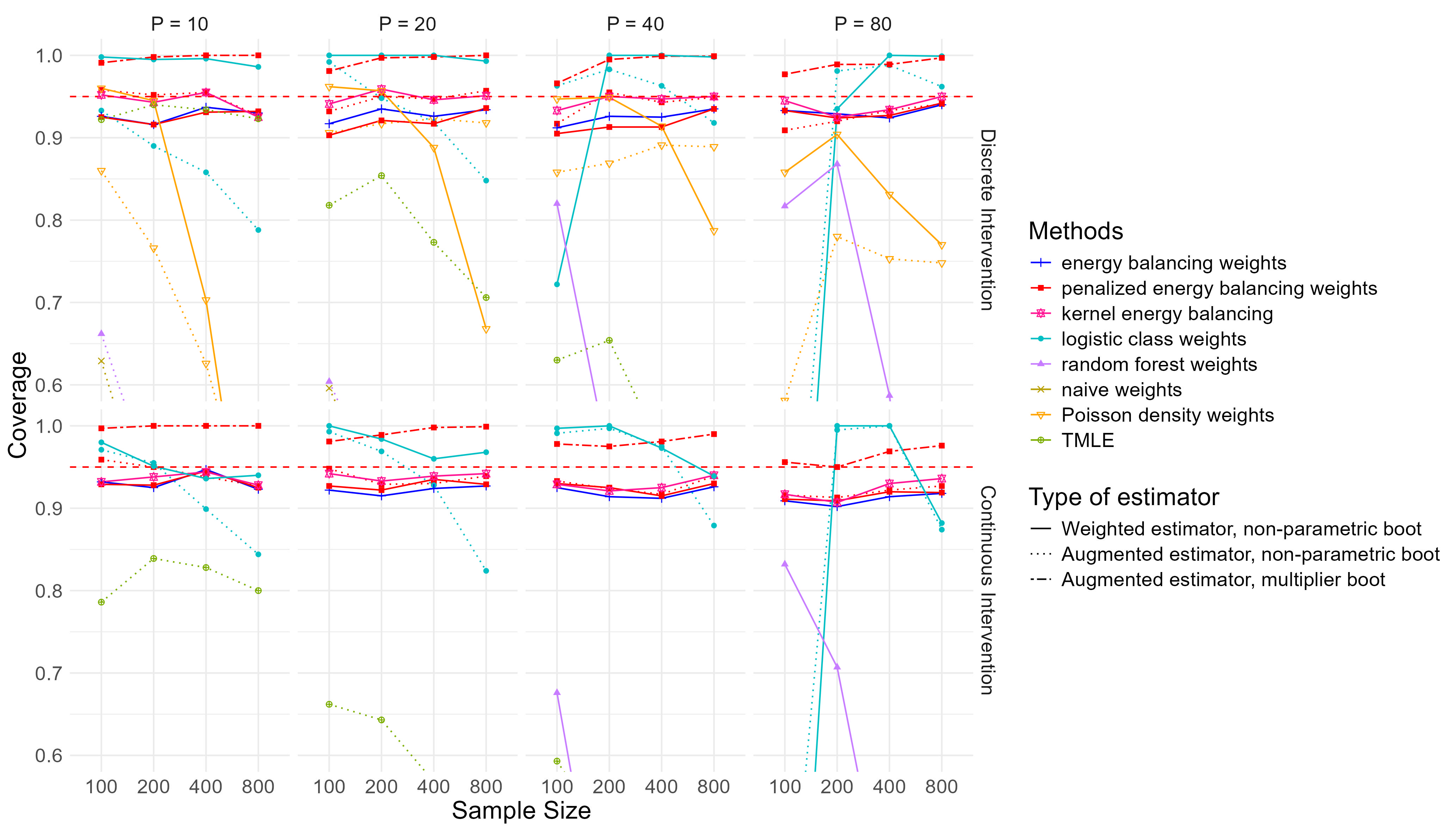}
    \caption{Simulation result for the coverage rate for 95\% confidence intervals for each method across different sample sizes and dimensionality of covariates. The data-generating mechanism is extremely complex (simulation \# 2). The red dashed line indicates the ideal 95\% coverage. The different weighting methods are displayed in different colors and shapes of points. Weighted estimators are displayed in solid lines and the augmented estimators are in dashed lines. The multiplier bootstrap method is displayed in a larger size of red points.}
    \label{fig:simulation2_2}
\end{figure}

The results of the second simulation study, displayed in Figures \ref{fig:simulation2_1}-\ref{fig:simulation2_3}, which has an extremely complex data-generating mechanism are similar to the moderately complex simulation study. Our proposed energy balancing methods consistently outperform other methods in most situations. The random forest method occasionally performs well with a small sample size, but its performance is not stable and seems to get worse with the increase in the sample size. In terms of the coverage rate of the 95\% confidence interval, the three energy balancing methods are closest to the nominal rate which indicates their good performance. Similar to the first simulation study, the logistic classification weights has too wide confidence interval in most situations. Since the data generating mechanism is more complex, the coverage rate of other balancing methods are very low, which further demonstrates the stable performance of our energy balancing methods.

\clearpage

\section{Technical Proofs}\label{suppsec:proof}
\subsection{Identification result}
Here we prove the identification result in Section \ref{sec:2.1}. By definition, we have:
\begin{equation*}
    \begin{split}
        \mu^{q}&:=\int_{(\mathcal{X},\mathcal{A})} \mathbb{E}[Y^{q(\x,a)}|\X=\x,A=a]dF_{\X,A}(\x,a)\\
        &=\int_{(\mathcal{X},\mathcal{A})} \sum_{j=1}^{J(\x)}\int_{a\in I_{j,\x}} \mathbb{E}(Y^{q_j(\x,a)}|\X=x,A=a) dF_{\X,A}(\x,a).
    \end{split}
\end{equation*}
By the definition of the integral, we can substitute the variable $a$ with $h_j(\x,a)$ for each $\int_{a\in I_j(\x)} E(Y^{q_j(\x,a)}|\X=x, A=a)dF(\x,a)$ where $h_j(\x,\cdot)$ be the inverse function of $q_j(\x,\cdot)$. Thus, we have:
\begin{equation*}
    \begin{split}
        \mu^{q}&=\int_{(\mathcal{X},\mathcal{A})} \sum_{j=1}^{J(\x)}\int_{a\in I_{j,\x}} \mathbb{E}(Y^{q_j(\x,a)}|\X=x,A=a) dF_{\X,A}(\x,a)\\
        &=\int_{(\mathcal{X},\mathcal{A})} \sum_{j=1}^{J(\x)}\int_{h_j(\x,a)\in I_{j,\x}} \mathbb{E}(Y^{q_j(\x,h_j(\x,a))}|\X=x,A=h_j(\x,a)) dF_{\X,A}(\x,h_j(\x,a))\\
        &=\int_{(\mathcal{X},\mathcal{A})} \sum_{j=1}^{J(\x)}\int_{h_j(\x,a)\in I_{j,\x}} \mathbb{E}(Y^{a}|\X=x,A=h_j(\x,a)) dF_{\X,A}(\x,h_j(\x,a))\\
        &=\int_{(\mathcal{X},\mathcal{A})} \sum_{j=1}^{J(\x)}\int_{h_j(\x,a)\in I_{j,\x}} \mathbb{E}(Y^{a}|\X=x,A=a) dF_{\X,A}(\x,h_j(\x,a))\\
        &=\int_{(\mathcal{X},\mathcal{A})} \sum_{j=1}^{J(\x)}\int_{h_j(\x,a)\in I_{j,\x}} \mathbb{E}(Y|\X=x,A=a) dF_{\X,A}(\x,h_j(\x,a)),
    \end{split}
\end{equation*}
where the third equality holds because $q_j(\x,h_j(\x,a))=a$. The second to last equation is due to the conditional exchangeability of the related population assumption (A2) which states that $Y^{a}|\X=\x,A=a$ and $Y^{a}|\X=\x,A=a'$ have the same distribution as long as $a=q(\x, a')$. The last equality is valid due to the consistency assumption (A0).

\subsection{Valid CDF function}\label{sec:validcdf}
Here we prove that $F_{\X,A}^{q} (\x,a)$ is the CDF of $(\X,q(\X,A))$.
\begin{proof}
Assume the cumulative distribution function (CDF) and the probability density function of $(\X,A)$ is $F_{\X,A}(\x,a)$ and $f_{\X,A}(\x,a)$, respectively. Since $q(\X,A)=\sum_{j=1}^{J(\X)}I_{j,\X}(A)q_j(\X,A)$ is piece-wise invertible. From the theorem of piece-wise invertible transformation of a random variable, the density function for the random variables $(\X,q(\X,A))$ is
\begin{equation}
    \begin{split}
        f_{\X,Q} (\x,q)= \sum_{j=1}^{J(\x)} I_{j,\x}(h_j(\x,q)) f_{\X,A}(\x,h_j(\x,q)) h'_j(\x,q).
    \end{split}
\end{equation}

On the other hand, by differentiating $F_{\X,\mathbf{A}}^{q} (\x,a)=\sum_{j=1}^{J(\x)} I_{j,\x}(h_j(\x,a)) F_{\X,A}(\x,h_j(\x,a))$, we have
\begin{equation*}
    \begin{split}
        \frac{d}{d\x da}&F_{\X,\mathbf{A}}^{q} (\x,a)=\frac{d}{d\x da} \sum_{j=1}^{J(\x)} I_{j,\x}(h_j(\x,a)) F_{\X,A}(\x,h_j(\x,a))\\
        &=\sum_{j=1}^{J(\x)} I_{j,\x}(h_j(\x,a)) f_{\X,A}(\x,h_j(\x,a)) h'_j(\x,a)\\
        &=f_{\X,Q} (\x,a).
    \end{split}
\end{equation*}
Note that, here, $f_{\X,Q} (\x,a)$ is identical to $f_{\X,Q} (\x,q)$ which is the density function of random variable $(\X,q(\X,A))$. Therefore, in order to prove  $F_{\X,\mathbf{A}}^{q} (\x,a)$ is the CDF of $(\X,q(\X,A))$, we only need to prove the limiting property:
\begin{equation*}
    \begin{split}
         \lim_{(\x,a)\to \boldsymbol{-\infty}} F_{\X,\mathbf{A}}^{q} (\x,a) = \lim_{(\x,a)\to \boldsymbol{-\infty}} \sum_{j=1}^{J(\x)} I_{j,\x}(h_j(\x,a)) F_{\X,A}(\x,h_j(\x,a))=0
    \end{split}
\end{equation*}
and 
\begin{equation*}
    \begin{split}
         \lim_{(\x,a)\to \boldsymbol{\infty}} F_{\X,\mathbf{A}}^{q} (\x,a) = \lim_{(\x,a)\to \boldsymbol{\infty}} \sum_{j=1}^{J(\x)} I_{j,\x}(h_j(\x,a)) F_{\X,A}(\x,h_j(\x,a))=1.
    \end{split}
\end{equation*}
We demonstrate that these limiting properties can be satisfied through standard assumptions, namely that $\lim_{a\to -\infty} q(\x,a)=-\infty$ and $\lim_{a\to \infty} q(\x,a)=\infty$ for $\forall \x$. 

Since $\forall j$, $q_j(\x,a)$ are invertible (strictly increasing or decreasing) functions and $h_j(\x,a)$ is the inverse function of $q_j(\x,a)$ for $\forall \x$, it is easy to prove that $\lim_{(\x,a)\to -\infty}h_j(\x,a)=-\infty$ and $\lim_{(\x,a)\to \infty}h_j(\x,a)=\infty$. Thus, since $F_{\X,A}(\x,a)$ is the CDF of the random vector $(\X,A)$, we have
\begin{equation*}
    \lim_{(\x,a)\to \boldsymbol{-\infty}}I_{j,\x}(h_j(\x,a))F_{\X,A}(\x,h_j(\x,a))=0
\end{equation*}
and
\begin{equation*}
    \lim_{(\x,a)\to \boldsymbol{\infty}}I_{j,\x}(h_j(\x,a))F_{\X,A}(\x,h_j(\x,a))=I_{j,\x}(h_j(\x,a))
\end{equation*}
for all $j$. These limits evaluate to
\begin{equation*}
   \lim_{(\x,a)\to \boldsymbol{-\infty}} F_{\X,\mathbf{A}}^{q} (\x,a) = \lim_{(\x,a)\to \boldsymbol{-\infty}} \sum_{j=1}^{J(\x)} I_{j,\x}(h_j(\x,a)) F_{\X,A}(\x,h_j(\x,a))=0
\end{equation*}
and
\begin{equation*}
      \lim_{(\x,a)\to \boldsymbol{\infty}} F_{\X,\mathbf{A}}^{q} (\x,a) = \lim_{(\x,a)\to \boldsymbol{\infty}} \sum_{j=1}^{J(\x)} I_{j,\x}(h_j(\x,a)) F_{\X,A}(\x,h_j(\x,a))=1.
\end{equation*}
Hence, the theorem is proved.

\end{proof}

\subsection{Proof of (\ref{the:3.1}) in the manuscript section \ref{sec:3} }
\begin{lemma} \label{slemma1}\citep{szekely2007measuring} If $0<\alpha<2$, then for all $\x$ in $\mathbf{R^d}$
 \begin{equation}
     \int_{\mathbb{R}^d}\frac{1-\cos(\boldsymbol{t}\cdot\x)}{|\boldsymbol{t}|_d^{d+\alpha}}d\boldsymbol{t}=C(d,\alpha)||\x||^\alpha
 \end{equation}
 where 
 \begin{equation*}
     C(d,\alpha)=\frac{2\pi^{d/2}\Gamma(1-\alpha/2)}{\alpha2^{\alpha}\Gamma((d+\alpha)/2)}.
 \end{equation*}
\end{lemma}
\begin{proof}

To prove (\ref{the:3.1}), we follow the proof from \citep{huling2020energy} Proposition 1 and the duality results of Theorem 1 of \citep{szekely2007measuring} with the definition of the energy distance. First, note that we have
\begin{equation*}
    \begin{split}
        |\varphi_{n,\w,\X,A}(\boldsymbol{t})-\varphi_{n,\X,A}^{q}(\boldsymbol{t})|^2 &=\varphi_{n,\w,\X,A}(\boldsymbol{t})\overline{\varphi_{n,\w,\X,A}(\boldsymbol{t})}-\varphi_{n,\w,\X,A}(\boldsymbol{t})\overline{\varphi_{n,\X,A}^{q}(\boldsymbol{t})}\\
        &-\varphi_{n,\X,A}^{q}(\boldsymbol{t})\overline{\varphi_{n,\w,\X,A}(\boldsymbol{t})}+\varphi_{n,\X,A}^{q}(\boldsymbol{t})\overline{\varphi_{n,\X,A}^{q}(\boldsymbol{t})},
    \end{split}
\end{equation*}
where the $\overline{\varphi_{n,\w,\X,A}(\boldsymbol{t})}$, and $\overline{\varphi_{n,\X,A}^{q}(\boldsymbol{t})}$ are the complex conjugates of ${\varphi_{n,\w,\X,A}(\boldsymbol{t})}$, and ${\varphi_{n,\X,A}^{q}(\boldsymbol{t})}$, respectively.
For the first component, we have
\begin{equation*}
    \begin{split}
        \varphi_{n,\w,\X,A}(\boldsymbol{t})\overline{\varphi_{n,\w,\X,A}(\boldsymbol{t})}&=\frac{1}{n^2}\sum_{i,k}w_i w_k \exp\{\boldsymbol{i}\boldsymbol{t}\cdot(\X_i-\X_k,A_i-A_k)\}\\
        &=\frac{1}{n^2}\sum_{i,k}w_i w_k \cos\{\boldsymbol{t}\cdot(\X_i-\X_k,A_i-A_k)\}+V_1
    \end{split}
\end{equation*}
where $V_1$ is an odd function which is zero when integrated over $\mathbb{R}^d$.

Similarly, since
\begin{equation}
    \varphi_{n,\X,A}^{q}(\boldsymbol{t})=\frac{1}{n}\sum_{i=1}^n\exp(\boldsymbol{i}\boldsymbol{t}^T(\X_i,Q_i)),
\end{equation}
where $Q_i=q(\X_i,A_i)$, we have
\begin{equation*}
    \begin{split}
        \varphi_{n,\X,A}^{q}(\boldsymbol{t})\overline{\varphi_{n,\w,\X,A}(\boldsymbol{t})}&=\frac{1}{n^2}\sum_{i=1}^n \sum_{k=1}^n w_i  \cos\{\boldsymbol{t}\cdot(\X_i-\X_k,A_i-Q_{k})\}+V_2
    \end{split}
\end{equation*}
and 
\begin{equation*}
    \begin{split}
        &\varphi_{n,\X,A}^{q}(\boldsymbol{t})\overline{\varphi_{n,\X,A}^{q}(\boldsymbol{t})}=\frac{1}{n^2}\sum_{i=1}^n \sum_{k=1}^n  \cos\{\boldsymbol{t}\cdot(\X_i-\X_k,Q_i-Q_k)\}+V_3,
    \end{split}
\end{equation*}
where $V_2$ and $V_3$ are also the odd functions that will evaluate to zero after integration over $\boldsymbol{t}$. Note that, due to the cancellation of a common term, we have:
\begin{equation*}
    \begin{split}
        &\varphi_{n,\w,\X,A}(\boldsymbol{t})\overline{\varphi_{n,\w,\X,A}(\boldsymbol{t})}-\varphi_{n,\w,\X,A}(\boldsymbol{t})\overline{\varphi_{n,\X,A}^{q}(\boldsymbol{t})}\\
        =&\frac{1}{n^2}\sum_{i=1}^n\sum_{k=1}^nw_i w_k \cos\{\boldsymbol{t}\cdot(\X_i-\X_k,A_i-A_k)\}\\
        &-\frac{1}{n^2}\sum_{i=1}^n \sum_{k=1}^n  w_i  \cos\{\boldsymbol{t}\cdot(\X_i-\X_k,A_i-Q_k)\}+V_1-V_2\\
        =&-\frac{1}{n^2}\sum_{i=1}^n\sum_{k=1}^nw_i w_k \bigg(1-\cos\{\boldsymbol{t}\cdot(\X_i-\X_k,A_i-A_k)\}\bigg)\\
        &+\frac{1}{n^2}\sum_{i=1}^n \sum_{k=1}^n  w_i  \bigg(1-\cos\{\boldsymbol{t}\cdot(\X_i-\X_k,A_i-Q_k)\}\bigg)+V_1-V_2.
    \end{split}
\end{equation*}
The last equation is due to the cancellation, as the summation of the weights are the same (note that $\sum_{i=1}^n w_i=n$):
\begin{equation*}
    \sum_{i=1}^n\sum_{k=1}^n w_i w_k=\sum_{i=1}^n \sum_{k=1}^n  w_i.
\end{equation*}
Similarly, we have:
\begin{equation*}
    \begin{split}
        &-\varphi_{n,\X,A}^{q}(\boldsymbol{t})\overline{\varphi_{n,\w,\X,A}(\boldsymbol{t})}+\varphi_{n,\X,A}^{q}(\boldsymbol{t})\overline{\varphi_{n,\X,A}^{q}(\boldsymbol{t})}\\
        =& \frac{1}{n^2}\sum_{i=1}^n \sum_{k=1}^n  w_i  \bigg(1-\cos\{\boldsymbol{t}\cdot(\X_i-\X_k,A_i-Q_k)\}\bigg) \\
        & - \frac{1}{n^2}\sum_{i=1}^n \sum_{k=1}^n   \cos\{\boldsymbol{t}\cdot(\X_i-\X_k,Q_i-Q_k)\} +V2-V3.
    \end{split}
\end{equation*}

Thus, when we integrate over $\mathbb{R}^d$, we have
\begin{equation*}
    \begin{split}
        &\int_{\mathbb{R}^p} \bigg( \varphi_{n,\w,\X,A}(\boldsymbol{t})\overline{\varphi_{n,\w,\X,A}(\boldsymbol{t})}-\varphi_{n,\w,\X,A}(\boldsymbol{t})\overline{\varphi_{n,\X,A}^{q}(\boldsymbol{t})}\bigg)v(\boldsymbol{t})d\boldsymbol{t}\\
        =&\int_{\mathbb{R}^p} -\frac{1}{n^2}\sum_{i=1}^n\sum_{k=1}^nw_i w_k \bigg(1-\cos\{\boldsymbol{t}\cdot(\X_i-\X_k,A_i-A_k)\}\bigg)v(\boldsymbol{t})d\boldsymbol{t}\\
        &+\int_{\mathbb{R}^p}\frac{1}{n^2}\sum_{i=1}^n \sum_{k=1}^n  w_i  \bigg(1-\cos\{\boldsymbol{t}\cdot(\X_i-\X_k,A_i-Q_k)\}\bigg)v(\boldsymbol{t})d\boldsymbol{t} \\
        =&-\frac{1}{n^2}\sum_{i=1}^n \sum_{k=1}^n w_i w_k ||(\X_i-\X_k,A_i-A_k)||_2+\frac{1}{n^2}\sum_{i=1}^n \sum_{k=1}^n  w_i  ||(\X_i-\X_k,A_i-Q_k)||_2\\
        =&-\frac{1}{n^2}\sum_{i=1}^n \sum_{k=1}^n w_i w_k ||(\X_i,A_i)-(\X_k,A_k)||_2+\frac{1}{n^2}\sum_{i=1}^n \sum_{k=1}^n  w_i  ||(\X_i, A_i)-(\X_k,Q_k)||_2.
    \end{split}
\end{equation*}
Note that the second equality is due to the result from Lemma \ref{slemma1}. Similarly, we have 
\begin{equation*}
    \begin{split}
        &\int_{\mathbb{R}^p} \bigg( -\varphi_{n,\X,A}^{q}(\boldsymbol{t})\overline{\varphi_{n,\w,\X,A}(\boldsymbol{t})}+\varphi_{n,\X,A}^{q}(\boldsymbol{t})\overline{\varphi_{n,\X,A}^{q}(\boldsymbol{t})}\bigg)v(\boldsymbol{t})d\boldsymbol{t}\\
        =&\int_{\mathbb{R}^p} \frac{1}{n^2}\sum_{i=1}^n \sum_{k=1}^n  w_i  \bigg(1-\cos\{\boldsymbol{t}\cdot(\X_i-\X_k,A_i-Q_k)\}\bigg)v(\boldsymbol{t})d\boldsymbol{t}\\
        &- \frac{1}{n^2}\sum_{i=1}^n \sum_{k=1}^n  \cos\{\boldsymbol{t}\cdot(\X_i-\X_k,Q_i-Q_k)\} v(\boldsymbol{t})d\boldsymbol{t} \\
        =&\frac{1}{n^2}\sum_{i=1}^n \sum_{k=1}^n  w_i  ||(\X_i, A_i)-(\X_k,Q_k)||_2-\frac{1}{n^2}\sum_{i=1}^n \sum_{k=1}^n ||(\X_i, Q_i)-(\X_k,Q_k)||_2.
    \end{split}
\end{equation*}

Therefore, by adding the two components in $\int_{\mathbb{R}^p} |\varphi_{n,\w,\X,A}(\boldsymbol{t})-\varphi_{n,\X,A}^{q}(\boldsymbol{t})|^2v(\boldsymbol{t})d\boldsymbol{t}$ and using the definition of $\mathcal{E}(F_{n,\w,\X,A},F_{n,\X,A}^{q})$, (\ref{the:3.1}) is proved.

\end{proof}
\subsection{Proof of Theorem \ref{the:3.2} in Main Manuscript Section \ref{sec:3}}
\begin{proof}
let $\{\Tilde{\X}_i,\Tilde{A}_i\}_{i=1}^n\stackrel{i.i.d}{\sim}F_{n,\w,\X,A}$ and let $\Tilde{F}_{n,\w,\X,A}$ and $\Tilde{\varphi}_{n,\w,\X,A}$ be the empirical cdf and characteristic function of $\{\Tilde{\X}_i,\Tilde{A}_i\}_{i=1}^n$. Same as \citep{huling2020energy}, by the Glivenko-Cantelli theorem for non-identically distributed random variables, we have
\begin{equation*}
    \lim_{n\to\infty} sup_{\x\in\mathcal{X},a\in \mathcal{A}} |\Tilde{F}_{n,\w,\X,A}(\x,a)-\Tilde{F}_{\X,A}(\x,a)|=0.
\end{equation*}
Now, we want to show
\begin{equation}\label{19}
    \lim_{n\to\infty}\mathcal{E}(\Tilde{F}_{n,\w,\X,A},F_{n,\X,A}^{q})=\mathcal{E}(\Tilde{F}_{\X,A},F_{\X,A}^{q}).
\end{equation}
Similar to \citep{huling2020energy} and \citep{szekely2007measuring}, define $D(\epsilon)=\{\boldsymbol{t}\in\mathbb{R}^d:\epsilon\leq|\boldsymbol{t}|_d\leq1/\epsilon\}$ and 
\begin{equation*}
    \mathcal{E}_{\epsilon}(\Tilde{F}_{n,\w,\X,A},F_{n,\X,A}^{q})=\int_{D(\epsilon)}|\Tilde{\varphi}_{n,\w,\X,A}(\boldsymbol{t})-\varphi_{n,\X,A}^{q}(\boldsymbol{t})|^2 v(\boldsymbol{t})d\boldsymbol{t}.
\end{equation*}
Then, by the strong law of large numbers for V-statistics 
\citep{csorgHo2013asymptotics,patterson1989strong} 
we have that the following holds almost surely
\begin{equation}
    \begin{split}
        \lim_{n\to\infty}\mathcal{E}_{\epsilon}(\Tilde{F}_{n,\w,\X,A},F_{n,\X,A}^{q})&=\mathcal{E}_{\epsilon}(\Tilde{F}_{\X,A},F_{\X,A}^{q})\\
        &=\int_{D(\epsilon)}|\Tilde{\varphi}_{\X,A}(\boldsymbol{t})-{\varphi}_{\X,A}^{q}(\boldsymbol{t})|^2 v(\boldsymbol{t})d\boldsymbol{t}.
    \end{split}
\end{equation}
Since we have $\lim_{\epsilon\to 0} \mathcal{E}_{\epsilon}(\Tilde{F}_{n,\w,\X,A},F_{n,\X,A}^{q})=\mathcal{E}(\Tilde{F}_{n,\w,\X,A},F_{n,\X,A}^{q})$, in order to prove (\ref{19}), we must show that
\begin{equation}
    \limsup_{\epsilon\to 0}\limsup_{n\to\infty}|\mathcal{E}_{\epsilon}(\Tilde{F}_{n,\w,\X,A},F_{n,\X,A}^{q})-\mathcal{E}(\Tilde{F}_{n,\w,\X,A},F_{n,\X,A}^{q})|=0.
\end{equation}
Following the proof of \citep{huling2020energy}, note that for each $\epsilon>0$, we have
\begin{equation*}
    \begin{split}
        \left|\mathcal{E}_{\epsilon}(\Tilde{F}_{n,\w,\X,A},F_{n,\X,A}^{q})-\mathcal{E}(\Tilde{F}_{n,\w,\X,A},F_{n,\X,A}^{q})\right|\leq & \int_{|\boldsymbol{t}|_p<\epsilon}\left|\Tilde{\varphi}_{n,\w,\X,A}(\boldsymbol{t})-{\varphi}_{n,\X,A}^{q}(\boldsymbol{t})\right|^2 v(\boldsymbol{t})d\boldsymbol{t}\\
        &+\int_{|\boldsymbol{t}|_p>1/\epsilon}\left|\Tilde{\varphi}_{n,\w,\X,A}(\boldsymbol{t})-{\varphi}_{n,\X,A}^{q}(\boldsymbol{t})\right|^2 v(\boldsymbol{t})d\boldsymbol{t}.
    \end{split}
\end{equation*}
Since,
\begin{equation*}
    \begin{split}
        &|\Tilde{\varphi}_{n,\w,\X,A}(\boldsymbol{t})-{\varphi}_{n,\X,A}^{q}(\boldsymbol{t})|^2\\
        &=\bigg|\frac{1}{n}\sum_{i=1}^n \exp(i\boldsymbol{t}^T\cdot(\Tilde{\X}_i,\Tilde{A}_i))-\frac{1}{n}\sum_{i=1}^n\sum_{j=1}^{J(\X_i)}I_{j,\X_i}(A_i)\exp(i\boldsymbol{t}^T\cdot(\X_i,q_j(\X_i,A_i)))\bigg|^2\\
        &=\bigg|\frac{1}{n}\sum_{i=1}^n(1- e^{i\boldsymbol{t}^T\cdot(\Tilde{\X}_i,\Tilde{A}_i)})-\frac{1}{n}\sum_{i=1}^n(1-\sum_{j=1}^{J(\X_i)}I_{j,\X_i}(A_i)e^{i\boldsymbol{t}^T\cdot(\X_i,q_j(\X_i,A_i))})\bigg|^2\\
        &=\bigg|\frac{1}{n}\sum_{i=1}^n(1- e^{i\boldsymbol{t}^T\cdot(\Tilde{\X}_i,\Tilde{A}_i)})-\frac{1}{n}\sum_{i=1}^n(1-e^{i\boldsymbol{t}^T\cdot(\X_i,Q_i)})\bigg|^2\\
        &\leq \frac{1}{n}\sum_{i=1}^n|1- e^{i\boldsymbol{t}^T\cdot(\Tilde{\X}_i,\Tilde{A}_i)}|^2+\frac{1}{n}\sum_{i=1}^n|1-e^{i\boldsymbol{t}^T\cdot(\X_i,Q_i)}|^2,
    \end{split}
\end{equation*}
where $Q_i=\sum_{j=1}^{J(\X_i)}I_{j,\X_i}(A_i) q_j(\X_i,A_i)$. Thus, we have
\begin{equation*}
    \begin{split}
        \int_{|\boldsymbol{t}|_p<\epsilon}|\Tilde{\varphi}_{n,\w,\X,A}(\boldsymbol{t})-{\varphi}_{n,\X,A}^{q}(\boldsymbol{t})|^2 v(\boldsymbol{t})d\boldsymbol{t} &\leq \frac{1}{n}\sum_{i=1}^n \int_{|\boldsymbol{t}|_p<\epsilon} |1- e^{i\boldsymbol{t}^T\cdot(\Tilde{\X}_i,\Tilde{A}_i)}|^2 v(\boldsymbol{t})d\boldsymbol{t}\\
        &+ \frac{1}{n}\sum_{i=1}^n \int_{|\boldsymbol{t}|_p<\epsilon} |1- e^{i\boldsymbol{t}^T\cdot(\X_i,Q_i)}|^2 v(\boldsymbol{t})d\boldsymbol{t}.
    \end{split}
\end{equation*}
Following the proof of Theorem 2 of \citep{szekely2007measuring}, define $G(\y)=\int_{|\boldsymbol{t}|_p<\y} \frac{1-\cos(t_1)}{|\boldsymbol{t}|^{1+p}}d\boldsymbol{t}$ where $t_1$ is the first element of $\boldsymbol{t}$, we have the following identity
\begin{equation*}
    \int_{|\boldsymbol{t}|_p<\epsilon} |1- e^{i\boldsymbol{t}^T\cdot(\Tilde{\X}_i,\Tilde{A}_i)}|^2 v(\boldsymbol{t})d\boldsymbol{t}=|(\Tilde{\X_i},\Tilde{A}_i)|G((\Tilde{\X_i},\Tilde{A}_i)\epsilon).
\end{equation*}
Note that $G(\y)$ is a function such that $\lim_{\y\to0}G(\y)=0$ and $G(\y)$ is bounded. Thus, by the strong law of large numbers, we have 
\begin{equation*}
   \begin{split}
        &\limsup_{n\to\infty} \int_{|\boldsymbol{t}|_p<\epsilon}|\Tilde{\varphi}_{n,\w,\X,A}(\boldsymbol{t})-{\varphi}_{n,\X,A}^{q}(\boldsymbol{t})|^2 v(\boldsymbol{t})d\boldsymbol{t} \\
        &\leq \mathbb{E}\{|(\Tilde{\X_i},\Tilde{A}_i)|G((\Tilde{\X_i},\Tilde{A}_i)\epsilon)\}+\mathbb{E}\{|({\X_i},{Q}_i)|G(({\X_i},{Q}_i)\epsilon)\}.
   \end{split}
\end{equation*}
Thus, by the Lebesgue bounded convergence theorem for integrals and expectations, we have 
\begin{equation}
    \limsup_{\epsilon\to0}\limsup_{n\to\infty} \int_{|\boldsymbol{t}|_p<\epsilon}|\Tilde{\varphi}_{n,\w,\X,A}(\boldsymbol{t})-{\varphi}_{n,\X,A}^{q}(\boldsymbol{t})|^2 v(\boldsymbol{t})d\boldsymbol{t}=0.
\end{equation}
By similar arguments, we have 
\begin{equation}
    \limsup_{\epsilon\to0}\limsup_{n\to\infty} \int_{|\boldsymbol{t}|_p>1/\epsilon}|\Tilde{\varphi}_{n,\w,\X,A}(\boldsymbol{t})-{\varphi}_{n,\X,A}^{q}(\boldsymbol{t})|^2 v(\boldsymbol{t})d\boldsymbol{t}=0,
\end{equation}
which proves (\ref{19}). To complete the proof, it remains to show that
\begin{equation}
    \limsup_{n\to\infty}|\mathcal{E}(\Tilde{F}_{n,\w,\X,A},F_{n,\X,A}^{q})-\mathcal{E}({F}_{n,\w,\X,A},F_{n,\X,A}^{q})|=0.
\end{equation}
We can decompose the above as 
\begin{equation*}
    \begin{split}
        &|\mathcal{E}(\Tilde{F}_{n,\w,\X,A},F_{n,\X,A}^{q})-\mathcal{E}({F}_{n,\w,\X,A},F_{n,\X,A}^{q})|\\
        =&\bigg|\int_{\mathbb{R}^p}\{2{\varphi}_{n,\X,A}^{q}(\boldsymbol{t})[\varphi_{n,\w,\X,A}(\boldsymbol{t})-\Tilde{\varphi}_{n,\w,\X,A}(\boldsymbol{t})]+\Tilde{\varphi}^2_{n,\w,\X,A}(\boldsymbol{t})-\varphi^2_{n,\w,\X,A}(\boldsymbol{t})\}dv\bigg|\\
        \leq& 2 \int_{\mathbb{R}^p} |{\varphi}_{n,\X,A}^{q}(\boldsymbol{t})|\bigg\{|\Tilde{\varphi}_{\X,A}-\varphi_{n,\w,\X,A}(\boldsymbol{t})|+|\Tilde{\varphi}_{\X,A}-\Tilde{\varphi}_{n,w,\X,A}(\boldsymbol{t})|\bigg\}dv\\
        & +\int_{\mathbb{R}^p} |\Tilde{\varphi}_{\X,A}+\Tilde{\varphi}_{n,\w,\X,A}(\boldsymbol{t})||\Tilde{\varphi}_{\X,A}-\Tilde{\varphi}_{n,\w,\X,A}(\boldsymbol{t})|dv\\
        & +\int_{\mathbb{R}^p} |\Tilde{\varphi}_{\X,A}+{\varphi}_{n,\w,\X,A}(\boldsymbol{t})||\Tilde{\varphi}_{\X,A}-{\varphi}_{n,\w,\X,A}(\boldsymbol{t})|dv\\
        =& \int_{\mathbb{R}^p}\bigg\{ 2|{\varphi}_{n,\X,A}^{q}(\boldsymbol{t})|+|\Tilde{\varphi}_{\X,A}+\Tilde{\varphi}_{n,\w,\X,A}(\boldsymbol{t})| \bigg\}|\Tilde{\varphi}_{\X,A}-\Tilde{\varphi}_{n,\w,\X,A}(\boldsymbol{t})|dv\\
        & + \int_{\mathbb{R}^p}\bigg\{2|{\varphi}_{n,\X,A}^{q}(\boldsymbol{t})|+|\Tilde{\varphi}_{\X,A}+{\varphi}_{n,\w,\X,A}(\boldsymbol{t})| \bigg\}|\Tilde{\varphi}_{\X,A}-{\varphi}_{n,\w,\X,A}(\boldsymbol{t})|dv\\
        \leq & \int_{\mathbb{R}^p}\bigg\{ 2|{\varphi}_{\X,A}^{q}|+2|{\varphi}_{\X,A}^{q}-{\varphi}_{n,\X,A}^{q}|+2|\Tilde{\varphi}_{\X,A}|+|\Tilde{\varphi}_{\X,A}-\Tilde{\varphi}_{n,\w,\X,A}| \bigg\}|\Tilde{\varphi}_{\X,A}-\Tilde{\varphi}_{n,\w,\X,A}|dv\\
        & + \int_{\mathbb{R}^p}\bigg\{2|{\varphi}_{\X,A}^{q}|+2|{\varphi}_{\X,A}^{q}-{\varphi}_{n,\X,A}^{q}|+2|\Tilde{\varphi}_{\X,A}|+|\Tilde{\varphi}_{\X,A}-{\varphi}_{n,\w,\X,A}| \bigg\}|\Tilde{\varphi}_{\X,A}-{\varphi}_{n,\w,\X,A}|dv,
    \end{split}
\end{equation*}
where ${\varphi}_{\X,A}^{q}$ is the characteristic function of $F_{\X,A}^{q}$, since $F_{n,\X,A}^{q} \stackrel{a.s.}{\to} F_{\X,\mathbf{A}}^{q}$, we have the convergence of the characteristic function: ${\varphi}_{n,\X,A}^{q}\to{\varphi}_{\X,A}^{q} $. Also, we have the almost sure convergence as $n\to\infty$ that $\varphi_{n,\w,\X,A}\to \Tilde{\varphi}_{\X,A} $ and $\Tilde{\varphi}_{n,\w,\X,A}\to \Tilde{\varphi}_{\X,A} $. Thus, let $g_n=2|{\varphi}_{\X,A}^{q}|+2|{\varphi}_{\X,A}^{q}-{\varphi}_{n,\X,A}^{q}|+2|\Tilde{\varphi}_{\X,A}|+|\Tilde{\varphi}_{\X,A}-\Tilde{\varphi}_{n,\w,\X,A}|$, we have its almost sure limit $g=2\{|{\varphi}_{\X,A}^{q}|+|\Tilde{\varphi}_{\X,A}|\}$.

We have the following inequality, 
\begin{equation*}
    \begin{split}
        0\leq & 2\bigg\{ 2|{\varphi}_{\X,A}^{q}|+2|{\varphi}_{\X,A}^{q}-{\varphi}_{n,\X,A}^{q}|+2|\Tilde{\varphi}_{\X,A}|+|\Tilde{\varphi}_{\X,A}-\Tilde{\varphi}_{n,\w,\X,A}| \bigg\}\{|\Tilde{\varphi}_{\X,A}|+|\Tilde{\varphi}_{n,\w,\X,A}|\}\\
        &-\bigg\{ 2|{\varphi}_{\X,A}^{q}|+2|{\varphi}_{\X,A}^{q}-{\varphi}_{n,\X,A}^{q}|+2|\Tilde{\varphi}_{\X,A}|+|\Tilde{\varphi}_{\X,A}-\Tilde{\varphi}_{n,\w,\X,A}| \bigg\}|\Tilde{\varphi}_{\X,A}-\Tilde{\varphi}_{n,\w,\X,A}|\\
        &= 2 g_n \{|\Tilde{\varphi}_{\X,A}|+|\Tilde{\varphi}_{n,\w,\X,A}|\} -g_n|\Tilde{\varphi}_{\X,A}-\Tilde{\varphi}_{n,\w,\X,A}|.
    \end{split}
\end{equation*}
Then, by Fatou's lemma, we have
\begin{equation*}
    \begin{split}
        4\int_{\mathbb{R}^p}g(\boldsymbol{t})|\Tilde{\varphi}_{\X,A}|dv\leq  & \liminf_{n\to\infty}\bigg\{2\int_{\mathbb{R}^p}g_n(\boldsymbol{t})|\Tilde{\varphi}_{n,\w,\X,A}|dv+2\int_{\mathbb{R}^p}g_n(\boldsymbol{t})|\Tilde{\varphi}_{\X,A}|dv\\
        &-\int_{\mathbb{R}^p}g_n(\boldsymbol{t})|\Tilde{\varphi}_{\X,A}-\Tilde{\varphi}_{n,\w,\X,A}|dv\bigg\}\\
        =& 4\int_{\mathbb{R}^p}g(\boldsymbol{t})|\Tilde{\varphi}_{\X,A}|dv-\limsup_{n\to\infty}\int_{\mathbb{R}^p}g_n(\boldsymbol{t})|\Tilde{\varphi}_{\X,A}-\Tilde{\varphi}_{n,\w,\X,A}|dv.
    \end{split}
\end{equation*}
Thus, we have $\limsup_{n\to\infty}\int_{\mathbb{R}^p}g_n(\boldsymbol{t})|\Tilde{\varphi}_{\X,A}-\Tilde{\varphi}_{n,\w,\X,A}|dv=0$. Similarly, we can prove that
\begin{equation}
    \limsup_{n\to\infty}\int_{\mathbb{R}^p}g_n(\boldsymbol{t})|\Tilde{\varphi}_{\X,A}-{\varphi}_{n,\w,\X,A}|dv=0,
\end{equation}
which concludes the proof.
\end{proof}

\subsection{Proof of Supplementary Material Theorem \ref{othe:5.1}}
\begin{proof}
We first prove the asymptotic property for the unpenalized energy balancing estimator. By \citep{amaral2017optimal} and \citep{huling2020energy}, the weights defined by the Radon-Nikodym derivative $\h(\x,a)=\frac{f^{q}_{\X,A}(\x,a)}{f_{\X,A}(\x,a)}$ is existed (although can not be determined). We then define
\begin{equation}
    \hat{h}(\X_i,A_i)=\frac{h(\X_i,A_i)}{\frac{1}{n}\sum_{j=1}^nh(\X_j,A_j)}
\end{equation}
and $\hat{\h}=(\hat{h}(\X_1,A_1),...,\hat{h}(\X_n,A_n))$ as the normalized Radon Nikodym weights.
Since we assume the positivity assumption holds, we have the absolute continuity of the target measure $F_{\X,A}^{q}$ with respect to the proposal measure $F_{\X,A}$, (same assumption as \citep{amaral2017optimal}).
Then, by the strong law of large numbers (SLLN), the Radon-Nikodym weights has the property that
\begin{equation}
    \lim_{n\to\infty} F_{n,\hat{\h}, \X,A}(\x,a)=F_{\X,A}^{q}(\x,a)
\end{equation}
almost everywhere for every continuity point $(\x,a)$. As in the proof of Theorem 2 in \citep{mak2018support} and Theorem 3.1 in \citep{huling2020energy}, by the Portmanteau and Lebesgue dominated convergence theorems, we have
\begin{equation}
    \lim_{n\to\infty} \mathbb{E}[|\varphi_{\X,A}^{q}(\boldsymbol{t})-\varphi_{n,\hat{\h},\X,A}(\boldsymbol{t})|^2]=0
\end{equation}
for all $\boldsymbol{t}$ and $\varphi_{n,\hat{\h},\X,A}(\boldsymbol{t})$ be the weighted ECHF with the Radon-Nikodym derivative weights.

Denote the expected weighted energy as 
\begin{equation}
   \begin{split}
        &\mathbb{E}[\mathcal{E}(F_{n,\hat{\h}, \X,A},F_{n,\X,A}^{q})]=\mathbb{E}\bigg[\int_{\mathbb{R}^p}|\varphi_{n,\X,A}^{q}(\boldsymbol{t})-\varphi_{n,\hat{\h},\X,A}(\boldsymbol{t})|^2v(\boldsymbol{t})d\boldsymbol{t}\bigg]\\
        &=\mathbb{E}\bigg[\int_{\mathbb{R}^p}|\varphi_{n,\X,A}^{q}(\boldsymbol{t})-\varphi_{\X,A}^{q}(\boldsymbol{t})+\varphi_{\X,A}^{q}(\boldsymbol{t})-\varphi_{n,\hat{\h},\X,A}(\boldsymbol{t})|^2v(\boldsymbol{t})d\boldsymbol{t}\bigg]\\
        &\leq \mathbb{E}\bigg[\int_{\mathbb{R}^p}|\varphi_{n,\X,A}^{q}(\boldsymbol{t})-\varphi_{\X,A}^{q}(\boldsymbol{t})|^2v(\boldsymbol{t})d\boldsymbol{t}+\int_{\mathbb{R}^p}|\varphi_{\X,A}^{q}(\boldsymbol{t})-\varphi_{n,\hat{\h},\X,A}(\boldsymbol{t})|^2v(\boldsymbol{t})d\boldsymbol{t}\bigg],
   \end{split}
\end{equation}
where the first inequality holds by the Minkowski inequality. Due to the Glivenko-Cantelli theorem, the first component of the integral converges to 0 as $n$ becomes infinite. Thus, by the same arguments as in \citep{mak2018support} and \citep{huling2020energy}, we have the following limiting property:
\begin{equation*}
    \lim_{n\to\infty}\mathbb{E}[\mathcal{E}(F_{n,\hat{\h}, \X,A},F_{n,\X,A}^{q})]=0.
\end{equation*}

Define $\varphi_{n,\w^e_n,\X,A}(\boldsymbol{t})=\frac{1}{n}\sum_{i=1}^n w_i^e \exp\{i\boldsymbol{t}\cdot(\X_i,A_i)\}$ as the weighted empirical characteristic function with energy balancing weights. By the definition of $\w_n^e$,
\begin{equation*}
    \begin{split}
        &\int_{\mathbb{R}^p}|\varphi_{\X,A}^{q}(\boldsymbol{t})-\varphi_{n,\w_n^e,\X,A}(\boldsymbol{t})|^2v(\boldsymbol{t})d\boldsymbol{t}\\
        &\leq \bigg(\bigg[\int_{\mathbb{R}^p}|\varphi_{n,\X,A}^{q}(\boldsymbol{t})-\varphi_{n,\w_n^e,\X,A}(\boldsymbol{t})|^2v(\boldsymbol{t})d\boldsymbol{t}\bigg]^{\frac{1}{2}}+\bigg[\int_{\mathbb{R}^p}|\varphi_{n,\X,A}^{q}(\boldsymbol{t})-\varphi_{\X,A}^{q}(\boldsymbol{t})|^2v(\boldsymbol{t})d\boldsymbol{t}\bigg]^{\frac{1}{2}}\bigg)^2\\
        &=\bigg([\mathcal{E}(F_{n,\X,A}^{q},F_{n,\w_n^e,\X,A})]^{\frac{1}{2}}+\bigg[\int_{\mathbb{R}^p}|\varphi_{n,\X,A}^{q}(\boldsymbol{t})-\varphi_{\X,A}^{q}(\boldsymbol{t})|^2v(\boldsymbol{t})d\boldsymbol{t}\bigg]^{\frac{1}{2}}\bigg)^2\\
        &\leq \bigg(\bigg[\mathbb{E}[\mathcal{E}(F_{n,\X,A}^{q},F_{n,\hat{\h},\X,A})]\bigg]^{\frac{1}{2}}+\bigg[\int_{\mathbb{R}^p}|\varphi_{n,\X,A}^{q}(\boldsymbol{t})-\varphi_{\X,A}^{q}(\boldsymbol{t})|^2v(\boldsymbol{t})d\boldsymbol{t}\bigg]^{\frac{1}{2}}\bigg)^2,
    \end{split}
\end{equation*}
where the first inequality holds by the Minkowski inequality. Due to the fact that 
\begin{equation*}
    \lim_{n\to\infty}\int_{\mathbb{R}^p}|\varphi_{n,\X,A}^{q}(\boldsymbol{t})-\varphi_{\X,A}^{q}(\boldsymbol{t})|^2v(\boldsymbol{t})d\boldsymbol{t}=0
\end{equation*}
holds almost surely, we have:
\begin{equation}
    \lim_{n\to\infty}\mathcal{E}(F_{n,\w_n^e,\X,A},F_{\X,A}^{q})=\lim_{n\to\infty}\mathcal{E}(F_{n,\w_n^e,\X,A},F_{n,\X,A}^{q})=0.
\end{equation}

For any subsequence $\{n_k\}_{k=1}^{\infty}$ of $\mathbf{N}_+$, we have the same that  $\lim_{n\to\infty}\mathcal{E}(F_{n_k,\w_{n_k}^e,\X,A},F_{\X,A}^{q}) = 0$. Following \citep{huling2020energy}, by the Riesz-Fischer Theorem, a sequence of functions $f_n$ which converge to $f$ in $L_2$ norm has a subsequence $f_{n_k}$ which converges almost everywhere to $f$, implying the existence of a subsubsequence $\{n'_k\}_{k=1}^{\infty}\subseteq \{n_k\}_{k=1}^{\infty}$ such that $\varphi_{n'_k,\w_{n'_k}^e,\X,A}(\boldsymbol{t})$ converges to $\varphi^q_{\X,A}(\boldsymbol{t})$ almost everywhere as $k\to\infty$. Since $\{n_k\}_{k=1}^{\infty}$ is an arbitrary subsequence, we have 
\begin{equation}
    \lim_{n\to\infty}\varphi_{n'_k,\w_{n'_k}^e,\X,A}(\boldsymbol{t})= \varphi^q_{\X,A}(\boldsymbol{t}) 
\end{equation}
holds almost everywhere. Therefore, we have $\lim_{n\to\infty} F_{n,\w_n^e, \X,A}(\x,a)=F_{X,A}^{q}(\x,a)$ holds almost surely.
\end{proof}

\subsection{Proof of Supplementary Material Theorem \ref{othe:5.2}}
\begin{proof}
From the error decomposition, we have
\begin{equation*}
    \begin{split}
        \mu^{q}-\hat{\mu}_{\w_n^e}^q = {} & \int_{(\mathcal{X},\mathcal{A})}\mu(\x,a)d(F^{q}_{n,\X,A}-F_{n,\w,\X,A})\\
        &+\int_{(\mathcal{X},\mathcal{A})}\mu(\x,a)d(F^{q}_{\X,A}-F^{q}_{n,\X,A})-\frac{1}{n}\sum_{i=1}^n \w^e_{i,n}\epsilon_i\\
        = {} & \int_{(\mathcal{X},\mathcal{A})}\mu(\x,a)d(F^{q}_{\X,A}-F_{n,\w,\X,A})-\frac{1}{n}\sum_{i=1}^n \w^e_{i,n}\epsilon_i.
    \end{split}
\end{equation*}
The absolute bias then can be written as:
\begin{equation}
    \begin{split}
        \left|\mathbb{E}[\hat{\mu}_{\w_n^e}^q]-\mu^{q}\right|&=\left|\int_{(\mathcal{X},\mathcal{A})}\mu(\x,a)d(F^{q}_{\X,A}-F_{n,\w,\X,A})\right|.
    \end{split}
\end{equation}
By (\ref{the:3.1}), we know that $F_{n,\w,\X,A}$ converges to $F^{q}_{\X,A}$ as $n$ goes to infinity. By the Portmanteau Theorem, it follows that
\begin{equation}
\lim_{n\to\infty}\int_{(\mathcal{X},\mathcal{A})}\mu(\x,a)dF_{n,\w,\X,A}=\int_{(\mathcal{X},\mathcal{A})}\mu(\x,a)dF^{q}_{\X,A}.
\end{equation}
Hence, we have 
\begin{equation}
    \lim_{n\to\infty}|\mathbb{E}[\hat{\mu}_{\w_n^e}^q]-\mu^{q}|=0,
\end{equation}
which demonstrates the asymptotic unbiasedness of the estimator $\hat{\mu}_{\w_n^e}^q$.
\color{black}
\end{proof}

\subsection{Proof of Supplementary Material Theorem \ref{othe:5.3}}\label{suppsec7.7}

\begin{lemma}
Suppose $\X_1,...,\X_n\stackrel{i.i.d.}{\sim}F$ Then $\mathbb{E}[\mathcal{E}(F,F_n)]={O}_p(1/n)$
\end{lemma}
\begin{proof}
This is the same result as Lemma A.1 in \citep{huling2020energy}. 
\end{proof}

\begin{lemma}\label{slemma3}
Assume that the causal assumptions of positivity and strong ignorability hold. Let $\w_n^e$ be the energy balancing weight. Under assumption \normalfont{\textbf{C4}},
we have 
\begin{equation}
    \mathcal{E}(F_{n,\w_n^e,\X,A},F^{q}_{n,\X,A})={O}_p(1/n).
\end{equation}
holds almost surely.

\end{lemma}

\begin{proof}
Consider the non-normalized Radon-Nikodym derivative weights $h^{nnrn}$, which is a function of $(\x,a)$ in the sense that 
\begin{equation}
    h_i^{nnrn}=h^{nnrn}(\X_i,A_i)=\frac{f^q_{\X,A}(\X_i,A_i)}{f_{\X,A}(\X_i,A_i)}.
\end{equation}
Denote $\h^{nnrn}=\{h_1^{nnrn},...,h_n^{nnrn}\}$ and $\{(\Tilde{\X}_1,\Tilde{A}_1),...,(\Tilde{\X}_n,\Tilde{A}_n)\} \sim F^q_{\X,A}$.

Then $\mathcal{E}(F_{n,\h^{nnrn},X,A},F^{q}_{n,\X,A})$ is a two sample V statistic with kernel $g(\cdot)$ and can be written as
\begin{equation*}
    \begin{split}
        \mathcal{E}(F_{n,\h^{nnrn},\X,A},F^{q}_{n,\X,A}) = {} & \frac{2}{n^2}\sum_l\sum_j h^{nnrn}_j||(\Tilde{\X}_i,\Tilde{A}_i)-(\X_j,A_j)||_2\\
        &-\frac{1}{n^2}\sum_i\sum_j h^{nnrn}_i h^{nnrn}_j||(\X_i,A_i)-(\X_j,A_j)||_2\\
        &-\frac{1}{n^2}\sum_l\sum_m ||(\Tilde{\X}_i,\Tilde{A}_i)-(\Tilde{\X}_j,\Tilde{A}_j)||_2\\
        = {} & \frac{1}{n^4}\sum_i\sum_j\sum_l\sum_m g(\W_i,\W_j;\Tilde{\W}_l,\Tilde{\W}_m),
    \end{split}
\end{equation*}
where $\W_i=(\X_i,A_i)$ and $\Tilde{\W}_i=(\Tilde{\X}_i,\Tilde{A}_i)$.
From positivity and strong ignorability, it can be shown that $\mathcal{E}(F_{n,\h^{nnrn},\X,A},F^{q}_{n,\X,A})$ is first-order degenerate in the sense that $\mathbb{E}g(\mathbf{W},\W_j;\Tilde{\W}_l,\Tilde{\mathbf{W}})=0$ for any $\Tilde{\mathbf{W}}$ and $\mathbf{W}$.
\begin{equation*}
    \begin{split}
    &\mathbb{E}g(\mathbf{W},\W_j;\Tilde{\W}_l,\Tilde{\mathbf{W}}) \\ 
    = {} & \mathbb{E}(h(\W)||\W-\Tilde{\W}_l||_2)+\mathbb{E}(h(\W_j)||\W_j-\Tilde{\W}||_2)\\
    &+\mathbb{E}(h(\W)h(\W_j)||\W-\W_j||_2)+ \mathbb{E}(||\Tilde{\W}-\Tilde{\W}_l||_2)\\
    ={} &\int_{\w,\Tilde{\w}_l} \frac{f^q_{\X,A}(\w)}{f_{\X,A}(\w)}||\w-\Tilde{\w}_l||_2 f_{\X,A}(\w) f^q_{\X,A}(\Tilde{\w}_l)d\w d\Tilde{\w}_l\\
    &+\int_{\w_j,\Tilde{\w}} \frac{f^q_{\X,A}(\w_j)}{f_{\X,A}(\w_j)}||\w_j-\Tilde{\w}||_2 f_{\X,A}(\w_j) f^q_{\X,A}(\Tilde{\w})d\w_j d\Tilde{\w}\\
    &-\int_{\w,\w_j} \frac{f^q_{\X,A}(\w)}{f_{\X,A}(\w)}\frac{f^q_{\X,A}(\w_j)}{f_{\X,A}(\w_j)}
    ||\w-\w_j||_2 f_{\X,A}(\w) f_{\X,A}(\w_j)d\w d\w_j\\
    &-\int_{\Tilde{\w},\Tilde{\w}_l} ||\Tilde{\w}-\Tilde{\w}_l||_2 f^q_{\X,A}(\Tilde{\w}) f^q_{\X,A}(\Tilde{\w}_l)d\Tilde{\w} d\Tilde{\w}_l\\
    ={} &\int_{\w,\Tilde{\w}_l} f^q_{\X,A}(\w) f^q_{\X,A}(\Tilde{\w}_l)||\w-\Tilde{\w}_l||_2d\w d\Tilde{\w}_l\\
    &+\int_{\w_j,\Tilde{\w}} f^q_{\X,A}(\w_j) f^q_{\X,A}(\Tilde{\w})||\w_j-\Tilde{\w}||_2d\w_j d\Tilde{\w}\\
    &-\int_{\w,\w_j} f^q_{\X,A}(\w) f^q_{\X,A}(\w_j)||\w-\w_j||_2d\w d\w_j\\
    &-\int_{\Tilde{\w},\Tilde{\w}_l}  f^q_{\X,A}(\Tilde{\w}) f^q_{\X,A}(\Tilde{\w}_l)||\Tilde{\w}-\Tilde{\w}_l||_2d\Tilde{\w} d\Tilde{\w}_l\\
    = {} & 0.
    \end{split}
\end{equation*}
 Thus, if $\mathbb{E}g^2<\infty$, then $\mathcal{E}(F_{n,\h^{nnrn},\X,A},F^{q}_{n,\X,A})={O}_p(n^{-1})$ by extensions of asymptotic results for one-sample V-statistics to multi-sample V-statistics 
\citep{serfling1980approximation, korolyuk1989theory, rizzo2002test}.
Note that this implies that $\mathcal{E}(F_{n,\hat{\h},\X,A},F^{q}_{n,\X,A})={O}_p(n^{-1})$ where $\hat{\h}$ is the normalized Radon Nikodym weights,
since the two weights differs only by a normalizing constant.

By the definition of the energy balancing weights $\w_n^e$, we have that
\begin{equation*}
    \mathcal{E}(F_{n,\w^{e}_n,\X,A},F^{q}_{n,\X,A})\leq\mathcal{E}(F_{n,\hat{\h},\X,A},F^{q}_{n,\X,A})
\end{equation*}
holds for each $n$, which leads to the result $\mathcal{E}(F_{n,\w^{e}_n,\X,A},F^{q}_{n,\X,A})={O}_p(n^{-1})$.
\end{proof}

The next lemma illustrates that, the sum of squared weights is upper bounded by $\mathcal{O}(n)$ if we impose a mild regularity condition to the weights.

\begin{lemma}\label{sumsqweights}
Let $\w^e_n$ be the energy balancing weights. Under assumption \normalfont{\textbf{C4}} and \normalfont{\textbf{C5}}, we have almost surely that:
\begin{equation}
    \sum_{i=1}^n \frac{(w_{i,n}^e)^2}{n}\leq B
\end{equation}
for all $n>n^*$ for some $n^*>1$ and some constant $B>0$ that does not depend on $n$.

\end{lemma}

\begin{proof}
Here, we basically follow the proof of Lemma A.3 in \citep{huling2020energy}. By the weighted energy distance duality, we have
\begin{equation*}
    \begin{split}
        &\mathcal{E}(F^{q}_{n,\X,A},F_{n,\w_n^e,\X,A})=\int_{(\mathcal{X},\mathcal{A})} |\varphi_{n,\X,A}^{q}(\boldsymbol{t})-\varphi_{n,\w^e_n,\X,A}(\boldsymbol{t})|^2v(\boldsymbol{t})d\boldsymbol{t}\\
        ={}&\int_{(\mathcal{X},\mathcal{A})} |\frac{1}{n}\sum_{i=1}^n\sum_{j=1}^{J(\X_i)}I_{j,\X_i}(A_i) \exp(i \boldsymbol{t}^T(\X_i,q_j(\X_i,A_i)))-\frac{1}{n}\sum_{i=1}^n w_{i,n}^e \exp(i \boldsymbol{t}^T(\X_i,A_i))|^2v(\boldsymbol{t})d\boldsymbol{t}\\
        ={}&\int_{(\mathcal{X},\mathcal{A})}\frac{1}{n^2}|\sum_{i=1}^n (\exp(i\boldsymbol{t}^T(\X_i,Q_i))-w_{i,n}^e\exp(i \boldsymbol{t}^T(\X_i,A_i))) |^2v(\boldsymbol{t})d\boldsymbol{t}\\
        ={}&\frac{1}{n^2}\int_{(\mathcal{X},\mathcal{A})}\sum_{i=1}^n\sum_{j=1}^n\Bigg\{\bigg(\exp(i\boldsymbol{t}^T(\X_i,Q_i))(\exp(i\boldsymbol{t}^T(\X_j,Q_j)\bigg)\\
        &-2\bigg(\exp(i\boldsymbol{t}^T(\X_i,Q_i))(w_{j,n}^e\exp(i \boldsymbol{t}^T(\X_j,A_j)))\bigg)\\
        &+\bigg((w_{i,n}^e\exp(i \boldsymbol{t}^T(\X_i,A_i))))(w_{j,n}^e\exp(i \boldsymbol{t}^T(\X_j,A_j))))\bigg) \Bigg\}v(\boldsymbol{t})d\boldsymbol{t}.
    \end{split}
\end{equation*}
Suppose the number of weights $w_{i,n}^e$ that attain the maximum $Cn^{1/3}$ is of order ${O}_p(n^{1/3})$, and denote the set of index of all such weights as $\mathcal{I}_n=\{i,w_{i,n}^e={O}_p(n^{1/3})\}$. We consider the worst case scenario that all terms in the above summation are positive. Then, we have
\begin{equation*}
    \begin{split}
        \mathcal{E}(F^{q}_{n,\X,\mathbf{A}},F_{n,\w_n^e,\X,\mathbf{A}})\geq{}&\frac{1}{n^2}\int\int \sum_{i\in \mathcal{I}_n}\sum_{j\in\mathcal{I}_n}\Bigg\{\bigg(\exp(i\boldsymbol{t}^T(\X_i,Q_i))(\exp(i\boldsymbol{t}^T(\X_j,Q_j)\bigg)\\
        &-2\bigg(\exp(i\boldsymbol{t}^T(\X_i,Q_i))(w_{j,n}^e\exp(i \boldsymbol{t}^T(\X_j,A_j)))\bigg)\\
        &+\bigg((w_{i,n}^e\exp(i \boldsymbol{t}^T(\X_i,A_i))))(\w_{j,n}^e\exp(i \boldsymbol{t}^T(\X_j,A_j))))\bigg)\Bigg\}v(\boldsymbol{t})d\boldsymbol{t}\\
        ={}&\frac{1}{n^2}\sum_{i\in \mathcal{I}_n}\sum_{j\in\mathcal{I}_n}\{{O}_p(n^{1/3})+{O}_p(n^{1/3}+{O}_p(n^{2/3})\}\\
        ={}&\frac{1}{n^2}\sum_{i\in \mathcal{I}_n}\{{O}_p(n^{2/3}+{O}_p(n^{2/3})+{O}_p(n^1)\}\\
        ={}&\frac{1}{n^2}\{{O}_p(n^{1}+{O}_p(n^{1})+{O}_p(n^{4/3})\}\\
        ={}&{O}_p(n^{-2/3}).
    \end{split}
\end{equation*}
Note that this is contradictory to Lemma \ref{slemma3} which states that $\mathcal{E}(F^{q}_{n,\X,A},F_{n,\w_n^e,\X,A})={O}_p(1/n)$ holds almost surely. Using similar arguments, we can show that the maximum size that $\mathcal{I}_n$ can take such that no contradiction to Lemma \ref{slemma3} arises is ${O}_p(n^{1/6})$.

Similar to the energy balancing paper, we consider the set $\mathcal{J}_n=\{i,i\notin \mathcal{I}_n,\, w_{i,n}^e={O}_p(r(n))$ where $\lim_{n\to\infty}r(n)=\infty$ and $\lim_{n\to\infty}r(n)/n^{\frac{1}{3}}=0$. We then denote the constant set of the weights as $\mathcal{K}_n=\{i:w_{i,n}^e={O}_p(1)\}$. Then, $\mathcal{I}_n\cup \mathcal{J}_n\cup \mathcal{K}_n$ is the total set of index $1$ to $n$.

We now think about how large $|\mathcal{J}_n|$ can be to avoid the contradiction above. Use the same identity, we find that we need ${O}_p(|\mathcal{J})n|r(n)n^{1/2})={O}_p(n)$ to avoid the contradiction, i.e. $|\mathcal{J})n|r(n)={O}_p(n^{1/2})$. Thus, we have the upper bound of the sum
\begin{equation*}
    \begin{split}
        \sum_{i=1}^n (w_{i,n}^e)^2&=\sum_{i\in\mathcal{K}_n} (w_{i,n}^e)^2+\sum_{i\in\mathcal{I}_n} (w_{i,n}^e)^2+\sum_{i\in\mathcal{J}_n} (w_{i,n}^e)^2\\
        &=\sum_{i\in\mathcal{K}_n} {O}_p(1)+\sum_{i\in\mathcal{I}_n} {O}_p(r^2(n))+\sum_{i\in\mathcal{J}_n}{O}_p(n^{2/3})\\
        &={O}_p(n)+{O}_p(n^{5/6})+{O}_p(|\mathcal{J}_n|r^2(n))={O}_p(n),
    \end{split}
\end{equation*}
where the last equality holds since $|\mathcal{J}_n|r(n)$ is at most ${O}_p(n^{1/2})$ and $r(n)$ cannot reach ${O}_p(n^{1/2})$.

With this identity, we have the desired result,
\begin{equation}
     \sum_{i=1}^n \frac{(\w_{i,n}^e)^2}{n}\leq B.
\end{equation}
\end{proof}
\begin{proof}[of Supplementary Material Theorem \ref{othe:5.3}]
We can express the expectation as conditional expectation:
\begin{equation}
    \mathbb{E}_{\X,A,Y}[(\hat{\mu}_{\w_n^e}^q-\mu^q)^2]=\mathbb{E}_{\X,A}\bigg[\mathbb{E}_{Y|\X,A}[(\hat{\mu}_{\w_n^e}^q-\mu^{q})^2]\bigg].
\end{equation}
and we have
\begin{equation*}
    \begin{split}
        &\mathbb{E}_{Y|\X,A}[(\hat{\mu}_{\w_n^e}^q-\mu^q)^2]\\
        ={}& \mathbb{E}_{Y|\X,A}\bigg[\int_{\mathcal{X}}\int_{\mathcal{A}}\mu(x,a)d(F^{q}_{n,\X,A}-F_{n,\w,\X,A})\\
        &+\int_{\mathcal{X}}\int_{\mathcal{A}}\mu(\x,a)d(F^{q}_{\X,A}-F^{q}_{n,\X,A})-\frac{1}{n}\sum_{i=1}^n w_i^e\epsilon_i\Bigg]^2\\
        ={}& \mathbb{E}_{Y|\X,A}[-\frac{1}{n}\sum_{i=1}^n w_i^e\epsilon_i]^2+\mathbb{E}_{Y|\X,A}[-\frac{1}{n}\sum_{i=1}^n w_i^e\epsilon_i]\times\\
        &\bigg(\int_{\mathcal{X}}\int_{\mathcal{A}}\mu(x,a)d(F^{q}_{n,\X,A}-F_{n,\w,\X,A})+\int_{\mathcal{X}}\int_{\mathcal{A}}\mu(\x,a)d(F^{q}_{\X,A}-F^{q}_{n,\X,A})\bigg)\\
        &+\bigg(\int_{\mathcal{X}}\int_{\mathcal{A}}\mu(x,a)d(F^{q}_{n,\X,A}-F_{n,\w,\X,A})+\int_{\mathcal{X}}\int_{\mathcal{A}}\mu(\x,a)d(F^{q}_{\X,A}-F^{q}_{n,\X,A})\bigg)^2\\
        ={}& \mathbf{V}_{Y|\X,A}\bigg[\frac{1}{n}\sum_{i=1}^n w_i^e\epsilon_i\bigg]+\bigg(\int_{\mathcal{X}}\int_{\mathcal{A}}\mu(\x,a)d(F^{q}_{n,\X,A}-F_{n,\w,\X,A})\\
        &+\int_{\mathcal{X}}\int_{\mathcal{A}}\mu(\x,a)d(F^{q}_{\X,A}-F^{q}_{n,\X,A})\bigg)^2\\
        \leq {} & \mathbf{V}_{Y|\X,A}\bigg[\frac{1}{n}\sum_{i=1}^n w_i^e\epsilon_i\bigg]+2\bigg(\int_{\mathcal{X}}\int_{\mathcal{A}}\mu(\x,a)d(F^{q}_{n,\X,A}-F_{n,\w,\X,A})\bigg)^2\\
        &+2\bigg(\int_{\mathcal{X}}\int_{\mathcal{A}}\mu(\x,a)d(F^{q}_{\X,A}-F^{q}_{n,\X,A})\bigg)^2\\
        ={}& \frac{1}{n^2} \sum_{i=1}^n (w_i^e)^2\sigma^2(Y|\X,A)+2\bigg(\int_{\mathcal{X}}\int_{\mathcal{A}}\mu(\x,a)d(F^{q}_{n,\X,A}-F_{n,\w,\X,A})\bigg)^2\\
        &+2\bigg(\int_{\mathcal{X}}\int_{\mathcal{A}}\mu(\x,a)d(F^{q}_{\X,A}-F^{q}_{n,\X,A})\bigg)^2,
    \end{split}
\end{equation*}
where the second to last inequality follows $(a+b)^2\leq2a^2+2b^2$.

Consider $\frac{1}{n^2} \sum_{i=1}^n (w_i^e)^2\sigma^2(Y|\X,\mathbf{A})$, since $\sigma^2(Y|\X,\mathbf{A})$ is assumed to be bounded over $(\mathcal{X},\mathcal{A})$, we can define $\sigma^2_{max}=\sup_{(\x,\mathbf{a})\in(\mathcal{X},\mathcal{A})}\sigma^2(Y|\X,\mathbf{A})$. Then, we have:
\begin{equation*}
    \begin{split}
        \mathbb{E}_{\X,\mathbf{A}}\bigg[\frac{1}{n^2} \sum_{i=1}^n (w_i^e)^2\sigma^2(Y|\X,\mathbf{A})\bigg] &\leq \sigma^2_{max} \mathbb{E}_{\X,\mathbf{A}}\bigg[\frac{1}{n^2} \sum_{i=1}^n (w_i^e)^2\bigg]\\
        &\leq \frac{B \sigma^2_{max}}{n}={O}_p\left(\frac{1}{n}\right),
    \end{split}
\end{equation*}
where the second to last inequality is due to the Lemma \ref{sumsqweights} we proved above.

For $(\int_{\mathcal{X}}\int_{\mathcal{A}}\mu(\x,\mathbf{a})d(F^{q}_{n,\X,A}-F_{n,\w,\X,\mathbf{A}}))^2$, we have:
\begin{equation*}
    \begin{split}
        &\mathbb{E}_{\X,\mathbf{A}}\bigg[\bigg(\int_{\mathcal{X}}\int_{\mathcal{A}}\mu(\x,\mathbf{a})d(F^{q}_{n,\X,A}-F_{n,\w,\X,\mathbf{A}})\bigg)^2\bigg]\\
        &\leq \mathbb{E}_{\X,\mathbf{A}}\bigg[ \sup_{\zeta\in\mathcal{H}:||\zeta||_{\mathcal{H}}\leq||\mu(\x,\mathbf{a})||_{\mathcal{H}}} \bigg( \int_{\mathcal{X}}\int_{\mathcal{A}}\zeta(\x,\mathbf{a})d(F^{q}_{n,\X,A}-F_{n,\w,\X,\mathbf{A}})\bigg)^2\bigg]\\
        &\leq C \mathbb{E}_{\X,\mathbf{A}} [\mathcal{E}(F^{q}_{n,\X,A},F_{n,\w,\X,\mathbf{A}})]\\
        &={O}_p\left(\frac{1}{n}\right),
    \end{split}
\end{equation*}
where the second to last inequality is due to Equation \eqref{equ:3.3} in the main text.

And for $(\int_{\mathcal{X}}\int_{\mathcal{A}}\mu(\x,a)d(F^{q}_{\X,A}-F^{q}_{n,\X,A}))^2$, we have
\begin{equation*}
    \begin{split}
        &\mathbb{E}_{\X,\mathbf{A}}\bigg[\bigg(\int_{\mathcal{X}}\int_{\mathcal{A}}\mu(\x,a)d(F^{q}_{\X,A}-F^{q}_{n,\X,A})\bigg)^2\bigg]\\
        ={}&\mathbb{E}_{\X,\mathbf{A}}\bigg[\bigg(\int_{\mathcal{X}}\int_{\mathcal{A}}\mu(\x,a)dF^{q}_{n,\X,A}-\int_{\mathcal{X}}\int_{\mathcal{A}}\mu(\x,a)dF^{q}_{\X,A}\bigg)^2 \bigg]\\
        ={}&\mathbb{E}_{\X,\mathbf{A}}\bigg(\int_{\mathcal{X}}\int_{\mathcal{A}}\mu(\x,a)dF^{q}_{n,\X,A}\bigg)^2-\bigg(\int_{\mathcal{X}}\int_{\mathcal{A}}\mu(\x,a)dF^{J}_{q,\X,\mathbf{A}} \bigg)^2\\
        ={}&\mathbb{E}_{\X,\mathbf{A}}\bigg(\int_{\mathcal{X}}\int_{\mathcal{A}}\mu(\x,a)dF^{q}_{n,\X,A}\bigg)^2-\bigg[\mathbb{E}_{\X,\mathbf{A}}\bigg(\int_{\mathcal{X}}\int_{\mathcal{A}}\mu(\x,a)dF^{q}_{n,\X,A} \bigg)\bigg]^2\\
        ={}& \frac{Var_{\X,\mathbf{A}}[\mu(\X,\mathbf{A})]}{n} ={O}_p\left(\frac{1}{n}\right).
    \end{split}
\end{equation*}
Thus, we have the desired result:
\begin{equation}
    \mathbb{E}_{\X,\mathbf{A},Y}[(\hat{\mu}_{\w_n^e}^q-\mu^q)^2]={O}_p\left(\frac{1}{n}\right).
\end{equation}
\end{proof}

\subsection{Proof of Theorem \ref{the:5.1} in the Main Manuscript}
\begin{proof}
The proof follows the arguments of the proof of Theorem 3.1 in the Supplementary Materials in \citep{huling2020energy}. The difference between the penalized EBW and the EBW is the inequality, for EBW we have:
\begin{equation*}
    \begin{split}
        &\int_{\mathbb{R}^p}|\varphi_{\X,A}^{q}(\boldsymbol{t})-\varphi_{n,\w_n^e,\X,A}(\boldsymbol{t})|^2v(\boldsymbol{t})d\boldsymbol{t}\\
        &\leq \bigg(\bigg[\int_{\mathbb{R}^p}|\varphi_{n,\X,A}^{q}(\boldsymbol{t})-\varphi_{n,\w_n^e,\X,A}(\boldsymbol{t})|^2v(\boldsymbol{t})d\boldsymbol{t}\bigg]^{\frac{1}{2}}+\bigg[\int_{\mathbb{R}^p}|\varphi_{n,\X,A}^{q}(\boldsymbol{t})-\varphi_{\X,A}^{q}(\boldsymbol{t})|^2v(\boldsymbol{t})d\boldsymbol{t}\bigg]^{\frac{1}{2}}\bigg)^2\\
        ={}&\bigg([\mathcal{E}(F_{n,\X,A}^{q},F_{n,\w_n^e,\X,A})]^{\frac{1}{2}}+\bigg[\int_{\mathbb{R}^p}|\varphi_{n,\X,A}^{q}(\boldsymbol{t})-\varphi_{\X,A}^{q}(\boldsymbol{t})|^2v(\boldsymbol{t})d\boldsymbol{t}\bigg]^{\frac{1}{2}}\bigg)^2\\
        &\leq \bigg(\bigg[\mathbb{E}[\mathcal{E}(F_{n,\X,A}^{q},F_{n,\hat{\h},\X,A})]\bigg]^{\frac{1}{2}}+\bigg[\int_{\mathbb{R}^p}|\varphi_{n,\X,A}^{q}(\boldsymbol{t})-\varphi_{\X,A}^{q}(\boldsymbol{t})|^2v(\boldsymbol{t})d\boldsymbol{t}\bigg]^{\frac{1}{2}}\bigg)^2,
    \end{split}
\end{equation*}
which uses the following inequality that holds by definition of the energy balancing weights
\begin{equation}
    \mathcal{E}(F_{n,\X,A}^{q},F_{n,\w_n^e,\X,A})\leq \mathbb{E}[\mathcal{E}(F_{n,\X,A}^{q},F_{n,\hat{\h},\X,A})].
\end{equation}

But for penalized EBW we have:
\begin{equation}
    \mathcal{E}(F_{n,\X,A}^{q},F_{n,\w_n^{p},\X,A})+\frac{\lambda}{n^2}\sum_{i=1}^n (w_i^{p})^2\leq \mathbb{E}[\mathcal{E}(F_{n,\X,A}^{q},F_{n,\hat{\h},\X,A})]+\frac{\lambda}{n^2}\sum_{i=1}^n h_i^2.
\end{equation}
Since the Radon-Nikodym weights are $h_i=\frac{f^q_{\X,A}(\X_i,A_i)}{f_{\X,A}(\X_i,A_i)}$, we have $\mathbb{E}(h^2)\leq\infty$. Thus, by the SLLN 
\begin{equation}
    \lim_{n\to\infty}\frac{\lambda}{n^2}\sum_{i=1}^n h_i^2=0 
\end{equation}
almost surely. Thus, we have:
\begin{equation*}
    \begin{split}
        &\int_{\mathbb{R}^p}|\varphi_{\X,A}^q(\boldsymbol{t})-\varphi_{n,\w_n^{p},\X,A}(\boldsymbol{t})|^2v(\boldsymbol{t})d\boldsymbol{t}\\
        &\leq \bigg(\bigg[\int_{\mathbb{R}^p}|\varphi_{n,\X,A}^q(\boldsymbol{t})-\varphi_{n,\w_n^{p},\X,A}(\boldsymbol{t})|^2v(\boldsymbol{t})d\boldsymbol{t}\bigg]^{\frac{1}{2}}+\bigg[\int_{\mathbb{R}^p}|\varphi_{n,\X,A}^q(\boldsymbol{t})-\varphi_{\X,A}^q(\boldsymbol{t})|^2v(\boldsymbol{t})d\boldsymbol{t}\bigg]^{\frac{1}{2}}\bigg)^2\\
        &=\bigg([\mathcal{E}(F_{n,\X,A}^{q},F_{n,\w_n^{p},\X,A})]^{\frac{1}{2}}+\bigg[\int_{\mathbb{R}^p}|\varphi_{n,\X,A}^{q}(\boldsymbol{t})-\varphi_{\X,A}^{q}(\boldsymbol{t})|^2v(\boldsymbol{t})d\boldsymbol{t}\bigg]^{\frac{1}{2}}\bigg)^2\\
        &\leq \bigg(\bigg[\mathbb{E}[\mathcal{E}(F_{n,\X,A}^{q},F_{n,\hat{\h},\X,A})]+\frac{\lambda}{n^2}\sum_{i=1}^n h_i^2\bigg]^{\frac{1}{2}}+\bigg[\int_{\mathbb{R}^p}|\varphi_{n,\X,A}^{q}(\boldsymbol{t})-\varphi_{\X,A}^{q}(\boldsymbol{t})|^2v(\boldsymbol{t})d\boldsymbol{t}\bigg]^{\frac{1}{2}}\bigg)^2,
    \end{split}
\end{equation*}
which converges to $0$ as $n\to \infty$.
\end{proof}

\subsection{Proof of Theorem \ref{the:5.2} in the Main Manuscript}
\begin{proof}
The proof of this theorem is similar to the EBW version (Supplementary Material Section \ref{suppsec7.7}). The difference of the penalized EBW and EBW is with regards to the proof that 
\begin{equation}
    \mathbb{E}_{\X,A}\bigg[\frac{1}{n^2} \sum_{i=1}^n (w_i)^2\sigma^2(Y|\X,A)\bigg]={O}_p\left(\frac{1}{n}\right).
\end{equation}
Note that the penalized objective function gives
\begin{equation}
    \mathcal{E}(F_{n,\X,A}^{q},F_{n,\w_n^{p},\X,A})+\frac{\lambda}{n^2}\sum_{i=1}^n (\w_i^{p})^2\leq \mathbb{E}[\mathcal{E}(F_{n,\X,A}^{q},F_{n,\hat{\h},\X,A})]+\frac{\lambda}{n^2}\sum_{i=1}^n h_i^2.
\end{equation}
Since we have $\mathbb{E}[\mathcal{E}(F_{n,\X,A}^{q},F_{n,\hat{\h},\X,A})]={O}_p(\frac{1}{n})$,  $\frac{\lambda}{n^2}\sum_{i=1}^n h_i^2={O}_p(\frac{1}{n})$, and the two terms in the left hand side are all positive, we have that the two terms in the left hand side are necessarily ${O}_p(\frac{1}{n})$: in particular $ \mathcal{E}(F_{n,\X,A}^{q},F_{n,\w_n^{p},\X,A})={O}_p(\frac{1}{n})$ and $\frac{\lambda}{n^2}\sum_{i=1}^n (w_i^{p})^2={O}_p\left(\frac{1}{n}\right)$.

Then, we have
\begin{equation}
    \mathbb{E}_{\X,A}\bigg[\frac{1}{n^2} \sum_{i=1}^n (w_i^{p})^2\sigma^2(Y|\X,A)\bigg]\leq\sigma^2_{max}\mathbb{E}_{\X,\mathbf{A}}\bigg[\frac{1}{n^2} \sum_{i=1}^n (w_i^{p})^2\bigg]={O}_p\left(\frac{1}{n}\right),
\end{equation}
which is the desired result.
\end{proof}
\subsection{Proof of Theorem \ref{the:6.1}}

\begin{lemma}\label{partiallycond CLT}

(Partially Conditional Central Limit Theorem of \citet{wong2017kernel}) Let $(W_1,\mathcal{B}_1),...,(W_n,\mathcal{B}_n)$ be indipendent and identically distributed where $W_1,...,W_n$ are random variables and $\mathcal{B}_1,...,\mathcal{B}_n$ are sets of random variables. Let $\{C_1,...,C_n\}$ be another set of random variables. Write $\mathcal{D}_n=\{W_1,...,W_n,\mathcal{B}_1,...,\mathcal{B}_n\}$. Assume these variables satisfy, for each $j=1,...,n$
\begin{equation*}
    E(W_j)=0
\end{equation*}
\begin{equation*}
    E(C_j|\mathcal{D}_n)=0
\end{equation*}
\begin{equation*}
    \sum_{j=1}^n var(W_j)=1
\end{equation*}
\begin{equation*}
    \sum_{j=1}^n var(C_j|\mathcal{D}_n)=1
\end{equation*}
and there exists $q>0$ such that $\sum_{j=1}^n E(|C_j|^{2+q}|\mathcal{D}_n)\to0$ in probability. Moreover, $C_1,...,C_n$ are conditionally independent given $\mathcal{D}_n$. Let $g$ be a (non-random) function mapping from the support of $\mathcal{D}_n$ to $\mathbb{R}^+$ such that there exists a constant $M>0$ such that $E(g^2(\mathcal{D}_n))\leq M$ and $g^2(\mathcal{D}_n)\leq M+o_p(1)$. For any positive real number $\tau$, consider two random variables:
\begin{equation*}
    Z_n=\tau\sum_{j=1}^n W_j+g(\mathcal{D}_n)\sum_{j=1}^n C_j
\end{equation*}
\begin{equation*}
    Z_n^*=\tau F+g(\mathcal{D}_n)\sum_{j=1}^n \{var(C_j|\mathcal{D}_j)\}^{1/2} G_j,
\end{equation*}
where $F,G_1,...,G_n$ are i.i.d. standard normal random variables independent of $C_1,...,C_n$ and $\mathcal{D}_n$. Let $\phi_n$ and $\phi_n^*$ be the corresponding characteristic function of $Z_n$ and $Z_n^*$ respectively. Then 
\begin{equation*}
    |\phi_n(t)-\phi_n^*(t)|\to0 
\end{equation*}
for every $t\in \mathbb{R}$. Moreover, $E(Z^{*2}_n)=\tau^2+E\{g^2(\mathcal{D}_n)\}\leq\tau^2+M$ and therefore $\phi^*_n$ is twice differentiable. 
\end{lemma}

Now we start to prove Theorem \ref{the:6.1}:
\begin{proof}
From the error decomposition, we have
\begin{equation*}
    \begin{split}
        \hat{\mu}_{AG}^q-\mu^{q}=&  \int_{(\mathcal{X},\mathcal{A})}[\hat{\mu}(\x,a)-\mu(\x,a)]d(F^{q}_{n,\X,A}-F_{n,\w^e_n,\X,A})\\
        &+ \int_{(\mathcal{X},\mathcal{A})}\mu(\x,a) d(F^{q}_{n,\X,A}-F_{\X,A}^q)+\frac{1}{n}\sum_{i=1}^n\w^e_i\epsilon_i.
    \end{split}
\end{equation*}

For the element $\int_{(\mathcal{X},\mathcal{A})}[\hat{\mu}(\x,a)-\mu(\x,a)]d(F_{n,\w^e_n,\X,A}-F^{q}_{n,\X,A})$, note that by Lemma 3.3, since $\hat{\mu}-\mu \in \mathcal{H}$,  we have
\begin{equation*}
   \bigg[ \int_{(\mathcal{X},\mathcal{A})}[\hat{\mu}(\x,a)-\mu(\x,a)]d(F_{n,\w^e_n,\X,A}-F^{q}_{n,\X,A})\bigg]^2\leq C_\alpha \mathcal{E}(F^{q}_{n,\X,A},F_{n,\w^e_n,\X,A}),
\end{equation*}
where $C_\alpha\geq0$ is a constant depending on only $\mu(x,a)$. From Lemma \ref{slemma3}, we have 
\begin{equation}
    C_\alpha \mathcal{E}(F^{q}_{n,\X,A},F_{n,\w^e_n,\X,A})\leq C_\alpha \mathcal{O}(1/n).
\end{equation}
Note that here $C_\alpha = \langle\hat{\mu}-\mu,\hat{\mu}-\mu\rangle$. Since $\hat{\mu}(\x,a)$ is a consistent estimator of $\mu(\x,a)$, we have $C_\alpha={O}_p(n^{-1})$. Thus, the first element would be $o_p(n^{-1/2})$ which does not influence the efficiency result.

Now, let's consider the term $\int_{(\mathcal{X},\mathcal{A})}\mu(\x,a) d(F^q_{\X,A}-F^{q}_{n,\X,A})+\frac{1}{n}\sum_{i=1}^n\w^e_i\epsilon_i$.
Its variance can be obtained as 
\begin{equation*}
\begin{split}
    \Var(J_n)={}&\Var(n^{1/2}(\hat{\mu}_{AG}^{q}-\mu^{q}))\\
    ={}& n \Var\bigg(\int_{(\mathcal{X},\mathcal{A})}\mu(\x,a) d(F^q_{\X,A}-F^{q}_{n,\X,A})+\frac{1}{n}\sum_{i=1}^n\w^{p}_i\epsilon_i\bigg)\\
    ={}& n\Var\bigg(\mu^{q}-\frac{1}{n}\sum_{i=1}^n\sum_{j=1}^{J(\X_i)}I_{j,\X_i}(A_i)\mu(\X_i,q_j(\X_i,A_i))+\frac{1}{n}\sum_{i=1}^n\w^{p}_i\epsilon_i\bigg)\\
    ={}& \frac{1}{n}\sum_{i=1}^n\Var\bigg(\sum_{j=1}^{J(\X_i)} I_{j,\X_i}(A_i)\mu(\X_i,Q_j)\bigg)+\frac{1}{n}\sigma^2\mathbb{E}\bigg(\sum_{i=1}^n (\w_i^{p})^2\bigg)\\
    \leq{}& \frac{1}{n}\sum_{i=1}^n\Var\{\sum_{j=1}^{J(\X_i)} I_{j,\X_i}(A_i)\mu(\X_i,Q_j)\}+ \sigma^2\bigg(\frac{1}{n}\sum_{i=1}^n h_i^2 +n {O}_p\left(\frac{1}{n}\right)\bigg).
\end{split}
\end{equation*}
where the second to last equality is obtained since $\mathbb{E}(\frac{1}{n}\sum_{i=1}^n\w^{p}_i\epsilon_i)=0$, the last inequality is due to the definition of the penalized energy balancing weights, see the proof of Theorem \ref{the:5.2}. 
Therefore, by the central limit theorem of the Radon-Nikodym weights, we have the following limiting property:
\begin{equation}
    \limsup_n \Var(J_n)\leq \Var(\mu(\X,Q))+\sigma^2\bigg(\E\left(\frac{f^q_{\X,A}(\X,A)}{f_{\X,A}(\X,A)}\right)^2+B\bigg),
\end{equation}
where $B$ is some real number.

Note that we have 
\begin{equation*}
    {W}_i=\frac{\sum_{j=1}^{J(\X_i)}I_{j,\X_i}(A_i)\mu(\X_i,Q_i)-E_{q}(\mu(\X,A))}{\bigg(n \Var(\mu(\X_i,Q_i))\bigg)^{1/2}}
\end{equation*}
\begin{equation*}
    \mathcal{B}_i=(\X_i,A_i)
\end{equation*}
\begin{equation*}
    C_i=\frac{w_i^e\epsilon_i}{(\sigma^2\sum_{j=1}^n (w^e_j)^2)^{1/2}}
\end{equation*}
\begin{equation*}
    \mathcal{D}_n=\{W_1,...,W_n,\mathcal{B}_1,...,\mathcal{B}_n\}
\end{equation*}
and let $\tau^2=Var(\mu(X,q(\X,A)))$, $g^2(\mathcal{D}_n)=\sigma^2\frac{1}{n}\sum_{j=1}^n (w^e_j)^2$, by assumption CA-2, we have $1\leq w_i^e\leq Bn^{1/3}$. Therefore, $(\sum_{i=1}^n (w^e_i)^2)^{-1}=\mathcal{O}(\frac{1}{n})$ and $\max_i|w_i^e|=o_p(n^{1/2})$. Moreover, by assumption CA-3, $\max_i E(|\epsilon_i|^3)<\infty$. Hence
\begin{equation*}
    \begin{split}
        0&\leq E\bigg(\sum_{i=1}^n|C_i|^3|\mathcal{D}_n\bigg)\leq\frac{\max_i E(|\epsilon_i|^3)\sum_{j=1}^n (w_j^e)^3}{\sigma^3(\sum_{j=1}^n (w^e_j)^2)^{3/2}}\\
        &\leq\frac{\max_i E(|\epsilon_i|^3)\sum_{j=1}^n (w_j^e)^2 \max_i |w^e_i|}{\sigma^3(\sum_{j=1}^n (w^e_j)^2)^{3/2}}=\frac{\max_i E(|\epsilon_i|^3)\max_i |w^e_i|}{\sigma^3(\sum_{j=1}^n (w^e_j)^2)^{1/2}}
        =o_p(1)
    \end{split}
\end{equation*}

By Lemma \ref{sumsqweights}, we have $E(g^2(\mathcal{D}_n))\leq M$ and $g^2(\mathcal{D}_n)\leq M+o_p(1)$ by taking $M=\sigma^2 B$. Then, let 
\begin{equation*}
   \begin{split}
        Z_n&=\tau\sum_{j=1}^n W_j+g(\mathcal{D}_n)\sum_{j=1}^nC_j\\
        &=n^{1/2}\bigg(\int_{(\mathcal{X},\mathcal{A})}\mu(\x,a) d(F^q_{X,A}-F^{q}_{n,X,A})+\frac{1}{n}\sum_{i=1}^n\w^e_i\epsilon_i\bigg)
   \end{split}
\end{equation*}
\begin{equation*}
    \begin{split}
        Z_n^*=\tau F+g(\mathcal{D}_n)\sum_{j=1}^n \Var(C_j|\mathcal{D}_n)^{1/2} G_j=[\Var(\mu(X,Q))]^{1/2}F+\sigma n^{-1/2}\sum_{j=1}^nw_j^e G_j,
    \end{split}
\end{equation*}
where $F, G_1,...,G_n$ are i.i.d. standard normal random variables independent of $C_1,...,C_n$ and $\mathcal{D}_n$. Let $\phi_n$ and $\phi_n^*$ be the corresponding characteristic function of $Z_n$ and $Z_n^*$ respectively. Applying the Partially Conditional Central Limit Theorem \ref{partiallycond CLT}, we have the desired result.
\end{proof}

\subsection{Proof of Theorem \ref{the:6.2}}
\begin{proof}
Since we have $Y_i=\mu(\X_i,A_i)+\epsilon_i$ Rewrite 
\begin{equation*}
    \begin{split}
        \hat{q}^{q}_{r}&=\frac{1}{\sqrt{n}}\sum_{i=1}^n \xi_i \hat{\Phi}_i =\frac{1}{\sqrt{n}}\sum_{i=1}^n\xi_i \bigg(w_i^{p}(Y_i-\hat{\mu}(\X_i,A_i))+  \hat{\mu}(\X_i,Q_i)-\hat{\mu}_{AG}^{q}\bigg)\\
        &=\frac{1}{\sqrt{n}}\sum_{i=1}^n \bigg(\xi_i w_i^{p}(Y_i-\mu(\X_i,A_i))+\xi_i w_i^{p}(\mu(\X_i,A_i)-\hat{\mu}(\X_i,A_i))+  \xi_i\hat{\mu}(\X_i,Q_i)-\xi_i\hat{\mu}_{AG}^{q}\bigg)\\
         &=\frac{1}{\sqrt{n}}\sum_{i=1}^n \bigg(\xi_i w_i^{p}\epsilon_i+\xi_i w_i^{p}(\mu(\X_i,A_i)-\hat{\mu}(\X_i,A_i))+  \xi_i\hat{\mu}(\X_i,Q_i)-\xi_i\hat{\mu}_{AG}^{q}\bigg)\\
         &=\frac{1}{\sqrt{n}}\sum_{i=1}^n \xi_i w_i^{p}\epsilon_i+\sum_{i=1}^n \xi_i\bigg( (\hat{\mu}(\X_i,Q_i)-w_i^{p}\hat{\mu}(\X_i,A_i))+ (w_i^{p} \mu(\X_i,A_i)-\mu(\X_i,Q_i))\bigg)\\
         &\quad\quad-\sum_{i=1}^n\xi_i\bigg(\hat{\mu}_{AG}^{q}-\mu(\X_i,Q_i)\bigg)\\
         &=\frac{1}{\sqrt{n}}\sum_{i=1}^n \xi_i w_i^{p}\epsilon_i+\frac{1}{\sqrt{n}}\sum_{i=1}^n ((\sum_{i=1}^n \xi_i) +1)\bigg(\frac{\xi_i+\frac{1}{n}}{(\sum_{i=1}^n \xi_i) +1}-\frac{\frac{1}{n}}{(\sum_{i=1}^n \xi_i) +1}\bigg)\\
         &\bigg( (\hat{\mu}(\X_i,Q_i)-w_i^{p}\hat{\mu}(\X_i,A_i))+ (w_i^{p} \mu(\X_i,A_i)-\mu(\X_i,Q_i))\bigg)
         -\frac{1}{\sqrt{n}}\sum_{i=1}^n\xi_i\bigg(\hat{\mu}_{AG}^{q}-\mu(\X_i,Q_i)\bigg)\\
         &=\bigg[\frac{1}{\sqrt{n}}\sum_{i=1}^n \xi_i w_i^{p}\epsilon_i-\frac{1}{\sqrt{n}}\sum_{i=1}^n\xi_i\bigg(\hat{\mu}_{AG}^{q}-\mu(\X_i,Q_i)\bigg)\bigg]\\
         &+\frac{1}{\sqrt{n}} (\sum_{i=1}^n \xi_i +1)\sum_{i=1}^n \bigg(\frac{\xi_i+\frac{1}{n}}{(\sum_{i=1}^n \xi_i) +1}\bigg)\bigg( (\hat{\mu}(\X_i,Q_i)-w_i^{p}\hat{\mu}(\X_i,A_i))+ (w_i^{p} \mu(\X_i,A_i)-\mu(\X_i,Q_i))\bigg)\\
         &+\frac{1}{\sqrt{n}}\sum_{i=1}^n \frac{1}{n}\bigg( (\hat{\mu}(\X_i,Q_i)-w_i^{p}\hat{\mu}(\X_i,A_i))+ (w_i^{p} \mu(\X_i,A_i)-\mu(\X_i,Q_i))\bigg),
    \end{split}
\end{equation*}
where, for the first row, since $\xi_i\epsilon_i$ can be seen as i.i.d. random variable with mean 0 and variance 1, from the partially conditional central limit theorem, $\frac{1}{\sqrt{n}}\sum_{i=1}^n \xi_i w_i^{p}\epsilon_i$ is distributed the same as 
\begin{equation*}
    \sigma n^{-1/2}\sum_{j=1}^n w_j^e G_j,
\end{equation*}
where $G_1,...,G_n$ are i.i.d. standard normal random variable. Similarly, the term 
\begin{equation*}
    \frac{1}{\sqrt{n}}\sum_{i=1}^n\xi_i\bigg(\hat{\mu}_{AG}^{q}-\mu(\X_i,Q_i)\bigg)
\end{equation*}
has the same distribution as $[\Var(\mu(\X,Q))]^{1/2}F$ where $F$ is a standard normal random variable. 

For the second term, let
\begin{equation*}
    \xi'_i=\frac{n\xi_i+1}{(\sum_{i=1}^n \xi_i) +1}.
\end{equation*} Note that since $\xi_i=\frac{1}{n}\xi'_i$ and $\sum_{i=1}^n\xi'_i=n$, we can view $\{\xi'_i\}_{i=1}^n$ as a new weight and apply the equation for weighted energy distance. The second term becomes:
\begin{equation*}
    \begin{split}
        &\frac{1}{\sqrt{n}} (\sum_{i=1}^n \xi_i +1)\frac{1}{n}\sum_{i=1}^n \xi_i'\bigg( (\hat{\mu}(\X_i,Q_i)-w_i^{p}\hat{\mu}(\X_i,A_i))+ (w_i^{p} \mu(\X_i,A_i)-\mu(\X_i,Q_i))\bigg)\\
        =&\frac{1}{\sqrt{n}}(\sum_{i=1}^n \xi_i +1)\int_{(\mathcal{X},\mathcal{A})}[\hat{\mu}(\x,a)-\mu(\x,a)]d(F_{n,\xi'\w^e_n,\X,A}-F^{q}_{n,\xi',\X,A}),
    \end{split}
\end{equation*}
where $F_{n,\xi'\w^e_n,\X,A}$ is the weighted empirical CDF of $\{\X_i,A_i\}_{i=1}^n$ with weights $\xi'\w^e_n$ and $F^{q}_{n,\xi',\X,A}$ is the weighted empirical CDF of $\{\X_i,Q_i\}_{i=1}^n$ with weights $\xi'$. Since $\{\xi_i\}_{i=1}^n$ has mean 0 and variance 1, by the central limit theorem, we have 
\begin{equation}
    \lim_{n\to\infty}\frac{1}{\sqrt{n}}(\sum_{i=1}^n \xi_i +1) =O_p(1),
\end{equation}
which is stochastically bounded. From (\ref{equ:3.3}) in the Main Manuscript Section \ref{sec:3}, we have 
\begin{equation*}
    \bigg(\int_{(\mathcal{X},\mathcal{A})}[\hat{\mu}(\x,a)-\mu(\x,a)]d(F_{n,\xi'\w^e_n,\X,A}-F^{q}_{n,\xi',\X,A})\bigg)^2\leq C_\alpha \mathcal{E}(F_{n,\xi'\w^e_n,\X,A},F^{q}_{n,\xi',\X,A}).
\end{equation*}
Similarly, since $\mathcal{E}(F_{n,\xi'\w^e_n,\X,A},F^{q}_{n,\xi',\X,A})=\mathcal{O}(n^{-1})$ from Lemma \ref{slemma3} and $C_\alpha={O}_p(n^{-1})$ due to the efficiency of $\hat{\mu}$, the second term is also $o_p(n^{-1/2})$ which vanishes when $n\to\infty$.

For the third term, note that 
\begin{equation*}
    \begin{split}
        &\frac{1}{\sqrt{n}}\sum_{i=1}^n \frac{1}{n}\bigg( (\hat{\mu}(X_i,Q_i)-w_i^{p}\hat{\mu}(\X_i,A_i))+ (w_i^{p} \mu(\X_i,A_i)-\mu(\X_i,Q_i))\bigg)\\
        &=\frac{1}{\sqrt{n}}\int_{(\mathcal{X},\mathcal{A})}[\hat{\mu}(\x,a)-\mu(\x,a)]d(F_{n,\w^{p}_n,\X,A}-F^{q}_{n,\X,A}),
    \end{split}
\end{equation*}
which, through the same arguments, is ${O}_p(n^{-1})$. 

Therefore, through the above arguments, we have demonstrated that $\hat{q}^{q}_{r}$ has the same distribution as $\hat{\mu}_{AG}^q-\mu^{q}$ in Theorem \ref{the:6.2}.

\end{proof}

%
%




\def\spacingset#1{\renewcommand{\baselinestretch}%
{#1}\small\normalsize} \spacingset{0.5}

\bibliographystyle{imsart-nameyear}
\bibliography{Bibliography}

\makeatletter\@input{xx.tex}\makeatother

%% file: section/section_01.tex
\section{Introduction}
\label{sec:1}
Understanding the causal effects of changes in doses or otherwise continuous values of a treatment is important in many scientific disciplines. 
A typical approach to quantifying the causal effect of varying a continuous treatment is to estimate the causal average dose response function (ADRF), which is the expected potential outcome as a function of all possible or likely treatment values. In our motivating application, the treatment is the mechanical power of a ventilator applied to patients in intensive care who have acute respiratory distress syndrome. In this setting, the ADRF is the expected in-hospital mortality as a function of the power of ventilation. While the ADRF provides an intuitive way to evaluate the causal effect of a continuous treatment, its estimation requires one to assume that every individual could hypothetically receive all possible values of the treatment, which is often unlikely in practice as some values may be clinically unlikely or impossible for subsets of the population. For example, in mechanically-ventilated patients with acute respiratory distress, ventilation guidelines restrict the values of mechanical ventilation settings based on measures of the patient's condition, making it impossible in practice for some patients to receive some ranges of settings. Thus, use of the ADRF in this setting may not have clinical meaning. Moreover, confounding is exacerbated in causal ADRF estimation because the characteristics of subjects with a higher treatment value may differ substantively from those with a lower value. For example, ventilation guidelines suggest explicit actions be taken with regards to mechanical ventilation settings based on patient characteristics, inducing strong and complex confounding. Additionally, the performance of causal ADRF estimators is often poor (i.e. cannot attain root-$n$ consistency nonparametrically), even with flexible, doubly robust methods \citep{kennedy2017non}.

Alternative definitions of causal effects in this setting have been introduced in part to help alleviate some of these problems while still providing clinically useful results. One of the alternatives is to imagine a hypothetical world where the treatment may be a random or deterministic function of the observed treatment value \citep{JamesStock1989Nonparametric, Robins2004EffectsOM}. \citet{munoz2012population} proposed stochastic interventions which focus on the effect of assigning each subject's intervention based on a random draw from a given distribution depending on their own characteristics. \citet{haneuse2013estimation} further proposed modified treatment policies (MTPs) which generalize the idea of stochastic interventions. Each subject's counterfactual treatment under a given MTP is defined as a function of their baseline characteristics and their \textit{observed} values of the treatment without the MTP (the ``natural'' value of the treatment); the MTP thus imagines a slight manipulation of each individual's treatment value from their actual values. The estimand is then defined as the expected potential outcome under the specified manipulation. For example, in our motivating application, one could estimate the effect of slightly reducing the mechanical power of ventilation, compared with standard of care, on in-hospital mortality among patients in the intensive care unit (ICU) with acute respiratory distress \citep{neto2018mechanical}.

MTPs can help policymakers evaluate hypotheses about the treatment and generate practical interventions that can be later tested via experiments. Another advantage of MTPs is that, by imagining counterfactual worlds where the treatment is modified slightly from reality, the corresponding counterfactual world is not substantially different from reality, making positivity hold by construction. This further allows for studying interventions that do not differ starkly from current practice, potentially helping clinical uptake of findings. 
 MTPs are an attractive tool for causal inference as they can be quite general and can be used in a wide variety of settings. For example, \citet{diaz2021nonparametric} generalized MTPs to time-varying treatment settings and \cite{diaz2020causal} developed stochastic interventions for causal mediation analysis.

Confounding is a major hurdle in causal inference from observational studies, both for standard estimands and for MTPs. 
Weighting approaches to confounding control re-weight each subject to balance the distribution of the covariates and do not require the use of outcome information. 
Traditionally, weights are generated through inverse-probability weighting (IPW) which models the treatment assignment mechanism given covariates and inverts it. However, the performance of IPW relies critically on the choice of the propensity score model and model misspecification can result in severe bias \citep{kang2007demystifying}. Alternative weighting methods have been proposed to mitigate this issue in the setting of estimating standard causal effects of discrete-valued treatments by using weights that encourage \citep{imai2014covariate} or enforce \citep{hainmueller2012entropy} balance of pre-specified moments of covariates. Particularly relevant to our article is the work by \citet{huling2020energy} who consider weights that minimize the energy distance between the weighted empirical distribution of the intervention groups and the joint population, aiming to balance the full joint distributions of covariates. This nonparametric distributional balancing approach has been shown to work empirically well with minimal need for careful modeling decisions.

Several challenging issues remain unresolved for weighting methods for MTPs. Existing weighting methods for MTPs have focused on the estimation of the conditional density of treatment given the covariates, also known as the generalized propensity score (GPS); weights are a ratio of the GPS at two different values. \citet{munoz2012population} and \citet{haneuse2013estimation} both proposed weighting estimators based on estimation of the GPS. \citet{diaz2021nonparametric} proposed weights based on direct modeling of the density ratio. These methods show great promise, but their success hinges on accurately specifying the conditional density or density ratio model. Yet, estimating a conditional density is highly challenging especially with an increasing number of covariates \citep{huling2023independence}. Even slight misspecification of the estimated density function from the truth can impact the performance of the IPW-style estimators of the ADRF \citep{naimi2014constructing}; these issues persist in the estimation of MTPs. As the role of covariate balance has not been explored for MTPs, no balancing weight methods exist for them, excacerbating these challenges. Moreover, the impact of weights on finite sample performance is not yet fully understood and no diagnostic tools are available to assess the performance of a given set of weights or to understand whether a given set of weights is likely to yield an unconfounded comparison for a given dataset. 
%
%
As MTPs involve investigating the effect of a shift of treatment values from their observed values, larger shifts are likely to yield analyses with more confounding and with more potential for positivity violations.
Yet no tools exist that can assess what range of shifts in a class of MTPs are ``safe'' (i.e. have confounding that can be fully adjusted using a given weighting method) for a given application. 
Clear guidance is thus needed for practitioners to determine an appropriate magnitude of the MTP shift. 

In this work, we introduce distance-based tools that provide solutions to the aforementioned issues. Based on the extension of the identification result of the modified treatment policy from \citet{haneuse2013estimation}, we clearly define the form of the target counterfactual population for an MTP (i.e., MTP-shifted population) and provide a clear mathematical expression of it. Even though the target population is purely hypothetical, we show its distribution can be estimated using the observed data, a fact used extensively in our methods. This builds the connection between the complex problem of MTPs and stochastic interventions and the well-understood problem of covariate balancing. By examining the role of the weights in finite sample error of a generic weighting estimator, we present a novel error decomposition to show that the estimation error of a weighted estimate for the causal effect of MTP is directly related to the imbalance of the weighted empirical distribution (referring to the empirical distribution of the weighted observed sample) and the MTP-shifted empirical distribution. 
We provide a measure of this distance with a modification of the energy distance \citep{szekely2013energy}, a computationally simple measure of the distance between two multivariate distributions, however other distributional measures can also be used. It has been used to measure distributional imbalance in the context of categorical treatments \citep{huling2020energy} and to quantify dependence between continuous treatments and confounders \citep{huling2023independence}. In this paper, we provide 1) a novel testing framework for the assessment of a feasible range of shifts for a class of MTPs for arbitrary weighting methods for a given dataset and 2) methods for comparison of different weighting methods for MTPs. Finally, we introduce energy balancing weights for MTPs, a weighting approach for MTP estimation which reduces \textit{distributional} imbalance to the target population, reducing sensitivity of findings to model misspecification of the conditional density function.  Although this paper primarily focuses on MTPs, we show that our error decomposition, methods, and setup are straightforwardly extended to stochastic interventions \citep{kennedy2019nonparametric, diaz2020causal, hejazi2023nonparametric}, allowing the use of our methods for a far broader class of settings beyond MTPs. See Section \ref{sec:stochastic} and Supplementary Material Section \ref{suppsec:SI} for more detail. We conclude our manuscript with an analysis of the impact of changes to mechanical power of ventilation from current practice on in-hospital mortality.

%% file: section/section_02.tex
\section{Setup}\label{sec:2}
\subsection{Causal framework for modified treatment policies (MTPs)}\label{sec:2.1}
Consider data $\{\boldsymbol{X}_i,A_i,Y_i\}_{i=1}^n$ collected from an observational study, where $\X_i \in \mathcal{X} \subseteq\mathbb{R}^p$ denotes a p-dimensional vector of pre-treatment covariates for subject $i$, $A_i \in \mathcal{A} \subseteq \mathbb{R}$ denotes the received value of the treatment, and $Y_i\in \mathbb{R}$ denotes the outcome. Here we assume $A_i$ is a continuous treatment variable (e.g., the taken dose of a treatment for individual $i$). We adopt the potential outcome framework in which we assume that for each subject $i$, there exists a potential outcome $Y^a_i$ defined as the outcome that subject $i$ would have if subject $i$ is intervened on to take the treatment at value $A=a$. An MTP analysis starts with an analyst-specified treatment policy $q(\boldsymbol{x}, a)$ that inputs patient characteristics and observed treatment values and outputs a modified value of the treatment.
The goal of an MTP analysis is to estimate the mean potential outcome $\mu^q \equiv \mathbb{E}[Y^{q(\X,A)}]$ under this policy; this can then be contrasted with the average \textit{observed} outcome $\mathbb{E}[Y^A] = \mathbb{E}[Y]$ to understand whether the policy improves or harms outcomes on average.
To ensure $Y^a_i$ is well-defined (i.e, $Y^a_i$ is unique for a given treatment level $a$ and subject $i$), we assume 
the following consistency condition:
\begin{itemize}\label{the:2.consistency}
    \item (A0 Consistency): For any subject $i$ in the population, if $A_i=a$ then $Y_i=Y^a_i$.
\end{itemize}

One of the main strengths of MTPs is that they can satisfy the positivity assumption by construction. In many scientific applications, it is not realistic for some patients to have extreme values of treatment, e.g. extremely low mechanical power of ventilation for patients who have serious respiratory symptoms (see Section \ref{sec:8} for the case study). As such, calculating the causal effect of an arbitrary treatment value $a$ may not be meaningful since this treatment value may not be feasible or applicable to the entire population. Alternatively, as in the work of \citet{haneuse2013estimation}, MTPs can be justified with a much weaker assumption on positivity (see assumption A1 below). Specifically, by carefully designing the MTP $q(\x,a)$, we can be more confident that for subject $i$ with covariate $\x_i$ and observed treatment $a_i$, there is a positive density to observe a subject with the same covariate $\boldsymbol{x}_i$ and treatment value $q(\x,a)$.

MTPs are pertinent to clinical practice, as they can provide insight into how modifications to clinical practice could have changed patient outcomes. By considering the effects of a collection of MTPs, researchers can gain insights into the underlying mechanisms that lead to different outcomes and identify potential strategies for improving patient outcomes. For example, to evaluate a care policy that encourages a slight reduction in intensity of mechanical ventilation, we could estimate the causal effect of the MTP $q(\boldsymbol{x},a)=a-2$ which reduced mechanical power for all subjects by 2 Joules/min compared with current standards. The MTP could be further refined to modify the amount of reduction differently for people with different characteristics or prior ventilation patterns.

For any $(\boldsymbol{x}, a) \in (\mathcal{X},\mathcal{A})$, where $(\mathcal{X},\mathcal{A})$ is the joint support of $\X$ and $A$, we can define $\mathbb{E}(Y^{q(\boldsymbol{x}, a)}|\X=\boldsymbol{x}, A=a)$ to be the expected potential outcome averaging over all subjects who have $\X=\boldsymbol{x}, A=a$. The mean potential outcome of the MTP $q(\boldsymbol{x},a)$ is then the mean of the conditional average potential outcome $\mathbb{E}(Y^{q(\boldsymbol{x}, a)}|\X=\boldsymbol{x}, A=a)$ over the entire population, i.e., 
\begin{equation}
    \mu^{q}:=\int_{(\mathcal{X},\mathcal{A})} \mathbb{E}[Y^{q(\boldsymbol{x},a)}|\X=\boldsymbol{x},A=a]dF_{\X,A}(\boldsymbol{x},a)
\end{equation}
where $F_{\X,A}(\boldsymbol{x},a)$ is the distribution function of $(\X,A)$. In order to make the integral well-defined, we follow \citet{haneuse2013estimation} to assume the continuity of $\mathbb{E}[Y^{q(\boldsymbol{x},a)}|\X=\boldsymbol{x}, A=a]$ over $(\boldsymbol{x},a)\in (\mathcal{X}, \mathcal{A})$ and continuity of $Pr(A|\X=\boldsymbol{x})$ over $\boldsymbol{x} \in \mathcal{X}$. In order to estimate the causal effect from observational data, $\mu^{q}$ needs to be causally-identifiable. In other words, we need to express $\mu^{q}$ as a function of the distribution of $(Y,\X,A)$. From \citet{haneuse2013estimation}, $\mu^q$ can be identified with the following assumptions:
\begin{itemize}
    \item (A1 Positivity): If $(\boldsymbol{x},a)\in (\mathcal{X},\mathcal{A})$ then $(\boldsymbol{x},q(\boldsymbol{x},a))\in (\mathcal{X},\mathcal{A})$.\label{the:2.positivity}
    \item (A2 Conditional exchangeability of related populations) For each $(\boldsymbol{x},a)\in (\mathcal{X},\mathcal{A})$, let $a'=q(\boldsymbol{x},a)$, then $Y^{a'}|\X=\boldsymbol{x},A=a$ and $Y^{a'}|\X=\boldsymbol{x},A=a'$ have the same distribution.\label{the:2.exchangeability}
\end{itemize}
The positivity assumption states that for each subject in the population, it is always possible to find some other subjects who have the same covariates and receive the modified treatment. The second assumption state that the potential outcome of the modified treatment will not be affected by the original treatment allocation. In other words, subjects with $\X=\boldsymbol{x}$ who received treatment $a$ could have received treatment $q(\boldsymbol{x}, a)$. The difference between A2 and the usual no-unmeasured confounders assumption is that the latter is much stronger, as it requires the subjects with $\X=\boldsymbol{x}$ who received treatment $a$ could have received any possible dose $a'\in \mathcal{A}_{\boldsymbol{x}}$ where $\mathcal{A}_{\boldsymbol{x}}:=\{a: (\boldsymbol{x}, a)\in(\mathcal{X}, \mathcal{A})\}$ be the support of $A$ given $\X=\boldsymbol{x}$.

In order to have a closed-form derivation of our proposed estimator, we further require the MTP to have a piecewise differentiable inverse function.

\begin{itemize}
    \item (C1 Piece-wise smooth invertibility): For each $\boldsymbol{x} \in \mathcal{X}$, there exists a partition $\{I_{j,\boldsymbol{x}}\}_{j=1}^{J(\boldsymbol{x})} $ of $\mathcal{A}_{\boldsymbol{x}}$ where $q(\boldsymbol{x},\cdot)$ is smooth and invertible within the partition. Specifically, let $q_j(\boldsymbol{x},\cdot)$ denote the function $q(\boldsymbol{x},\cdot)$ on the $I_{j,\boldsymbol{x}}$ part. Then $q_j(\boldsymbol{x},\cdot)$ has differentiable inverse function $h_j(\boldsymbol{x},\cdot)$ on the interior of $I_{j,\boldsymbol{x}}$, such that $h_j(\boldsymbol{x},q_j(\boldsymbol{x},a))=a$\label{the:2.piecewise}
\end{itemize}
Under Condition C1, for a given $\boldsymbol{x} \in \mathcal{X}$, the MTP $q(\boldsymbol{x},a)$ can be expressed as the sum of the functions $\{q_j(\boldsymbol{x},a)\}_{j=1}^{J(\boldsymbol{x})}$ on each interval $\{I_{j,\boldsymbol{x}}\}_{j=1}^{J(\boldsymbol{x})}$, $ q(\boldsymbol{x},a)=\sum_{j=1}^{J(\boldsymbol{x})}I_{j,\boldsymbol{x}}(a) q_j(\boldsymbol{x},a)$
where $I_{j,\boldsymbol{x}}(a)$ be the indicator function s.t. $I_{j,\boldsymbol{x}}(a)=1$ if $a\in I_{j,\boldsymbol{x}}$ and $I_{j,\boldsymbol{x}}(a)=0$ otherwise. Condition C1 is important, as it allows us to cleanly express the expected potential outcome under the MTP as a function of the observed data distribution. With C1, we have the following identification result that extended from \citet{haneuse2013estimation},
\begin{equation}\label{eq:identification}
    \mu^{q}=\int_{\mathcal{X}} \sum_{j=1}^{J(\boldsymbol{x})}\int_{h_j(\boldsymbol{x},a)\in I_{j,\boldsymbol{x}}} \mathbb{E}(Y|\X=\boldsymbol{x},A=a)dF_{\X,A}(\boldsymbol{x},h_j(\boldsymbol{x},a)).
\end{equation}
The proof is in the Supplementary Material Section \ref{suppsec:proof}.

\subsection{Novel error decomposition for weighted MTP estimators}\label{sec:2.2}

We present a novel error decomposition that allows a clear inspection of the role of sample weights in finite sample errors when estimating $\mu^q$ using weighting methods. This decomposition provides insights that enable us to provide a measure of confounding bias as a function of the sample weights. This measure then allows us to provide tools that compare weights and further tools to assess when a given MTP is so far from the observed treatments that measured confounding is excessive and/or too difficult to control with weights.

Denoting $\mu(\boldsymbol{x},a) \equiv \mathbb{E}(Y|\X=\boldsymbol{x},A=a)$, we have $Y_i=\mu(\X_i,A_i)+{\epsilon}_i$
where $\epsilon_i \equiv Y_i - \mu(\X_i, A_i)$ is the error term with mean zero. 
We can express the estimand \eqref{eq:identification} as  
\begin{equation}\label{eqn:identification_obs}
    \mu^q=\int_{(\mathcal{X},\mathcal{A})} \mu(\boldsymbol{x},a)dF_{\X,A}^q (\boldsymbol{x},a), \text{ where}
\end{equation}
\begin{equation}
    F_{\X,A}^q (\boldsymbol{x},a)= \sum_{j=1}^{J(\boldsymbol{x})} I_{j,\boldsymbol{x}}(h_j(\boldsymbol{x},a)) F_{\X,A}(\boldsymbol{x},h_j(\boldsymbol{x},a))
\end{equation}
and $F_{\X,A}(\boldsymbol{x},a)$ is the CDF of $(\X,A)$. In the Supplementary Materials Section \ref{sec:validcdf}, we prove that $F_{\X,A}^q (\boldsymbol{x},a)$ is the CDF of $(\X,q(\X,A))$, as long as 
$q(\boldsymbol{x}, a)$ has the property that
$\lim_{(\boldsymbol{x},a)\to \boldsymbol{-\infty}}q(\boldsymbol{x},a)=\boldsymbol{-\infty}$ and $\lim_{(\boldsymbol{x},a)\to \boldsymbol{\infty}}q(\boldsymbol{x},a)=\boldsymbol{\infty}$. The significance of \eqref{eqn:identification_obs} will be made clear in our error decomposition below.

In this paper, we focus on weighted estimators of $\mu^{q}$ which do not require one to use outcome information, allowing the analyst to conduct objective causal inference. A weighting estimator with arbitrary sample weights $\w=(w_1, \dots, w_n)$ can be expressed as: 
$$\hat{\mu}^{q}_{\w}=\frac{1}{n}\sum_{i=1}^{n} w_i Y_i=\frac{1}{n}\sum_{i=1}^{n} w_i\mu(\X_i,A_i)+\frac{1}{n}\sum_{i=1}^{n}w_i\epsilon_i.$$

For the following, we further require that $\sum_{i=1}^n w_i=n$ so that $F_{n,\w,\X,\A}=\frac{1}{n}\sum_{i=1}^n w_iI(\X_i\leq \boldsymbol{x},A_i\leq a)$ can be interpreted as a weighted empirical CDF of the the covariates and treatment values.
Now, we can express the error of a weighted estimator $\hat{\mu}^{q}_{\w}$ with arbitrary weights as:
\begin{align}
       \hat{\mu}^{q}_{\w}-\mu^q={} & \underbrace{\int_{(\mathcal{X},\mathcal{A})}\mu(\boldsymbol{x},a)d(F_{n,\w,\X,A}-F_{n,\X,A}^{q})(\boldsymbol{x},a)}_{\text{error due to confounding}} \label{equ:errordecomp_bias}\\
        &+\underbrace{\int_{(\mathcal{X},\mathcal{A})}\mu(\boldsymbol{x},a)d(F_{n,\X,A}^{q}-F_{\X,A}^{q})}_{\text{sampling error}} +\frac{1}{n}\sum_{i=1}^n w_i\epsilon_i, \label{equ:errordecomp_error}
\end{align}
where $F_{n,\X,A}^{q}(\boldsymbol{x},a)=\frac{1}{n}\sum_{i=1}^{n}\sum_{j=1}^{J(\X_i)} I_{j,\X_i}(A_i)I(\X_i\leq \boldsymbol{x},q_j(\X_i,A_i)\leq a)$ is the empirical estimator of the shifted distribution $F_{\X,A}^{q}$. The error decomposition emphasizes the importance of sample weights in mitigating bias. The second term is sampling error. The third term is affected by the variance of the weights, however, it always has mean $0$. Thus, bias can be mitigated by minimizing the first term $\int_{(\mathcal{X},\mathcal{A})}\mu(\boldsymbol{x},a)d(F_{n,\w,\X,A}-F_{n,\X,A}^{q})(\boldsymbol{x},a)$ which depends on the distance between the two distribution functions and the form of the outcome function $\mu(\x, a)$. 

Note that
$F_{n,\X,A}^{q}(\boldsymbol{x},a)$ is the empirical CDF of the MTP-shifted random variable $(\X,  q(\X,A))$ and $F_{n,\w,\X,A}(\boldsymbol{x},a)=\frac{1}{n}\sum_{i=1}^n w_iI(\X_i\leq \boldsymbol{x},A_i\leq a)$ is the weighted empirical CDF of random variable $(\X,A)$. Thus, the term $\int_{(\mathcal{X},\mathcal{A})}\mu(\boldsymbol{x},a)d(F_{n,\w,\X,A}-F_{n,\X,A}^{q})(\boldsymbol{x},a)$ depends on both the weights $\w$ and the pre-specified modified treatment policy $q(\boldsymbol{x},a)$. Larger shifts on the MTP will tend to lead to the larger difference between the distribution of $(\X, q(\X,A))$ and $(\X,A)$ which can make the confounding bias term larger, generally speaking. 

The significance of the error decomposition above is that both the key distributional terms, $F_{n,\w,\X,A}(\boldsymbol{x},a)$ and $F_{n,\X,A}^{q}(\boldsymbol{x},a)$, contributing to the error term in \eqref{equ:errordecomp_bias} are entirely functions of the observed data. Thus, while the outcome function is unknown, it is possible to compare $F_{n,\w,\X,A}(\boldsymbol{x},a)$ with $F_{n,\X,A}^{q}(\boldsymbol{x},a)$ to characterize bounds on bias due to measured confounding.

\subsection{Extension to stochastic interventions}\label{sec:stochastic}

Although our paper is focused on the modified treatment policy effect, most of the assertions can be easily extended to the context of stochastic intervention which is an alternative causal framework for estimating the effect of continuous intervention \citep{munoz2012population,diaz2021nonparametric,kennedy2019nonparametric}. Instead of fixing the shifted intervention $q(\X,A)$ to a single, deterministic value, stochastic interventions take the shifted intervention value as a random draw from a modified intervention distribution. In Supplementary Material Section \ref{suppsec:SI}, we establish an error decomposition result analogous to \eqref{equ:errordecomp_bias} for the estimation of the stochastic intervention effects. As a result, all of our proposed tools are directly applicable to stochastic interventions, allowing for a far more general use of our methods.

%% file: section/section_03.tex
\section{Weighted Energy Distance for MTPs}\label{sec:3}
With the key component of confounding bias in weighted estimates of the causal effect of an MTP, the next step is devoloping an outcome-agnostic manner measure or bound of the magnitude of \eqref{equ:errordecomp_bias} for a given set of weights. Doing so would enable evaluation of a set of weights in the specific context of a given MTP for a given dataset. A challenge in quantifying the bias term is that it depends on 1) an unknown mean function and 2) the distance between a weighted and an MTP-shifted distribution. Since the bias component $\int_{(\mathcal{X},\mathcal{A})}\mu(\boldsymbol{x},a)d(F_{n,\w,\X,A}-F_{n,\X,A}^{q})(\boldsymbol{x},a)$ is an integral over the difference of the two empirical distributions, a metric that can measure the distance between these two distributions can be used for this purpose. 
In this paper, we adopt the energy distance \citep{szekely2013energy,huling2020energy}, a metric based on powers of the Euclidean distance, for use in estimating weights that minimize the key component of the confounding bias. Although there are other measures of distributional distance, we focus on the energy distance due to its simplicity and lack of need for tuning parameters, making it a widely applicable tool for a broad range of investigators. However, other distances can be used in its place, such as the maximum mean discrepancy (MMD) which is the distance of two distributions embedded to the reproducing kernel Hilbert spaces (RKHS). MMDs with the distance kernel have been shown to be equivalent to the energy distance when it is calculated with a semimetric of negative type \citep{sejdinovic2013equivalence}. In Supplementary Material Section \ref{suppsec2.2}, we show an MMD formulation as an alternative measure. 

Following \citet{huling2020energy}, we generalize the energy distance (see Supplementary Material Section \ref{suppsec2.1} for the definition) to the weighted energy distance which measures the distance between the weighted empirical CDF $F_{n,\w,\X,A}$ and the MTP-shifted empirical CDF $F_{n,\X,A}^{q}$,
\begin{align}
&\mathcal{E}(F_{n,\w,\X,A},F_{n,\X,A}^{q}) \nonumber \\ 
&= {}  \frac{1}{n^2}\bigg\{2\times\sum_{i=1}^n\sum_{k=1}^n\sum_{j=1}^{J(\X_k)}I_{j,\X_k}(A_k)w_i||(\X_i,A_i)-(\X_k,q_j(\X_k,A_k))||_2 \nonumber\\
        &\quad\quad-\sum_{i=1}^n\sum_{k=1}^n w_i w_k ||(\X_i,A_i)-(\X_k,A_k)||_2 \nonumber\\
        &\quad\quad-\sum_{i=1}^n\sum_{j=1}^{J(\X_i)}\sum_{k=1}^n\sum_{j'=1}^{J(\X_k)} I_{j,\X_i}(A_i) I_{j',\X_k}(A_k)||(\X_i,q_j(\X_i,A_i))-(\X_k,q_{j'}(\X_k,A_k))||_2 \bigg\}.
\end{align}
Since the MTP-shifted empirical CDF $F_{n,\X,A}^{q}$ can be viewed as the empirical CDF for the sample $\{\X_i,q(\X_i,A_i)\}_{i=1}^n$, we can define its empirical characteristic function as
\begin{equation}
      \begin{split}
          \varphi_{n,\X,A}^{q}(\boldsymbol{t})&=\frac{1}{n}\sum_{i=1}^n \sum_{j=1}^{J(\X_i)} I_{j,\X_i}(A_i)\exp(\boldsymbol{i}\boldsymbol{t}^T(\X_i,q_j(\X_i,A_i))) =\frac{1}{n}\sum_{i=1}^n\exp(\boldsymbol{i}\boldsymbol{t}^T(\X_i,Q_i)),
      \end{split}
\end{equation}
where $Q_i=q(\X_i,A_i)$. \citet{szekely2013energy} and \citet{huling2020energy} demonstrate that the energy distance is some (predefined) norm about the two distribution's characteristic functions. We extend the result from \citet{huling2020energy} and prove that, in the context of MTP, the defined weighted energy distance is a distance between the intended distribution, i.e., 
 \begin{equation}\label{the:3.1}
     \mathcal{E}(F_{n,\w,\X,A},F_{n,\X,A}^{q})=\int_{\mathbb{R}^p} |\varphi_{n,\w,\X,A}(\boldsymbol{t})- 
     \varphi_{n,\X,A}^{q}(\boldsymbol{t})|^2v(\boldsymbol{t})d\boldsymbol{t},
 \end{equation}
 where $\w$ be a vector of weights such that $\sum_{i=1}^n w_i=n$ and $w_i>0$ for $\forall i$, $v(\boldsymbol{t})=1/(C_d||\boldsymbol{t}||_{1+d}^{1+d})$, $C_d$ is a constant, $d=p+1$ is the dimension of the variate $(\X,A)$, $\varphi_{n,\X,A}^{q}(\boldsymbol{t})$ is the characteristic function of $F_{n,\X,A}^{q}$, and $\varphi_{n,\w,\X,A}(\boldsymbol{t})=\frac{1}{n}\sum_{i=1}^n w_i e^{i\boldsymbol{t}^T(\X_i,A_i)}$ is the weighted empirical characteristic function (ECHF) of sample $\{\X_i,A_i\}_{i=1}^n$. Therefore, the weighted energy distance can be used to determine whether the weighted distribution is approaching the MTP-shifted distribution (i.e., the empirical distribution of $(\X,q(\X,A))$). 

The following shows that the weighted energy distance converges to the energy distance between the limiting CDF and the MTP-shifted CDF $F_{\X,A}^{q}$.
\begin{theorem}\label{the:3.2}
Assume $\lim_{n\to\infty} \varphi_{n,\w,\X,A}(\boldsymbol{t})=\tilde{\varphi}_{\X,A}(\boldsymbol{t})$ almost surely for $\forall \boldsymbol{t}$, where $\tilde{\varphi}_{\X,A}(\boldsymbol{t})$ is some integrable characteristic function with associated CDF $\tilde{F}_{\X,A}$. Then we have almost surely 
\begin{equation}
    \lim_{n\to\infty} \mathcal{E}(F_{n,\w,\X,A},F_{n,\X,A}^{q})=\mathcal{E}(\tilde{F}_{\X,A},F_{\X,A}^{q}).
\end{equation}
\end{theorem}

We now build the connection between the key error term \eqref{equ:errordecomp_bias} and the energy distance. Suppose that $\mu(\cdot)\in\mathcal{H}$ where $\mathcal{H}$ is the native space induced by the radial kernel $\Phi(\cdot)=-||\cdot||_2$ on $(\mathcal{X},\mathcal{A})$, for any weights $\w$ satisfying $\sum_{i=1}^n w_i=n$, $w_i\geq0$, then from the results of \citet{mak2018support} and \citet{huling2020energy}, we have
\begin{equation}\label{equ:3.3}
\bigg[\int_{(\mathcal{A},\mathcal{X})}\mu(\boldsymbol{x},a)d(F^{q}_{n,\X,A}-F_{n,\w,\X,A})(\boldsymbol{x},a)\bigg]^2\leq C_\alpha\mathcal{E}(F^{q}_{n,\X,A},F_{n,\w,\X,A}),
\end{equation}
where $0 \leq C_\alpha = \langle\mu,\mu\rangle$ is a constant depending only on $\mu(\boldsymbol{x},a)$; here, $\langle\cdot\rangle$ is the inner product in $\mathcal{H}$.

Since \eqref{equ:errordecomp_bias} is the key component of the bias, this result therefore connects the bias with the distributional balance. It states that under the condition that the conditional mean function $\mu(\boldsymbol{x},a)$ belongs to $\mathcal{H}$, the key component of the bias is bounded by the energy distance. Here we emphasize that $\mathcal{H}$ contains a wide class of functions. For extended discussion and details of the specific form of the native space, readers can refer to \citet{mak2018support}.

Of import for practical use, we note that the weighted energy distance depends on the scales of both $\X$ and $A$. Therefore, in order to make the distance unaffected by the measurement scale of $\X$ or $A$, we suggest standardizing covariates and treatment to have unit variance.

%% file: section/section_04.tex
\section{Using weighted energy distance for MTPs}\label{sec:4}

In this section, we introduce several important tools that facilitate and enhance the application of MTP techniques. In particular, these tools help provide critical and rigorous guidance on important decisions necessary within the MTP framework. Each of the tools is based on the weighted energy distance due to its connection with estimation bias for MTPs. The first tool centers on determining a reasonable range of treatment policies that can be reliably addressed with a given weighting approach. The second involves assessing arbitrary weighting approaches in their ability to control for measured confounding for a specific dataset and MTP, enabling researchers to determine the most effective weighting method for confounding control.

\subsection{A framework for choosing a feasible MTP scale for a given weighting method}\label{sec:4.1}

When exploring the causal effect of an MTP, 
the magnitude of the modification then determines how different the treatment under the MTP is compared with assigned treatments in reality. Larger magnitudes or scales of an MTP's modification to treatment values may allow researchers to explore a wider variety of effects, yet more extreme policies differ more starkly from reality, making confounding more severe and thus more difficult to control adequately.
Thus, the scale of the intervention or modification must be chosen carefully. In this section we develop a formal hypothesis testing framework to provide explicit guidance for this choice.

From the error decomposition in Section \ref{sec:2.2}, bias operates only through the term \eqref{equ:errordecomp_bias} which depends on the difference between the weighted empirical CDF for the observed population and the MTP-shifted empirical CDF for the target population. If the MTP scale is too large, a given set of weights or weighting approach may not properly balance these two populations to have the same distribution, leading to a greater potential for bias in the estimator. To address this issue, we propose a method based on the weighted energy distance to identify a scale of intervention for a given class of MTPs for which a given weighting approach can adequately balance the two populations and thus adequately control for confounding. By doing so, researchers can ensure that the MTP scale is not too large to be estimated well, reducing potential for bias.

\begin{figure}
    \centering
    \caption{ An illustration of our method for selecting a feasible MTP scale based on four simulated datasets. The X-axis represents the MTP scale, with larger values of $\tau$ indicating larger shifts in the treatment value. The blue curve shows the energy distance after weighting, while the green curve represents the calculated permutation threshold. The yellow area of the plot indicates values of $\tau$ for which the weighted energy distance exceeds the permutation threshold, indicating uncontrolled measured confounding. The estimation error is plotted as a function of $\tau$ (red curve). The error tends to increase substantially when employing MTP magnitudes with identified risk of bias.}
     \includegraphics[width=\textwidth]{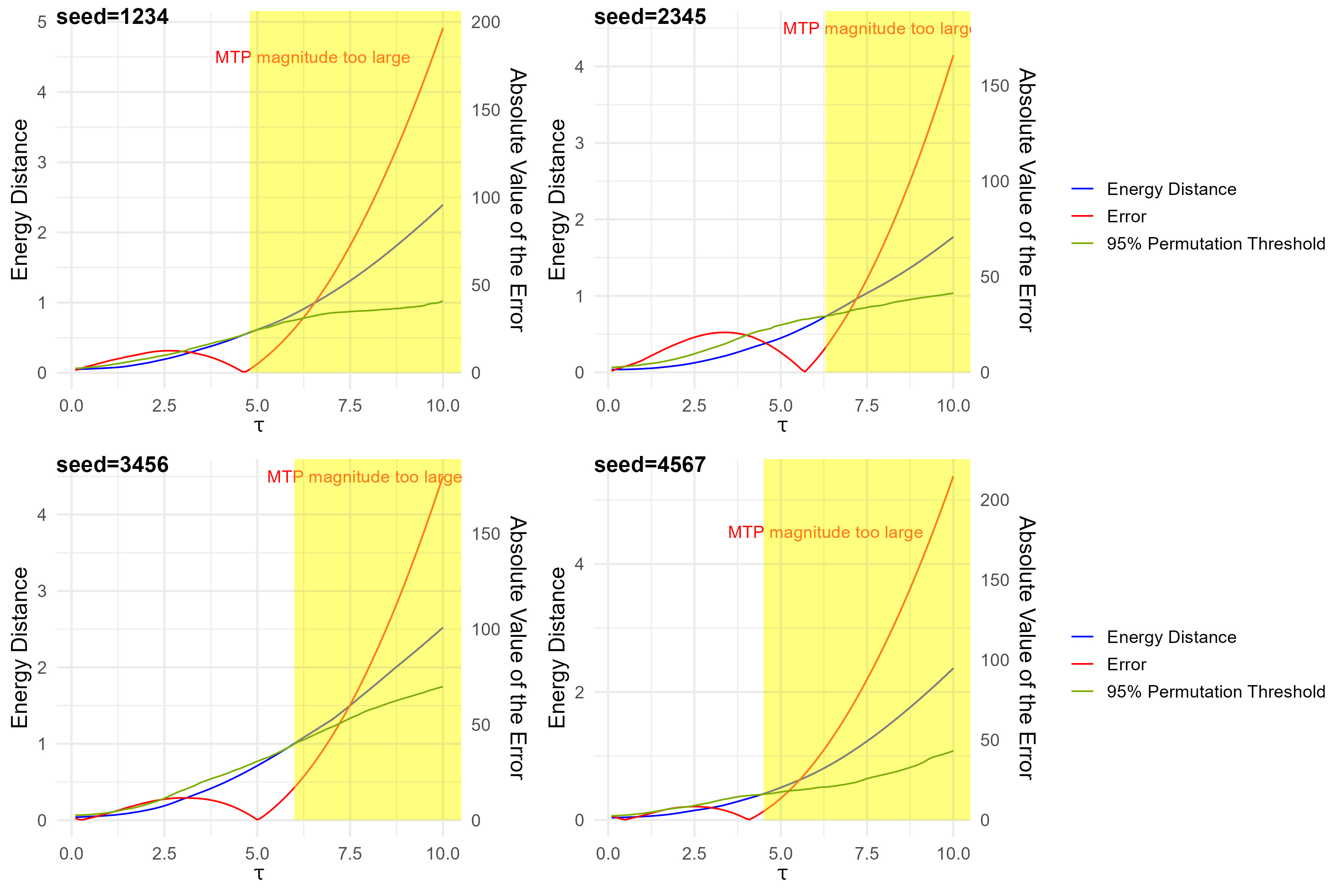}
    \label{fig:application1}
\end{figure}

In the ideal scenario that the balancing weights perfectly balance the observed population $\{\w_i\cdot (\X_i,A_i)\}_{i=1}^n$ to the MTP-shifted population $\{\X_i,q(\X_i,A_i)\}_{i=1}^n$ for an MTP, measured confounding is completely controlled for. 
Thus, denoting 
$F_{\w,\X,A}(\mathbf{x},a)=\mathbb{E}[F_{n,\w,\X,A}(\x,a)]$
our method therefore focuses on the following null and alternative hypotheses:
\begin{equation*}
    \begin{split}
        H_0:F_{\w,\X,A}(\mathbf{x},a)=F_{\X,A}^q(\mathbf{x},a) \textnormal{ } \forall (\mathbf{x},a)\in(\mathcal{X},\mathcal{A})
        \textnormal{ v.s.   } H_\alpha: F_{\w,\X,A}(\mathbf{x},a)\neq F_{\X,A}^q(\mathbf{x},a).
    \end{split}
\end{equation*}
Rejecting $H_0$ indicates that the given balancing weights do not perfectly balance the observed population's joint distribution of covariates and treatment to that of the counterfactual MTP-shifted population; thus, the alternative hypothesis represents scenarios where confounding has not been adequately controled, because the given weights are inadequate for confounding control for the given MTP. Some weighting methods may work well for smaller MTP magnitudes, but fail when the MTP's modification is more extreme. Carrying out a test to assess the above hypotheses is especially challenging, as unlike \citet{rizzo2016energy} where they use the energy statistic to test the equal distribution of two \textit{independent} samples, in our scenario, the observed sample and the MTP-shifted sample are pair-to-pair correlated (since they are derived from the same sample points). To handle this issue, we randomly partition the sample data $\{\X_i,A_i\}_{i=1}^n$ into two subsamples $\{\X_{I_k},A_{I_k}\}_{k=1}^{n/2}$ and $\{\X_{J_k},A_{J_k}\}_{k=1}^{n/2}$ where $I$ and $J$ are complementary index sets with equal size, without loss of generality. We use the first subsample to form the weighted observed subsample $w_{I_k}\cdot\{\X_{I_k},A_{I_k}\}_{k=1}^{n/2}$ and the second to form the MTP-shifted subsample $\{\X_{J_k},q(\X_{J_k},A_{J_k})\}_{k=1}^{n/2}$. Due to the random sampling, we have that $\int_{(\mathcal{X,A})} F_{n,\w,\X,A}(\x,a)dF_{\X,A}=\int_{(\mathcal{X,A})} F_{n/2,\w,\X_I,A_I}(\x,a)dF_{\X,A}$ and $F_{\X,A}^q=\int_{(\mathcal{X,A})} F_{n,\X,A}^q(\x,a)dF_{\X,A}=\int_{(\mathcal{X,A})} F_{n/2,\X_J,A_J}^q(\x,a)dF_{\X,A}$. We can then use the permutation test described in \citet{lehmann1986testing} Section 15.2 using the weighted observed subsample and the MTP-shifted subsample. Since for any $k_1\neq k_2$, $(\X_{I_{k_1}},A_{I_{k_1}})$ and $(\X_{I_{k_2}},A_{I_{k_2}})$ are independent, our partition ensures that the two samples are mutually independent, thus satisfying the ``randomization hypothesis'' required in Theorem 15.2.1 from \citet{lehmann1986testing} under permutation.

We use the weighted energy distance between the two subsamples as the test statistic $T=\mathcal{E}(F_{n/2,\w_I,\X_I,A_I}, F_{n/2,\X_J,A_J}^q)$. \citep{szekely2007measuring} have shown that, for two independent samples, this test statistic converges in distribution to a quadratic form of independent standard normal random variables under the null hypothesis of equal distribution and tends to infinity stochastically under the alternative hypothesis. From the Theorem 15.2.1 in \citet{lehmann1986testing}, the permutation test will have level $\alpha$ under $H_0$. The critical value for the hypothesis testing is determined by the permutation randomization described in the following algorithm:

\begin{algorithm}
Determining whether the MTP scale is feasible for a given weighting approach
\begin{tabbing}
   \qquad \textbf{Input:} Data $\{\X_i,A_i\}_{i=1}^n$, MTP $q$, and balancing weights $\w$.\\
   
   \qquad Randomly partition the data into two parts indexed as $I$ and $J$ and calculate the test statistic: \\ 
   \qquad \qquad  $T=\mathcal{E}(F_{n/2,\w_I,\X_I,A_I}, F_{n/2,\X_J,A_J}^q)$\\
   \qquad Combine the data as $\{\Tilde{w}_i, \Tilde{\X}_i, \Tilde{A}_i\}_{i=1}^n = (\{w_{I_k}, \X_{I_k},A_{I_k}\}_{k=1}^{n/2}, \{w_{J_k}, \X_{J_k},q(\X_{J_k}, A_{J_k})\}_{k=1}^{n/2})$ \\
   \qquad \qquad where $w_{J_k}=1$ for $k=1,...,n/2$.\\
   \qquad \textbf{For} $b=1$ \textbf{to} $B$-th iteration: \\
   \qquad \qquad \qquad Randomly partition the combined data into two parts indexed as $I^b$ and $J^b$ \\
   \qquad \qquad \qquad Calculate the test statistic:  $T_b=\mathcal{E}(F_{n/2,\Tilde{w}_{I^b},\Tilde{\X}_{I^b},\Tilde{A}_{I^b}}, F_{n/2,\Tilde{w}_{J^b},\Tilde{\X}_{J^b},\Tilde{A}_{J^b}})$ \\
    \qquad \textbf{End For}\\
   \qquad Calculate the critical value $T^p$ as the upper 0.95th quantile of the permutation sample $\{T_b\}_{b=1}^{B}$. \\
    \qquad Reject $H_0$ if $T^p<T$.
\end{tabbing}
\end{algorithm}

We illustrate this new approach through a simulated example using our novel energy balancing weights (described in more detail in Sections \ref{sec:5} and \ref{sec:6}). Simulated data are generated using the first condition of our simulation experiments described in Section \ref{sec:7} with $n=200$ and dimensionality $p=10$. We adopt the shift function $q(\boldsymbol{x},a)=\tau \Tilde{q}(\mathbf{x},a)$ where $\Tilde{q}(\mathbf{x},a)$ is the shift function we used in our simulation study and $\tau$ ranges from $[0, 10]$ in increments of 0.1 and controls the magnitude of our MTP. Larger values of $\tau$ represent the larger discrepancies from the original treatment. For each $\tau$, we calculate the weighted energy distance as the test statistic, the error of the energy balancing weights estimator, and the $95\%$ upper quantiles of the permutation sample. Figure \ref{fig:application1} displays four general scenarios with different random seeds. The energy distance for the energy balancing weights increases with a larger MTP scale $\tau$; along with this increase, the absolute error of the estimator also increases mostly monotonically. The vertical dashed line indicates the intersection of the energy distance (the test statistic) and the permutation threshold of the null distribution, to the right of which the values of $\tau$ are large enough that there is evidence that the weights do not adequately control for confounding. We can see that for $\tau$ values to the left of the vertical lines, the estimation error is generally small, indicating that this procedure is a reasonable criterion for choosing a feasible MTP scale. 

We also conduct a full simulation study to verify correct type I error rate control under $H_0$ and further explore alternative permutation strategies for the test. These results are presented in the Supplementary Material Section \ref{sec:suppapplication1}.

\subsection{Evaluation of arbitrary weights for a given MTP and dataset}\label{sec:4.2}

In the previous section, we evaluate different MTP scales with a fixed weighting method. Similarly, we can also use the weighted energy distance to evaluate the performance of different weighting methods for a fixed MTP. Since smaller weighted energy distances indicate a better balance between the weighted sample and the shifted sample, a natural criterion is to select the method that yields the smallest weighted energy distance. We demonstrate this approach to weight selection using the simulation results with sample sizes $n=800$ and covariate dimensionality $p=80$ under the first simulation scenario described in Section \ref{sec:7}.

\begin{figure}
    \centering
     \includegraphics[width=0.75\textwidth]{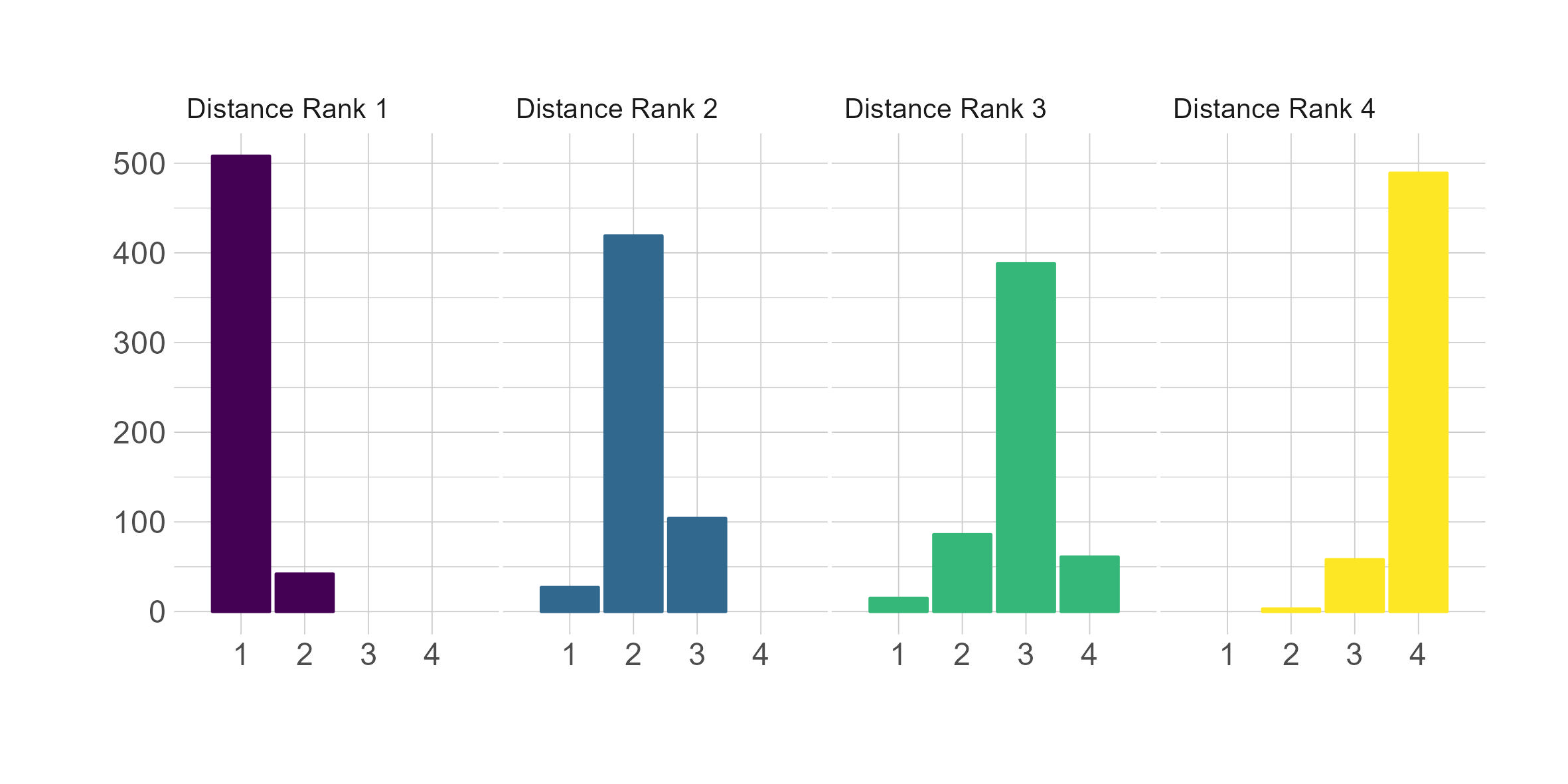}
    \caption{The histogram depicts the association between the rank of estimation error and the rank of weighted energy distance using the same data as in Figure 2. The x-axis of each plot is the rank of each method in terms of performance. The figure shows that, in most cases, the method with the lowest weighted energy distance yields the best performance.}
    \label{fig:application2_2}
\end{figure}

Figure \ref{fig:application2_2} displays the distribution of the rank of the absolute estimation error for a given dataset compared with the rank of the weighted energy distance for the four weighting methods. From this, the weighting method with the smallest energy distance (rank 1) tends to perform best in most cases. In the Supplementary Material Section \ref{sec:additionalillu}, we further explore the validity of this method under broader conditions.

%% file: section/section_05.tex
\section{Energy balancing weights for MTP estimation}\label{sec:5}

As demonstrated in the error decomposition in Section \ref{sec:2.2}, weighted distributional imbalance is a critical component of the estimation bias of weighted estimators for MTPs. Further, we showed in (\ref{the:3.1}) and Theorem \ref{the:3.2} that the weighted energy distance characterizes weighted distributional imbalance. In this section we propose a new set of weights designed as the optimizer of this measure. By minimizing the distance between the weighted empirical distribution and target distribution, our proposed weights, which we call energy balancing weights (EBWs), control for confounding in a robust and flexible manner.
Specifically, these weights minimize the energy distance between the observed sample and the MTP-shifted sample. 
 Thus, the energy balancing weights balance the weighted empirical distribution to the MTP-shifted empirical distribution. 

Extensively highlighted in the weighting literature, extreme weights can inflate the variance of the estimator \citep{li2018balancing,chattopadhyay2020balancing}. To ensure root-$n$ consistency of our energy balancing estimator, it is necessary to impose a restriction to avoid extreme weights. One approach is to introduce an additional penalization term for the sum of squared weights, resulting in penalized energy balancing weights. In this section, we introduce the penalized energy balancing weights $\w^{p}_n$. Under mild conditions, we show these weights result in root-$n$ consistency in estimating the mean potential outcome under an MTP.

\subsection{Penalized energy balancing estimator}\label{sec:5.2}
The penalized energy balancing weights account for the magnitude of the weights in addition to the weighted energy distance. Specifically, we define the penalized energy balancing weights $\w^{p}_n$ with a user-specified parameter $\lambda>0$ as follows:
\begin{equation}\label{equ:optipen}
         \begin{split}
        \w_n^{p}&\in \underset{\w=(w_1,...,w_n)}{\arg\min} \mathcal{E}(F_{n,\w,\X,A},F_{n,\X,A}^{q})+\frac{\lambda}{n^2}\sum_{i=1}^nw_i^2 \textnormal{ s.t. } \sum_{i=1}^n w_i=n\,, w_i\geq0.
    \end{split}
\end{equation}
In the Supplementary Material Section \ref{suppSec: codingdetail}, we showed that (\ref{equ:optipen}) is equivalent to a quadratic optimization problem that can be solved efficiently by the existing algorithms. The corresponding penalized energy balancing estimator is then $\hat{\mu}^{q}_{\w^{p}_n}=n^{-1}\sum_{i=1}^n \w_{i,n}^{p}Y_i$. Analogous to \citet{huling2020energy}, we prove the following properties of the energy balancing weights. Theorem \ref{the:5.1} shows that the penalized energy balancing weights make the weighted empirical CDF of the observed data converge to the true MTP-shifted CDF. 

\begin{theorem}\label{the:5.1}
Assume the assumptions in the previous theorems hold. Let $w_n^{p}$ be the penalized energy balancing weights. Then, we have $\lim_{n\to\infty} F_{n,\w_n^{p}, \X,A}(\boldsymbol{x},a)=F_{\X,A}^{q}(\boldsymbol{x},a)$
almost surely for every continuity point $(\boldsymbol{x},a)\in \mathcal{X}$. Furthermore, the following holds almost surely
\begin{equation}
    \lim_{n\to\infty}\mathcal{E}(F_{n,\w_n^{p}, \X,A},F_{n,\X,A}^{q})=0.
\end{equation}
\end{theorem}

We now show that the penalized energy balancing estimator achieves root-$n$ consistency.
\begin{theorem}\label{the:5.2}
Assume the conditions in Theorem 2. Let $\mathcal{H}$ be the native space induced by the radial kernel $\Phi(.)=-||.||_2$ on $(\mathcal{X}, \mathcal{A})$. Suppose the following mild conditions hold:
\begin{itemize}
    \item \textnormal{\textbf{CP-1:}} $\mu(\cdot,\cdot)\in\mathcal{H}$
    \item \textnormal{\textbf{CP-2:}} Var $[\mu(\X,A)]<\infty$ 
    \item \textnormal{\textbf{CP-3:}} Var $[Y|X=\boldsymbol{x},A=a]$ are bounded over $(\boldsymbol{x},a)\in(\calX,\calA)$.
    \item \textnormal{\textbf{CP-4:}} $\mathbb{E}[g^2(\W,\W',\W'',\W''')]\leq\infty$ where $\W,\W',\W'',\W'''\stackrel{i.i.d.}{\sim}F_{\X,A}^{q}$ be a vector of $(\X,A)$ and, with $h(\w)=\frac{f^q_{\X,A}(\boldsymbol{x},a)}{f_{\X,A}(\boldsymbol{x},a)}$, the kernel function $g(.)$ is defined as:
\begin{align*}
    g(\w,\w',\w'',\w''')= {} & h(\w)||\w-\w''||_2+h(\w')||\w'-\w'''||_2 \\
    & -h(\w)h(\w')||\w-\w'||_2-||\w'''-\w''||_2
\end{align*}
\end{itemize}
Then, the proposed penalized EBW estimator $\hat{\mu}_{q,\w_n^{p}}$ is root-$n$ consistent in that
\begin{equation}
    \sqrt{\mathbb{E}_{\X,A,Y}[(\hat{\mu}^{q}_{\w^{p}_n}-\mu^{q})^2]}={O}_p(n^{-1/2}).
\end{equation}
\end{theorem}

Theorem \ref{the:5.2} shows that the penalized energy balancing weights yield a root-$n$ consistent estimate of $\mu^q$ under mild conditions on the data-generating process even though the energy balancing weights are not shown to be consistent for the true density ratio weights in any sense.

We briefly comment on the conditions required for Theorem \ref{the:5.2}. Condition \textbf{CP-1} requires that the outcome regression function be contained in the native space $\mathcal{H}$. As demonstrated in \citet{huling2020energy}, this condition is a smoothness assumption on the true outcome regression function $\mu(\boldsymbol{x},a)$. \textbf{CP-2} requires that the outcome regression function have finite variance and \textbf{CP-3} requires that the conditional variance function be bounded. These two conditions are both mild and fairly
weak in practice. \textbf{CP-4} is a requirement that certain moments of the covariates and treatments be bounded.

The selection of $\lambda$ can be important in practice, as it determines the trade-off between bias and variance. Larger $\lambda$ values result in estimators with reduced variance but increased bias, due to greater imbalance between the two distributions after weighting. In this context, bias cannot ever be exactly characterized, so choosing an optimal value for $\lambda$ is a major challenge in practice. However, as bias is related to the imbalance of covariates, we can at least characterize what range of values of $\lambda$ still result in acceptable covariate balance. Based on the method proposed in Section 4, which tests whether the weights effectively balance the two distributions, one approach for choosing $\lambda$ is to select the largest $\lambda$ for which the permutation test indicates no significant imbalance remains after weighting.

%% file: section/section_06.tex
\section{Augmented energy balancing estimator}\label{sec:6}
Augmented estimator is a widely-used approach in the causal inference literature that combine balancing weights and an estimated outcome model to improve the estimation of a causal effect, potentially increasing efficiency and reducing sensitivity to model misspecification. In this context, we propose an augmented version of the energy balancing estimator that incorporates an estimated outcome model. Because the energy balancing weights were shown to yield consistent estimates of $\mu^q$ without the need for an outcome regression model, this allows the analyst to utilize an outcome regression model primarily for the purpose of variance reduction.

Let $\hat{\mu}(\boldsymbol{x},a)$ be an estimate of the outcome regression function $\mu(\boldsymbol{x},a)$. The augmented estimator is constructed by subtracting $\int_{\mathcal{X}}\int_{\mathcal{A}}\hat{\mu}(\boldsymbol{x},a)d(F^{q}_{n,\X,A}-F_{n,\w^e_n,\X,A})(\boldsymbol{x},a)$ from the weighted estimator $\hat{\mu}_{\w_n^{p}}^{q}=\sum_{i=1}^n w_i^{p} Y_i$. The resulting augmented energy balancing estimator is
\begin{equation}
\begin{split}
        \hat{\mu}_{AG}^{q}&=\hat{\mu}_{\w_n^{p}}^{q}-\int_{\mathcal{X}}\int_{\mathcal{A}}\hat{\mu}(\boldsymbol{x},a)d(F^{q}_{n,\X,A}-F_{n,w^{p}_n,\X,A})(\boldsymbol{x},a)\\
        &=\frac{1}{n}\sum_{i=1}^nw^{p}_i(Y_i-\hat{\mu}(\X_i,A_i))+\frac{1}{n}\sum_{i=1}^n\sum_{j=1}^{J(\X_i)}I_{j,\X_i}(A_i)\hat{\mu}(\X_i,q_j(\X_i,A_i)).
\end{split}
\end{equation}
In the second form, the weighted residuals can be view as a bias correction term for the outcome regression-based estimate of the MTP effect. We show its asymptotic normality and propose a statistically-valid and computationally-efficient multiplier bootstrap for inference. 
\subsection{Asymptotic normality}\label{sec:6.1}
In this section we demonstrate the asymptotic distribution of augmented energy balancing estimator $\hat{\mu}_{AG}^{q}$. The required conditions for asymptotic normality are:
\begin{itemize}
    \item \textbf{CA-1:} $\hat{\mu}(\boldsymbol{x},a)$ is such that $\hat{\mu}-\mu\in\mathcal{H}$.
    \item \textbf{CA-2:} $\w^e_n$ satisfy $1\leq \w_i^e \leq BN^{1/3}$ for some constant $B$ where $N$ is the total sample size.
    \item \textbf{CA-3:} $\max_i E(|\epsilon_i|^3)<\infty$. 
    \item \textbf{CA-4:} $\hat{\mu}(\boldsymbol{x},a)$ is a consistent estimator of $\mu(\boldsymbol{x},a)$.
\end{itemize}
Condition \textbf{CA-1} requires that the outcome regression model minus the true outcome regression model be contained in the native space $\mathcal{H}$. This condition somewhat limits the complexity of the error function of the estimated outcome regression model. Condition \textbf{CA-2} has been used before in the literature; see \citet{athey2018approximate, huling2023independence}. This condition can be imposed directly in the weight optimization procedure without changing the empirical performance of the weights or any of the asymptotic results of the weights, as in \citet{huling2023independence}.  It should be noted that, if we use the penalized energy balancing weights, this condition will no longer be needed as adding the penalty term in the optimization problem helps to sufficiently restrict the variability of the weights.  Condition \textbf{CA-3} indicates that the error term has finite third moments, and condition \textbf{CA-4} requires consistency of the outcome regression model; \textbf{CA-4} is only required for studying the asymptotic distribution of $\hat{\mu}^q_{AG}$. 
With these conditions, we have the following theorem:
\begin{theorem}\label{the:6.1}
Under Conditions CA1-CA4, let $J_n=n^{1/2}(\hat{\mu}_{AG}^{q}-\mu^{q})$ and $J_n^*=[Var(\mu(\X,A))]^{1/2}F+\sigma n^{-1/2}\sum_{j=1}^nw_j^e G_j$
where $F,G_1,...,G_n$ are independently and identically distributed standard normal random variables independent of $\X_1,...,\X_n,A_1,...,A_n$ and $\epsilon_1,...,\epsilon_n$. Let $\psi_n$ and $\psi_n^*$ be the corresponding characteristic functions of $J_n$ and $J_n^*$. Then as $n\to \infty$, $ |\psi_n(t)-\psi_n^*(t)|\to0, t\in \mathbb{R}$ where $\psi_n^*(t)$ is twice differentiable, and 
\begin{equation}
    \limsup_n \Var(J_n)\leq \Var(\mu(\X,Q))+\sigma^2V,
\end{equation}
where $\mu(\X, Q)=\sum_{j=1}^{J(X)}I_{j,\X}(A)\mu(\X,q_j(\X,A))$ is the conditional mean under the MTP and $V=\E\left(\frac{f^q_{\X,A}(\X,A)}{f_{\X,A}(\X,A)}\right)^2+D$ is the variance of the Radon-Nikodym weights plus a positive number.
\end{theorem}
Theorem \ref{the:6.1} demonstrates that the energy balancing weights, when paired with a consistent outcome regression model, achieve asymptotic normality and may be efficient, however, an exploration of the conditions under which efficiency is guaranteed requires additional exploration. 

\subsection{Inference}\label{sec:6.2}
The asymptotic normality of our augmented estimator $\hat{\mu}_{AG}^{q}$ allows us to use a standard bootstrap method to estimate its uncertainty and conduct statistical inference for the estimator. However, in practice, the standard nonparametric bootstrap method has some limitations. First, it can be computationally intensive because both the penalized energy balancing weights and the outcome regression estimator must be re-computed during each bootstrap iteration $r=1,...,R$. This can be especially time-consuming for large values of $R$ (e.g., $R=10000$).

We thus propose a multiplier bootstrap method originally presented by \citet{wu1986jackknife} that allows for quick computation. Similar to \citet{matsouaka2023variance} (which they name a wild bootstrap), the method perturbs the influence function of the estimator to estimate its variance. The influence function for our augmented estimator is 
\begin{equation*}
    \varphi_i= \frac{f^q_{\X,A}(\X,A)}{f_{\X,A}(\X,A)}(Y_i-\mu(\X_i,A_i))+\sum_{i=1}^{J(\X_i)}I_{j,\X_i}(A_i)\mu(\X_i,q_j(\X_i,A_i))-\mu^{q}.
\end{equation*}
We estimate the influence function by plugging in the estimators $\hat{\mu}^{q}$, $\hat{\mu}(\X,A)$, and $w_i^{p}$ for their population counterparts:
\begin{equation}
    \hat{\varphi}_i= w_i^{p}(Y_i-\hat{\mu}(\X_i,A_i))+\sum_{i=1}^{J(\X_i)}I_{j,\X_i}(A_i)\hat{\mu}(\X_i,q_j(\X_i,A_i))-\hat{\mu}^{q}.
\end{equation}
Even though $w_i^{p}$ does not necessarily estimate the true inverse propensity score, we show that the following procedure is still valid.
Then, the multiplier bootstrap estimator $\hat{\Sigma}$ can be constructed as displayed in Algorithm \ref{Alg:2}. A $95\%$ Wald-type confidence interval of $\hat{\mu}_{AG}^{q}$ can be calculated as $\hat{\mu}_{AG}^{q} \pm 1.96\times n^{-1/2} \hat{\Sigma}^{1/2}$. 
\begin{algorithm}\label{Alg:2}
Multiplier bootstrap
\begin{tabbing}
   \qquad \textbf{For} $r=1$ \textbf{to} $r=R$ \\
   \qquad \qquad Generate the random variate $\boldsymbol{\xi}=({\xi}_1,...,{\xi}_n)$ with ${\xi}_i\sim \exp(1)$ for $i=1,...,n$.\\
   \qquad \qquad Compute $\hat{q}^{q}_{r}=\frac{1}{\sqrt{n}}\sum_{i=1}^n {\xi}_i \hat{\varphi}_i$.\\
   \qquad Calculate $\hat{\Sigma}^{1/2}=$ IQR$(\hat{q}^{q})/(z_{0.75}-z_{0.25})$ where IQR$(\hat{q}^{q})$ is the interquartile range of $\{\hat{q}^{q}_r\}_{r=1}^R$ \\
   \qquad and $z_{0.75}-z_{0.25}=1.349$ be the interquartile range of standard normal distribution. \\
\end{tabbing}
\end{algorithm}

Here, we prove the validity of the proposed multiplier bootstrap in the following theorem.
\begin{theorem}\label{the:6.2}
    Under Conditions CA1-CA4, and $\hat{q}^{q}_{r}$ defined in Algorithm 1, we have that $\hat{q}^{q}_{r}=\frac{1}{\sqrt{n}}\sum_{i=1}^n \xi_i \hat{\varphi}_i \to n^{1/2}(\hat{\mu}_{AG}^{q}-\mu^{q})$ as $n\to \infty$.
\end{theorem}

Theorem \ref{the:6.2} shows that one can estimate the asymptotic distribution of $\hat{\mu}^q_{AG}$ using the multiplier bootstrap procedure in Algorithm 1. An advantage of the multiplier bootstrap procedure over the standard bootstrap is that the weights and outcome regression model only need to be estimated once, not $R$ times, resulting in a substantial computational speedup.

%% file: section/section_07.tex
\section{Simulation studies}\label{sec:7}
In this section, we study the empirical properties of the proposed energy balancing weights and further to assess their relative performance to other commonly used estimators for MTPs. The aim of the simulation study is to evaluate the performance of our proposed estimators under different simulation conditions.

\subsection{Data generating mechanisms}\label{sec:7.2}
We have two main simulation conditions: a moderately complex data-generating mechanism (simulation study \# 1) and a highly complex (simulation study \# 2) data-generating mechanism. For each simulation setting, we vary the following settings: the sample size $(n=100, 200, 400, 800)$, the number of covariates $(p=10, 20, 40, 80)$ for each individual, and the treatment type. This allows us to inspect the joint effects of dimensionality and increasing sample size on the empirical performance of all methods. We replicate all simulation experiments under two different conditions on the treatment variable: one condition where the treatment is continuously-distributed, and the other where the treatment has discrete, but countably infinite, support. Here, we briefly illustrate the data-generating mechanism. Readers are referred to Section \ref{sec:datagene} of the Supplementary Material for a detailed description of the data-generating mechanism. 1) The covariates are generated randomly following either a uniform distribution or a binomial distribution. 2) The treatment mean for a participant follows a cubic function of the covariates. 3) The moderately complex simulation condition assumes the outcome is a function of a quadratic term of the treatment times a cubic term of the covariates. The highly complex simulation condition further assumes interaction terms of the covariates. 4) The MTP function $q$ shifts the treatment more aggressively when the observed treatment is small. We emphasize that the estimation error (i.e., the observed sample mean effect minus the true MTP effect) is designed to be relatively stable with the increase of the number of covariates $p$, so that the complexity of the data-generating processes does not explode with dimension.

\subsection{Estimands and estimators under evaluation}\label{sec:7.3}
In each simulation replication, the estimand of interest is the mean potential outcome under the MTP. We evaluate three proposed energy balancing weight methods, including the penalized energy balancing weights described in Section \ref{sec:5.2}, an unpenalized version of the energy balancing weights, and a kernelized energy balancing weights approach using the Gaussian kernel (see Supplementary Material Section \ref{suppsec2.2}). The penalty factors $\lambda$ for the penalized energy balancing weights are set to be $\lambda=1$ for all simulation scenarios. Various alternative estimators for balancing weights are considered, such as the naive, unadjusted method that assigns equal weights to all subjects, density ratio weights (IPW) with generalized propensity score estimated using a Poisson density (for the discrete intervention settings), and the classification method proposed by (\cite{diaz2021nonparametric} Section 5.4) with classification models estimated by either 1) logistic regression or 2) a random forest. The corresponding augmented estimators for each weighting method are implemented using the same outcome model, the ensemble method SuperLearner \citep{van2007super}, which incorporates the lasso regularized generalized linear models (\texttt{SL.glmnet}). Also included is a targeted minimum-loss based estimator (TMLE) method \citep{diaz2021nonparametric} with all nuisance parameters estimated using SuperLearner.

\subsection{Simulation results}\label{sec:7.5}
\begin{figure}[ht]
    \centering
     \includegraphics[width=\textwidth]{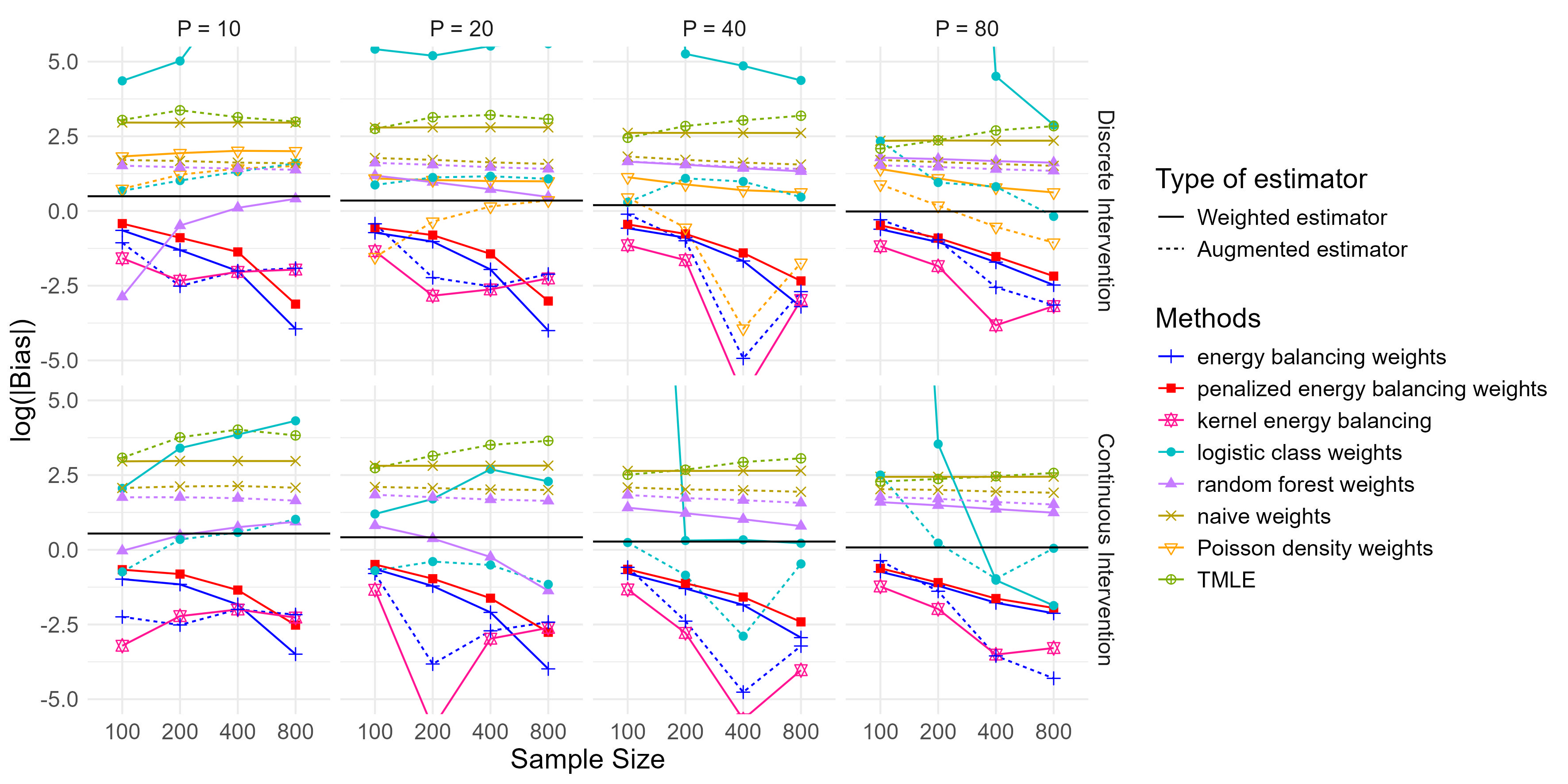}
    \caption{Simulation results in terms of the logarithm of the absolute value of bias across different sample sizes, type of treatment/intervention variable, and the dimensionality of covariates. The data-generating mechanism is moderately complex (simulation \# 1). Balancing methods are displayed in different colors and shapes of points. Weighted estimators are displayed in solid lines and the augmented estimators are in dashed lines.}
    \label{fig:simulation1_1}
\end{figure}

\begin{figure}[ht]
    \centering
     \includegraphics[width=\textwidth]{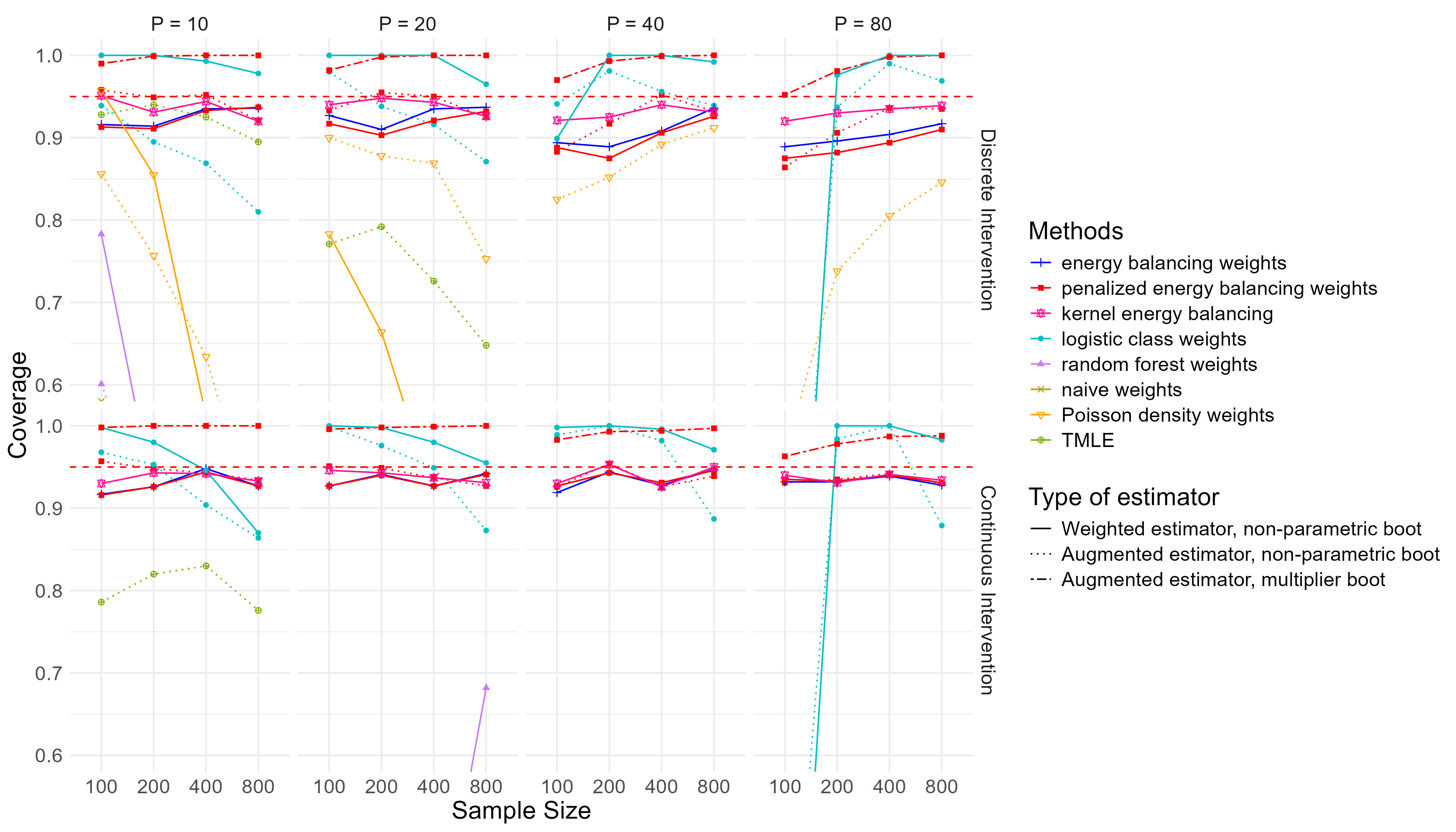}
    \caption{Simulation result for the 95\% coverage rate across different sample sizes and dimensionality of covariates. The data-generating mechanism is moderately complex (simulation \# 1). The red dashed line indicates nominal 95\% coverage. Different weighting methods are displayed in different colors and shapes of points. Weighted estimators are displayed in solid lines and the augmented estimators are in dashed lines. The multiplier bootstrap method is displayed in a larger size of the red points.}
    \label{fig:simulation1_2}
\end{figure}

In every simulation scenario, the true causal effect is determined using Monte Carlo with a sample size of 100,000. For each simulation setting, we repeat the experiment independently 1,000 times, applying each comparator method, allowing for the calculation of the mean squared error (MSE), bias, and coverage rate for 95\% confidence intervals. The coverage rate is calculated based on Wald-type intervals with SE for each estimator estimated by the non-parametric bootstrap with 100 replications. Additionally, we also use the multiplier bootstrap method proposed in Section \ref{sec:6.2} for the augmented energy balancing estimator. The simulation results are displayed in Figures \ref{fig:simulation1_1} and \ref{fig:simulation1_2}. Since the results of the two simulation studies are similar, we only display the MSE and coverage rate for the 95\% confidence interval for the first simulation study. Readers can refer to Supplementary Material Section \ref{sec:simresult} for the results of the highly complex setting.

In all simulation conditions, our three energy balancing methods (energy balancing, penalized energy balancing, and kernel energy balancing method) consistently outperform other methods in terms of both the bias and the coverage rate, indicating their robustness and ability to handle various situations. This pattern holds both for pure weighting (not augmented) estimators and for augmented estimators. The penalized energy balancing weights exhibit slightly worse performance in terms of bias compared to the energy balancing weights, which is probably due to the additional penalty term in the estimation. The kernel energy balancing weights (using the MMD described in the Supplementary Material Section \ref{suppsec2.2}) have the smallest bias in most cases when the sample size is relatively small; however, their performance varies significantly among different conditions and the bias does not improve as much with the increase of the sample sizes. This unstable performance may be due to the choice of bandwidth parameters in the Gaussian kernel function, which is the median heuristic that is known to not necessarily perform well in all scenarios. Therefore, although we believe the kernel energy balancing method has the potential to be very flexible and perform well in practice, its performance may depend critically on the choice of tuning parameters. Consequently, we recommend using penalized energy balancing methods that are more stable across different conditions.

The augmented energy balancing estimator generally has a smaller bias with a similar MSE (See Supplementary Material Figure \ref{fig:simulation1_3}) in most of the simulation conditions, which indicates the benefit of the inclusion of the outcome model. 
Although using the same outcome model, the augmented estimators with other balancing weights perform significantly worse than the energy balancing estimators. 
The coverage rate results (Figure \ref{fig:simulation1_2}) align with our observations for bias and MSE. All three energy balancing methods exhibit good coverage rate performance. Owing to its small bias, the coverage rate for the kernel energy balancing method approaches 95\% most closely.

%% file: section/section_08.tex
\section{Analysis of Mechanical Ventilation Data}\label{sec:8}
Critically ill patients often require mechanical ventilation to support their breathing. The ventilator delivers a controlled amount of air (tidal volume) at a specific pressure and rate to ensure adequate gas exchange in the lungs. In order to overcome the resistance of the airway and expand the thorax wall, the ventilator needs to transfer a certain amount of energy to the patient's respiratory system \citep{neto2018mechanical,serpa2016dissipated,cressoni2016mechanical}. Mechanical power (MP) is a comprehensive parameter that quantifies the amount of energy transferred to the patient's respiratory system during each breath. It takes into account tidal volume, respiratory rate, peak airway pressure, and positive end-expiratory pressure (PEEP) \citep{gattinoni2016ventilator}. Excessive ventilatory settings can result in ventilator-induced lung injury (VILI), which can worsen the patient's clinical condition and increase the risk of mortality \citep{serpa2016dissipated,cressoni2016mechanical}. Previous research has suggested that factors like high tidal volume, high airway pressures, and high respiratory rates could be associated with an increased risk of lung injury and poor outcomes \citep{nieman2016lung}. However, these individual factors do not capture the overall impact of ventilation settings on the lungs. Mechanical power can provide a more comprehensive assessment of the risk associated with different ventilatory settings. 

A recent study \citep{neto2018mechanical} examined data from the high-resolution database, the Medical Information Mart for Intensive Case (MIMIC-III) \citep{johnson2016mimic}. The MIMIC-III data contain information on critically ill patients who required mechanical ventilation in the intensive care unit (ICU). All included patients received invasive ventilation for at least 48 consecutive hours. Time-varying variables were collected at 8-hour intervals for the entire 48-hour period. \citep{neto2018mechanical} defined the exposure as the mean between the highest and lowest value of MP in the second 24 hours and concluded that high MP is independently associated with higher in-hospital mortality. In this case study, we reanalyze data from \citet{neto2018mechanical} to explore the causal relationship between a postulated MTP of MP and in-hospital mortality. 

We excluded patients with zero MP in Joules/min or extreme MP values (MP greater than 150 Joules/min) from the analysis. The exposure of interest is the mean value between the highest and lowest MP during the second 24 hours. The potential confounding variables we aim to balance are measured during or prior to the first 24 hours. As in \citet{neto2018mechanical}, our primary focus is on the in-hospital mortality of the included participants. The dataset contains a total of 5,011 participants with 97 identified covariates; among these covariates are many of the important factors such as the PaO2/FiO2 Ratio that are used in guidelines for determining mechanical ventilator settings. We use the following MTP to explore the causal effect of MTP that decreases the MP value based on its original value. This MTP reduces MP for individuals with high MP more aggressively, as earlier studies have indicated potential harms of high MP. The shifted MP value $q(\boldsymbol{x},a)$ is defined as $q(\boldsymbol{x},a)=a$ if $0<a<5$, $q(\boldsymbol{x},a)=a-5\tau$ if $5<a<10$, $q(\boldsymbol{x},a)=a-10\tau$ if $10<a<20$, $q(\boldsymbol{x},a)=a-15\tau$ if $20<a<40$, $q(\boldsymbol{x},a)=a-30\tau$ if $40<a$. Here $a$ is the original MP value and $\tau$ is the parameter that controls the magnitude of the MTP (a larger value of $\tau$ results in a more dramatic decrease of MP).

\begin{figure}[ht]
 \centering
    \includegraphics[width=\textwidth]{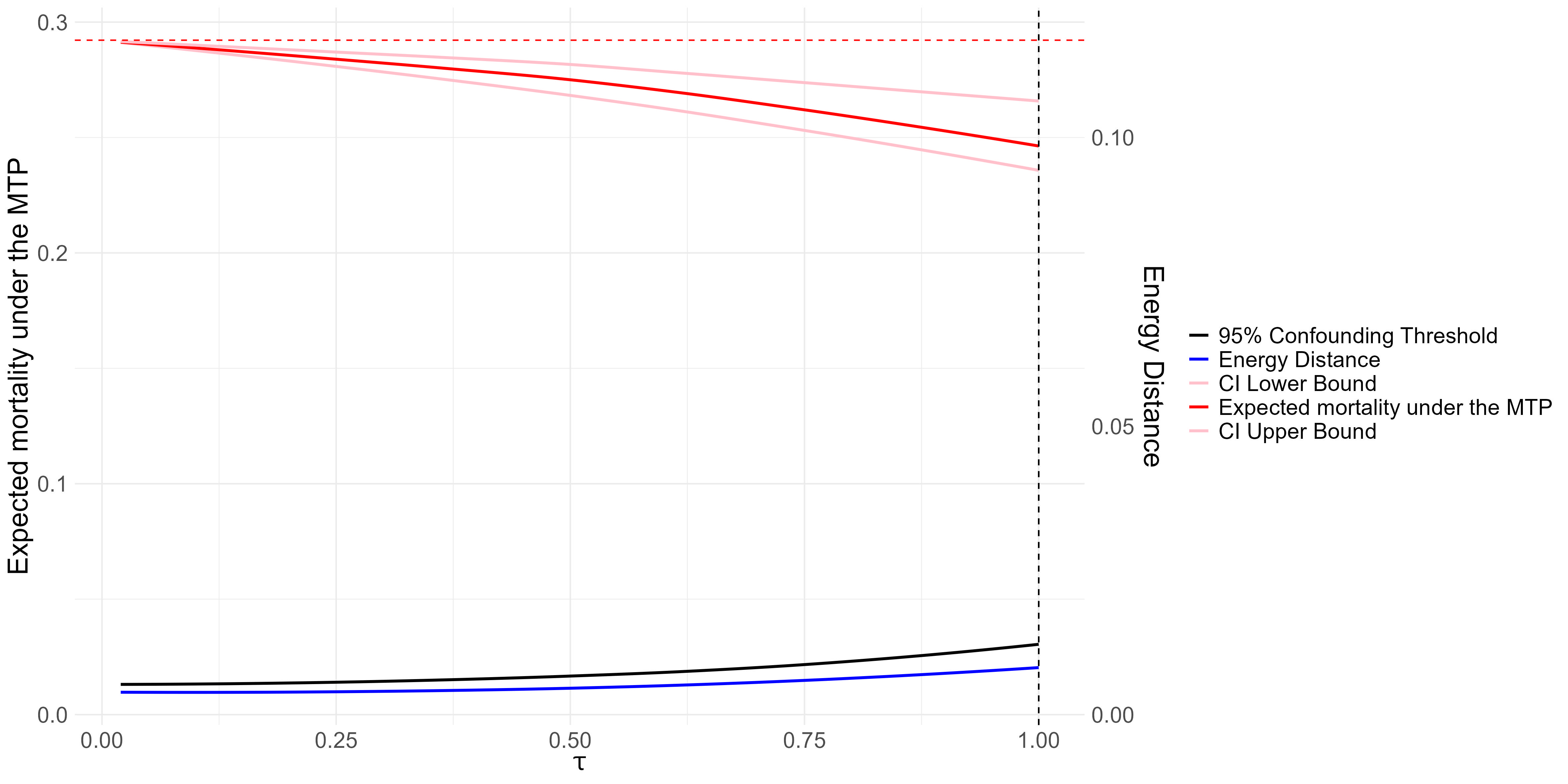}
    \caption{Exploring different magnitudes of MTP shifts to mechanical power of ventilation. The X-axis represents the $\tau$ value which controls the amount of the MTP shift. For each MTP shift, the blue curve displays the energy distance of the penalized energy balancing weights described in Section \ref{sec:5}. 
    The black line represents the 95\% Permutation Threshold estimated using the method described in Section \ref{sec:4.1}. The red curve along with the pink curves is the estimated mortality and the corresponding 95\% bootstrap confidence interval.}
    \label{fig:casestudy}
\end{figure}

With the $\tau$ ranging from 0 to 1 for the MTPs, our analysis results are displayed in Figure \ref{fig:casestudy}.  We adopt the unpenalized version of the energy balancing estimator. 
To identify a range of the shift magnitude $\tau$ for which confounding control via our energy balancing weights is feasible, we calculate the sampling variation of the energy distance under the corresponding MTP as described in Section \ref{sec:4.1}. The upper 95\% percentiles of the permutation sample are plotted with smoothed lines in the figure. Again, this value represents the intrinsic tails of variation of the energy distance when the two distributions are identical (i.e., the population is perfectly balanced). If the actual energy distance after balancing is larger than these upper bounds, the validity of the hypothesis that the population is balanced after weighting becomes questionable. The energy distance between the weighted sample and the target sample under the MTP is displayed as the blue curve for each value of $\tau$.  From the results, all MTPs with $\tau$ range between [0, 1] have energy distances below the 95\% permutation threshold, indicating that the shifts are within a reasonable range and have a low risk of bias due to measured confounding after weighting.

The red curve in Figure 5 represents the estimated potential in-hospital mortality rate under the MTP. The pink curves around it are the pointwise upper and lower bound for the 95\% confidence interval of the difference between the observed mortality and the estimated mortality under the MTP using the non-parametric bootstrap. To make the confidence interval more visually apparent, we have centered it around the estimated mortality under teh MTP. From Figure 5, we can see that the MTP results in a significant reduction in mortality even when the shift of treatment is small. 
When the MTP shift increases, it is more challenging to balance the distributions of covariates, which makes the confidence interval wider. However, the overall trend still shows that the mortality rate decreases with increases in the MTP shift. 
From Figure 5, as $\tau$ ranges between 0 to 1, the energy distances after balancing are consistently lower than the permutation thresholds, suggesting that measured confounding has been adequately controlled after weighting. This strengthens the validity of the result that a lower MP than is used in practice would likely decrease in-hospital mortality.

Compared to the results from \citep{neto2018mechanical,hong2021individualized}, we adopted an MTP analysis, strengthening the evidence about the potential harms of high MP due to the weakened assumptions under which our MTP estimators operate. Our set of tools provides a more convincing conclusion in terms of 1) the permutation test confirms that the measured confounding is not an issue after covariate balancing, and 2) the causal effect is significant even when the shift of MP is small, which further confirms the overall trend in our results.

%

%% file: section/section_09.tex
\section{Discussion}\label{sec:9}
Estimating causal effects of treatments from observational data is particularly challenging due to confounding. This issue is further compounded when the treatment takes continuous values. An MTP analysis is well-suited for this scenario and defines a hypothetical intervention on a treatment variable based on the baseline characteristics and the observed treatment value. However, the choice of the hypothetical intervention in an MTP is still subjective and it faces the dilemma that, 1) if the definition of MTP is too conservative (i.e., the shift of the intervention is very small), the causal effect of the MTP may be too small to have material clinical implications and on the other hand 2) if the definition of MTP is too aggressive, then the observed population and the population under MTP will be intrinsically different, which makes confounding a major issue again, negating many of the conceptual and practical benefits of MTPs. There is a lack of methods that can measure the magnitude of the potential for measured confounding and help to define a reasonable MTP that can be reliably estimated from the data at hand. 

In this paper, we demonstrate the connection between the estimation bias of weighted estimators of the causal effect under an MTP and weighted distributional imbalance. By explicitly defining the target population under MTP, we propose an error decomposition that highlights the role of covariate balance as a critical component of the estimation bias. We then introduce a distance, based on the weighted energy distance of \citet{huling2020energy}, that explicitly characterizes this imbalance. As a result, the performance of any arbitrary set of weights in controlling for confounding for MTPs can be gauged through the weighted energy distance. Two methods are proposed to enhance the estimation of MTP effects: 1) we propose a method for assessing the performance of various balancing weights according to their weighted energy distance; this approach enables researchers to choose different balancing techniques suitable for specific datasets, 2) we then propose a method to detect whether the MTP shift is too large for a balancing method. We use a permutation method to measure the variability of the energy distance under the null hypothesis of no measured confounding. Comparing this with the energy distance after applying the balancing weights enables an assessment of whether the weights can effectively balance the covariates under the current MTP and thereby control for measured confounding. This comparison also sheds light on the extent of confounding challenges associated with an MTP paired with a given dataset.  For the proposed permutation test, it is important to note that, as with all statistical tests, the power to reject the null hypothesis (and conclude that the two populations are unbalanced) decreases with smaller sample sizes. Therefore, we recommend that researchers relax the permutation threshold to 90\% or lower when sample sizes are small. Since there is no strong reason to strictly control the type I error rate to a small level in such cases, lowering the threshold can increase the power to detect imbalance after weighting.

Our second contribution is a novel covariate balancing weight method based on the energy distance. These energy balancing weights are generated by minimizing the weighted energy distance, thereby minimizing a measure of distributional imbalance and thus potential for bias. We additionally propose an augmented energy balancing estimator that integrates an outcome model with the balancing weights to help further reduce variance in the estimates. We establish the validity of our methods by demonstrating the root-$n$ consistency of these estimators and further asymptotic normality of the augmented estimator. Through simulation studies involving both moderately complex and highly complex scenarios, we showcase the stability and performance of our methods in terms of bias, mean squared error (MSE), and coverage rate across various conditions, reflecting their capacity to handle diverse real-world situations. 

We further proposed a multiplier bootstrap method to estimate the uncertainty of the augmented energy balancing estimator. The multiplier bootstrap approach offers a more computationally efficient means for statistical inference compared to the classical nonparametric bootstrap, as it does not require calculating weights within each bootstrap iteration. We prove the large-sample properties of this multiplier bootstrap and, through simulation, illustrate its similarity to the nonparametric bootstrap in terms of the coverage rates.

There remain some statistical properties that warrant further exploration. The first is to show whether the energy balancing weights converge to the true density ratio, as doing so may enable clearer studies of the potential efficiency of our proposed estimators. Secondly, as suggested in the simulation study, the coverage rate for the energy balancing estimator and the penalized energy balancing estimator is approximately 95\%, implying that both estimators exhibit asymptotic normal distribution. However, establishing the asymptotic distribution for the estimators using EBWs and penalized EBWs is highly challenging. 
Moreover, our methods can be extended by incorporating alternative distance measures in the estimation of energy balancing weights and studying their corresponding statistical properties. 
While this paper has focused on a broad set of tools for dealing with measured confounding, unmeasured confounding is always a possibility for any observational data analysis. To our knowledge, as of yet no tools to help assess sensitivity of findings to unmeasured confounding exist in the context of MTPs and we believe this is a critical area of future work, as it would help further enhance MTP analyses.
Finally, for treatments that take ordinal values but are not continuous, our method (along with the corresponding R code) can be applied directly without modification. An ordinal treatment with finite support inherently satisfies the `piece-wise smooth invertibility' condition outlined in Section 2, making it applicable to all theoretical results presented in this manuscript. However, for ordinal treatments with infinite support, this condition does not hold as stated, though it may be possible to extend our results to such cases.

%% file: manuscript.bbl
\begin{thebibliography}{34}

\bibitem[\protect\citeauthoryear{Athey, Imbens and
  Wager}{2018}]{athey2018approximate}
\begin{barticle}[author]
\bauthor{\bsnm{Athey},~\bfnm{Susan}\binits{S.}},
  \bauthor{\bsnm{Imbens},~\bfnm{Guido~W}\binits{G.~W.}} \AND
  \bauthor{\bsnm{Wager},~\bfnm{Stefan}\binits{S.}}
(\byear{2018}).
\btitle{Approximate residual balancing: debiased inference of average treatment
  effects in high dimensions}.
\bjournal{Journal of the Royal Statistical Society: Series B (Statistical
  Methodology)}
\bvolume{80}
\bpages{597--623}.
\end{barticle}
\endbibitem

\bibitem[\protect\citeauthoryear{Chattopadhyay, Hase and
  Zubizarreta}{2020}]{chattopadhyay2020balancing}
\begin{barticle}[author]
\bauthor{\bsnm{Chattopadhyay},~\bfnm{Ambarish}\binits{A.}},
  \bauthor{\bsnm{Hase},~\bfnm{Christopher~H}\binits{C.~H.}} \AND
  \bauthor{\bsnm{Zubizarreta},~\bfnm{Jos{\'e}~R}\binits{J.~R.}}
(\byear{2020}).
\btitle{Balancing vs modeling approaches to weighting in practice}.
\bjournal{Statistics in Medicine}
\bvolume{39}
\bpages{3227--3254}.
\end{barticle}
\endbibitem

\bibitem[\protect\citeauthoryear{Cressoni
  et~al.}{2016}]{cressoni2016mechanical}
\begin{barticle}[author]
\bauthor{\bsnm{Cressoni},~\bfnm{Massimo}\binits{M.}},
  \bauthor{\bsnm{Gotti},~\bfnm{Miriam}\binits{M.}},
  \bauthor{\bsnm{Chiurazzi},~\bfnm{Chiara}\binits{C.}},
  \bauthor{\bsnm{Massari},~\bfnm{Dario}\binits{D.}},
  \bauthor{\bsnm{Algieri},~\bfnm{Ilaria}\binits{I.}},
  \bauthor{\bsnm{Amini},~\bfnm{Martina}\binits{M.}},
  \bauthor{\bsnm{Cammaroto},~\bfnm{Antonio}\binits{A.}},
  \bauthor{\bsnm{Brioni},~\bfnm{Matteo}\binits{M.}},
  \bauthor{\bsnm{Montaruli},~\bfnm{Claudia}\binits{C.}},
  \bauthor{\bsnm{Nikolla},~\bfnm{Klodiana}\binits{K.}} \betal{et~al.}
(\byear{2016}).
\btitle{Mechanical power and development of ventilator-induced lung injury}.
\bjournal{Anesthesiology}
\bvolume{124}
\bpages{1100--1108}.
\end{barticle}
\endbibitem

\bibitem[\protect\citeauthoryear{D{\'\i}az and Hejazi}{2020}]{diaz2020causal}
\begin{barticle}[author]
\bauthor{\bsnm{D{\'\i}az},~\bfnm{Iv{\'a}n}\binits{I.}} \AND
  \bauthor{\bsnm{Hejazi},~\bfnm{Nima~S}\binits{N.~S.}}
(\byear{2020}).
\btitle{Causal mediation analysis for stochastic interventions}.
\bjournal{Journal of the Royal Statistical Society Series B: Statistical
  Methodology}
\bvolume{82}
\bpages{661--683}.
\end{barticle}
\endbibitem

\bibitem[\protect\citeauthoryear{D{\'{i}}az
  et~al.}{2021}]{diaz2021nonparametric}
\begin{barticle}[author]
\bauthor{\bsnm{D{\'{i}}az},~\bfnm{Iv{\'{a}}n}\binits{I.}},
  \bauthor{\bsnm{Williams},~\bfnm{Nicholas}\binits{N.}},
  \bauthor{\bsnm{Hoffman},~\bfnm{Katherine~L}\binits{K.~L.}} \AND
  \bauthor{\bsnm{Schenck},~\bfnm{Edward~J}\binits{E.~J.}}
(\byear{2021}).
\btitle{Nonparametric causal effects based on longitudinal modified treatment
  policies}.
\bjournal{Journal of the American Statistical Association}
\bpages{1--16}.
\end{barticle}
\endbibitem

\bibitem[\protect\citeauthoryear{Gattinoni
  et~al.}{2016}]{gattinoni2016ventilator}
\begin{barticle}[author]
\bauthor{\bsnm{Gattinoni},~\bfnm{L}\binits{L.}},
  \bauthor{\bsnm{Tonetti},~\bfnm{T}\binits{T.}},
  \bauthor{\bsnm{Cressoni},~\bfnm{M}\binits{M.}},
  \bauthor{\bsnm{Cadringher},~\bfnm{P}\binits{P.}},
  \bauthor{\bsnm{Herrmann},~\bfnm{P}\binits{P.}},
  \bauthor{\bsnm{Moerer},~\bfnm{O}\binits{O.}},
  \bauthor{\bsnm{Protti},~\bfnm{A}\binits{A.}},
  \bauthor{\bsnm{Gotti},~\bfnm{M}\binits{M.}},
  \bauthor{\bsnm{Chiurazzi},~\bfnm{C}\binits{C.}},
  \bauthor{\bsnm{Carlesso},~\bfnm{E}\binits{E.}} \betal{et~al.}
(\byear{2016}).
\btitle{Ventilator-related causes of lung injury: the mechanical power}.
\bjournal{Intensive Care Medicine}
\bvolume{42}
\bpages{1567--1575}.
\end{barticle}
\endbibitem

\bibitem[\protect\citeauthoryear{Hainmueller}{2012}]{hainmueller2012entropy}
\begin{barticle}[author]
\bauthor{\bsnm{Hainmueller},~\bfnm{Jens}\binits{J.}}
(\byear{2012}).
\btitle{Entropy balancing for causal effects: A multivariate reweighting method
  to produce balanced samples in observational studies}.
\bjournal{Political Analysis}
\bvolume{20}
\bpages{25--46}.
\end{barticle}
\endbibitem

\bibitem[\protect\citeauthoryear{Haneuse and
  Rotnitzky}{2013}]{haneuse2013estimation}
\begin{barticle}[author]
\bauthor{\bsnm{Haneuse},~\bfnm{Sebastian}\binits{S.}} \AND
  \bauthor{\bsnm{Rotnitzky},~\bfnm{Andrea}\binits{A.}}
(\byear{2013}).
\btitle{Estimation of the effect of interventions that modify the received
  treatment}.
\bjournal{Statistics in Medicine}
\bvolume{32}
\bpages{5260--5277}.
\end{barticle}
\endbibitem

\bibitem[\protect\citeauthoryear{Hejazi et~al.}{2023}]{hejazi2023nonparametric}
\begin{barticle}[author]
\bauthor{\bsnm{Hejazi},~\bfnm{Nima~S}\binits{N.~S.}},
  \bauthor{\bsnm{Rudolph},~\bfnm{Kara~E}\binits{K.~E.}}, \bauthor{\bsnm{Van
  Der~Laan},~\bfnm{Mark~J}\binits{M.~J.}} \AND
  \bauthor{\bsnm{D{\'\i}az},~\bfnm{Iv{\'a}n}\binits{I.}}
(\byear{2023}).
\btitle{Nonparametric causal mediation analysis for stochastic interventional
  (in) direct effects}.
\bjournal{Biostatistics}
\bvolume{24}
\bpages{686--707}.
\end{barticle}
\endbibitem

\bibitem[\protect\citeauthoryear{Hong et~al.}{2021}]{hong2021individualized}
\begin{barticle}[author]
\bauthor{\bsnm{Hong},~\bfnm{Yucai}\binits{Y.}},
  \bauthor{\bsnm{Chen},~\bfnm{Lin}\binits{L.}},
  \bauthor{\bsnm{Pan},~\bfnm{Qing}\binits{Q.}},
  \bauthor{\bsnm{Ge},~\bfnm{Huiqing}\binits{H.}},
  \bauthor{\bsnm{Xing},~\bfnm{Lifeng}\binits{L.}} \AND
  \bauthor{\bsnm{Zhang},~\bfnm{Zhongheng}\binits{Z.}}
(\byear{2021}).
\btitle{Individualized mechanical power-based ventilation strategy for acute
  respiratory failure formalized by finite mixture modeling and dynamic
  treatment regimen}.
\bjournal{EClinicalMedicine}
\bvolume{36}.
\end{barticle}
\endbibitem

\bibitem[\protect\citeauthoryear{Huling, Greifer and
  Chen}{2023}]{huling2023independence}
\begin{barticle}[author]
\bauthor{\bsnm{Huling},~\bfnm{Jared~D}\binits{J.~D.}},
  \bauthor{\bsnm{Greifer},~\bfnm{Noah}\binits{N.}} \AND
  \bauthor{\bsnm{Chen},~\bfnm{Guanhua}\binits{G.}}
(\byear{2023}).
\btitle{Independence weights for causal inference with continuous treatments}.
\bjournal{Journal of the American Statistical Association}
\bvolume{in press}
\bpages{1--25}.
\end{barticle}
\endbibitem

\bibitem[\protect\citeauthoryear{Huling and Mak}{2024}]{huling2020energy}
\begin{barticle}[author]
\bauthor{\bsnm{Huling},~\bfnm{Jared~D.}\binits{J.~D.}} \AND
  \bauthor{\bsnm{Mak},~\bfnm{Simon}\binits{S.}}
(\byear{2024}).
\btitle{Energy Balancing of Covariate Distributions}.
\bjournal{Journal of Causal Inference}
\bvolume{12}
\bpages{20220029}.
\bdoi{doi:10.1515/jci-2022-0029}
\end{barticle}
\endbibitem

\bibitem[\protect\citeauthoryear{Imai and Ratkovic}{2014}]{imai2014covariate}
\begin{barticle}[author]
\bauthor{\bsnm{Imai},~\bfnm{Kosuke}\binits{K.}} \AND
  \bauthor{\bsnm{Ratkovic},~\bfnm{Marc}\binits{M.}}
(\byear{2014}).
\btitle{Covariate balancing propensity score}.
\bjournal{Journal of the Royal Statistical Society: Series B (Statistical
  Methodology)}
\bvolume{76}
\bpages{243--263}.
\end{barticle}
\endbibitem

\bibitem[\protect\citeauthoryear{Johnson et~al.}{2016}]{johnson2016mimic}
\begin{barticle}[author]
\bauthor{\bsnm{Johnson},~\bfnm{Alistair~EW}\binits{A.~E.}},
  \bauthor{\bsnm{Pollard},~\bfnm{Tom~J}\binits{T.~J.}},
  \bauthor{\bsnm{Shen},~\bfnm{Lu}\binits{L.}},
  \bauthor{\bsnm{Li-Wei},~\bfnm{H~Lehman}\binits{H.~L.}},
  \bauthor{\bsnm{Feng},~\bfnm{Mengling}\binits{M.}},
  \bauthor{\bsnm{Ghassemi},~\bfnm{Mohammad}\binits{M.}},
  \bauthor{\bsnm{Moody},~\bfnm{Benjamin}\binits{B.}},
  \bauthor{\bsnm{Szolovits},~\bfnm{Peter}\binits{P.}},
  \bauthor{\bsnm{Celi},~\bfnm{Leo~Anthony}\binits{L.~A.}} \AND
  \bauthor{\bsnm{Mark},~\bfnm{Roger~G}\binits{R.~G.}}
(\byear{2016}).
\btitle{{MIMIC-III}, a freely accessible critical care database}.
\bjournal{Scientific Data}
\bvolume{3}
\bpages{1--9}.
\end{barticle}
\endbibitem

\bibitem[\protect\citeauthoryear{Kang and Schafer}{2007}]{kang2007demystifying}
\begin{barticle}[author]
\bauthor{\bsnm{Kang},~\bfnm{Joseph~DY}\binits{J.~D.}} \AND
  \bauthor{\bsnm{Schafer},~\bfnm{Joseph~L}\binits{J.~L.}}
(\byear{2007}).
\btitle{Demystifying double robustness: A comparison of alternative strategies
  for estimating a population mean from incomplete data}.
\bjournal{Statistical Science}
\bvolume{22}
\bpages{523--539}.
\end{barticle}
\endbibitem

\bibitem[\protect\citeauthoryear{Kennedy}{2019}]{kennedy2019nonparametric}
\begin{barticle}[author]
\bauthor{\bsnm{Kennedy},~\bfnm{Edward~H}\binits{E.~H.}}
(\byear{2019}).
\btitle{Nonparametric causal effects based on incremental propensity score
  interventions}.
\bjournal{Journal of the American Statistical Association}
\bvolume{114}
\bpages{645--656}.
\end{barticle}
\endbibitem

\bibitem[\protect\citeauthoryear{Kennedy et~al.}{2017}]{kennedy2017non}
\begin{barticle}[author]
\bauthor{\bsnm{Kennedy},~\bfnm{Edward~H}\binits{E.~H.}},
  \bauthor{\bsnm{Ma},~\bfnm{Zongming}\binits{Z.}},
  \bauthor{\bsnm{McHugh},~\bfnm{Matthew~D}\binits{M.~D.}} \AND
  \bauthor{\bsnm{Small},~\bfnm{Dylan~S}\binits{D.~S.}}
(\byear{2017}).
\btitle{Non-parametric methods for doubly robust estimation of continuous
  treatment effects}.
\bjournal{Journal of the Royal Statistical Society: Series B (Statistical
  Methodology)}
\bvolume{79}
\bpages{1229--1245}.
\end{barticle}
\endbibitem

\bibitem[\protect\citeauthoryear{Lehmann, Romano and
  Casella}{1986}]{lehmann1986testing}
\begin{bbook}[author]
\bauthor{\bsnm{Lehmann},~\bfnm{Erich~Leo}\binits{E.~L.}},
  \bauthor{\bsnm{Romano},~\bfnm{Joseph~P}\binits{J.~P.}} \AND
  \bauthor{\bsnm{Casella},~\bfnm{George}\binits{G.}}
(\byear{1986}).
\btitle{Testing statistical hypotheses}
\bvolume{3}.
\bpublisher{Springer}.
\end{bbook}
\endbibitem

\bibitem[\protect\citeauthoryear{Li, Morgan and
  Zaslavsky}{2018}]{li2018balancing}
\begin{barticle}[author]
\bauthor{\bsnm{Li},~\bfnm{Fan}\binits{F.}},
  \bauthor{\bsnm{Morgan},~\bfnm{Kari~Lock}\binits{K.~L.}} \AND
  \bauthor{\bsnm{Zaslavsky},~\bfnm{Alan~M}\binits{A.~M.}}
(\byear{2018}).
\btitle{Balancing covariates via propensity score weighting}.
\bjournal{Journal of the American Statistical Association}
\bvolume{113}
\bpages{390--400}.
\end{barticle}
\endbibitem

\bibitem[\protect\citeauthoryear{Mak and Joseph}{2018}]{mak2018support}
\begin{barticle}[author]
\bauthor{\bsnm{Mak},~\bfnm{Simon}\binits{S.}} \AND
  \bauthor{\bsnm{Joseph},~\bfnm{V~Roshan}\binits{V.~R.}}
(\byear{2018}).
\btitle{Support points}.
\bjournal{The Annals of Statistics}
\bvolume{46}
\bpages{2562--2592}.
\end{barticle}
\endbibitem

\bibitem[\protect\citeauthoryear{Matsouaka, Liu and
  Zhou}{2023}]{matsouaka2023variance}
\begin{barticle}[author]
\bauthor{\bsnm{Matsouaka},~\bfnm{Roland~A}\binits{R.~A.}},
  \bauthor{\bsnm{Liu},~\bfnm{Yi}\binits{Y.}} \AND
  \bauthor{\bsnm{Zhou},~\bfnm{Yunji}\binits{Y.}}
(\byear{2023}).
\btitle{Variance estimation for the average treatment effects on the treated
  and on the controls}.
\bjournal{Statistical Methods in Medical Research}
\bvolume{32}
\bpages{389-403}.
\bdoi{10.1177/09622802221142532}
\end{barticle}
\endbibitem

\bibitem[\protect\citeauthoryear{Mu{\~n}oz and Van
  Der~Laan}{2012}]{munoz2012population}
\begin{barticle}[author]
\bauthor{\bsnm{Mu{\~n}oz},~\bfnm{Iv{\'a}n~D{\'i}az}\binits{I.~D.}} \AND
  \bauthor{\bsnm{Van Der~Laan},~\bfnm{Mark}\binits{M.}}
(\byear{2012}).
\btitle{Population intervention causal effects based on stochastic
  interventions}.
\bjournal{Biometrics}
\bvolume{68}
\bpages{541--549}.
\end{barticle}
\endbibitem

\bibitem[\protect\citeauthoryear{Naimi et~al.}{2014}]{naimi2014constructing}
\begin{barticle}[author]
\bauthor{\bsnm{Naimi},~\bfnm{Ashley~I}\binits{A.~I.}},
  \bauthor{\bsnm{Moodie},~\bfnm{Erica~EM}\binits{E.~E.}},
  \bauthor{\bsnm{Auger},~\bfnm{Nathalie}\binits{N.}} \AND
  \bauthor{\bsnm{Kaufman},~\bfnm{Jay~S}\binits{J.~S.}}
(\byear{2014}).
\btitle{Constructing inverse probability weights for continuous exposures: a
  comparison of methods}.
\bjournal{Epidemiology}
\bpages{292--299}.
\end{barticle}
\endbibitem

\bibitem[\protect\citeauthoryear{Neto, Amato and
  Schultz}{2016}]{serpa2016dissipated}
\begin{barticle}[author]
\bauthor{\bsnm{Neto},~\bfnm{AS}\binits{A.}},
  \bauthor{\bsnm{Amato},~\bfnm{MBP}\binits{M.}} \AND
  \bauthor{\bsnm{Schultz},~\bfnm{MJ}\binits{M.}}
(\byear{2016}).
\btitle{Dissipated energy is a key mediator of VILI: rationale for using low
  driving pressures}.
\bjournal{Annual update in intensive care and emergency medicine 2016}
\bpages{311--321}.
\end{barticle}
\endbibitem

\bibitem[\protect\citeauthoryear{Neto et~al.}{2018}]{neto2018mechanical}
\begin{barticle}[author]
\bauthor{\bsnm{Neto},~\bfnm{Ary~Serpa}\binits{A.~S.}},
  \bauthor{\bsnm{Deliberato},~\bfnm{Rodrigo~Octavio}\binits{R.~O.}},
  \bauthor{\bsnm{Johnson},~\bfnm{Alistair~EW}\binits{A.~E.}},
  \bauthor{\bsnm{Bos},~\bfnm{Lieuwe~D}\binits{L.~D.}},
  \bauthor{\bsnm{Amorim},~\bfnm{Pedro}\binits{P.}},
  \bauthor{\bsnm{Pereira},~\bfnm{Silvio~Moreto}\binits{S.~M.}},
  \bauthor{\bsnm{Cazati},~\bfnm{Denise~Carnieli}\binits{D.~C.}},
  \bauthor{\bsnm{Cordioli},~\bfnm{Ricardo~L}\binits{R.~L.}},
  \bauthor{\bsnm{Correa},~\bfnm{Thiago~Domingos}\binits{T.~D.}},
  \bauthor{\bsnm{Pollard},~\bfnm{Tom~J}\binits{T.~J.}} \betal{et~al.}
(\byear{2018}).
\btitle{Mechanical power of ventilation is associated with mortality in
  critically ill patients: an analysis of patients in two observational
  cohorts}.
\bjournal{Intensive Care Medicine}
\bvolume{44}
\bpages{1914--1922}.
\end{barticle}
\endbibitem

\bibitem[\protect\citeauthoryear{Nieman et~al.}{2016}]{nieman2016lung}
\begin{barticle}[author]
\bauthor{\bsnm{Nieman},~\bfnm{Gary~F}\binits{G.~F.}},
  \bauthor{\bsnm{Satalin},~\bfnm{Joshua}\binits{J.}},
  \bauthor{\bsnm{Andrews},~\bfnm{Penny}\binits{P.}},
  \bauthor{\bsnm{Habashi},~\bfnm{Nader~M}\binits{N.~M.}} \AND
  \bauthor{\bsnm{Gatto},~\bfnm{Louis~A}\binits{L.~A.}}
(\byear{2016}).
\btitle{Lung stress, strain, and energy load: engineering concepts to
  understand the mechanism of ventilator-induced lung injury (VILI)}.
\bjournal{Intensive Care Medicine Experimental}
\bvolume{4}
\bpages{1--6}.
\end{barticle}
\endbibitem

\bibitem[\protect\citeauthoryear{Rizzo and Sz{\'e}kely}{2016}]{rizzo2016energy}
\begin{barticle}[author]
\bauthor{\bsnm{Rizzo},~\bfnm{Maria~L}\binits{M.~L.}} \AND
  \bauthor{\bsnm{Sz{\'e}kely},~\bfnm{G{\'a}bor~J}\binits{G.~J.}}
(\byear{2016}).
\btitle{Energy distance}.
\bjournal{wiley interdisciplinary reviews: Computational statistics}
\bvolume{8}
\bpages{27--38}.
\end{barticle}
\endbibitem

\bibitem[\protect\citeauthoryear{Robins, Hern{\'a}n and
  Siebert}{2004}]{Robins2004EffectsOM}
\begin{barticle}[author]
\bauthor{\bsnm{Robins},~\bfnm{James~M.}\binits{J.~M.}},
  \bauthor{\bsnm{Hern{\'a}n},~\bfnm{Miguel~A.}\binits{M.~A.}} \AND
  \bauthor{\bsnm{Siebert},~\bfnm{Uwe}\binits{U.}}
(\byear{2004}).
\btitle{Effects of multiple interventions}.
\bjournal{Comparative quantification of health risks: global and regional
  burden of disease attributable to selected major risk factors}
\bvolume{1}
\bpages{2191-2230}.
\end{barticle}
\endbibitem

\bibitem[\protect\citeauthoryear{Sejdinovic
  et~al.}{2013}]{sejdinovic2013equivalence}
\begin{barticle}[author]
\bauthor{\bsnm{Sejdinovic},~\bfnm{Dino}\binits{D.}},
  \bauthor{\bsnm{Sriperumbudur},~\bfnm{Bharath}\binits{B.}},
  \bauthor{\bsnm{Gretton},~\bfnm{Arthur}\binits{A.}} \AND
  \bauthor{\bsnm{Fukumizu},~\bfnm{Kenji}\binits{K.}}
(\byear{2013}).
\btitle{Equivalence of distance-based and {RKHS}-based statistics in hypothesis
  testing}.
\bjournal{The Annals of Statistics}
\bvolume{41}
\bpages{2263--2291}.
\bdoi{10.1214/13-AOS1140}
\end{barticle}
\endbibitem

\bibitem[\protect\citeauthoryear{Stock}{1989}]{JamesStock1989Nonparametric}
\begin{barticle}[author]
\bauthor{\bsnm{Stock},~\bfnm{James~H.}\binits{J.~H.}}
(\byear{1989}).
\btitle{Nonparametric Policy Analysis}.
\bjournal{Journal of the American Statistical Association}
\bvolume{84}
\bpages{567-575}.
\bdoi{10.1080/01621459.1989.10478805}
\end{barticle}
\endbibitem

\bibitem[\protect\citeauthoryear{Sz{\'e}kely, Rizzo and
  Bakirov}{2007}]{szekely2007measuring}
\begin{barticle}[author]
\bauthor{\bsnm{Sz{\'e}kely},~\bfnm{G{\'a}bor~J}\binits{G.~J.}},
  \bauthor{\bsnm{Rizzo},~\bfnm{Maria~L}\binits{M.~L.}} \AND
  \bauthor{\bsnm{Bakirov},~\bfnm{Nail~K}\binits{N.~K.}}
(\byear{2007}).
\btitle{Measuring and testing dependence by correlation of distances}.
\bjournal{The Annals of Statistics}
\bvolume{35}
\bpages{2769--2794}.
\end{barticle}
\endbibitem

\bibitem[\protect\citeauthoryear{Sz{\'e}kely and
  Rizzo}{2013}]{szekely2013energy}
\begin{barticle}[author]
\bauthor{\bsnm{Sz{\'e}kely},~\bfnm{G{\'a}bor~J}\binits{G.~J.}} \AND
  \bauthor{\bsnm{Rizzo},~\bfnm{Maria~L}\binits{M.~L.}}
(\byear{2013}).
\btitle{Energy statistics: A class of statistics based on distances}.
\bjournal{Journal of Statistical Planning and Inference}
\bvolume{143}
\bpages{1249--1272}.
\end{barticle}
\endbibitem

\bibitem[\protect\citeauthoryear{Van~der Laan, Polley and
  Hubbard}{2007}]{van2007super}
\begin{barticle}[author]
\bauthor{\bparticle{Van~der} \bsnm{Laan},~\bfnm{Mark~J}\binits{M.~J.}},
  \bauthor{\bsnm{Polley},~\bfnm{Eric~C}\binits{E.~C.}} \AND
  \bauthor{\bsnm{Hubbard},~\bfnm{Alan~E}\binits{A.~E.}}
(\byear{2007}).
\btitle{Super learner}.
\bjournal{Statistical applications in genetics and molecular biology}
\bvolume{6}.
\end{barticle}
\endbibitem

\bibitem[\protect\citeauthoryear{Wu}{1986}]{wu1986jackknife}
\begin{barticle}[author]
\bauthor{\bsnm{Wu},~\bfnm{Chien-Fu~Jeff}\binits{C.-F.~J.}}
(\byear{1986}).
\btitle{Jackknife, bootstrap and other resampling methods in regression
  analysis}.
\bjournal{The Annals of Statistics}
\bvolume{14}
\bpages{1261--1295}.
\end{barticle}
\endbibitem

\end{thebibliography}


\begin{thebibliography}{18}

\bibitem[\protect\citeauthoryear{Amaral, Allaire and
  Willcox}{2017}]{amaral2017optimal}
\begin{barticle}[author]
\bauthor{\bsnm{Amaral},~\bfnm{Sergio}\binits{S.}},
  \bauthor{\bsnm{Allaire},~\bfnm{Douglas}\binits{D.}} \AND
  \bauthor{\bsnm{Willcox},~\bfnm{Karen}\binits{K.}}
(\byear{2017}).
\btitle{Optimal $L_2$-norm empirical importance weights for the change of
  probability measure}.
\bjournal{Statistics and Computing}
\bvolume{27}
\bpages{625--643}.
\end{barticle}
\endbibitem

\bibitem[\protect\citeauthoryear{Csorgo and
  Nasari}{2013}]{csorgHo2013asymptotics}
\begin{barticle}[author]
\bauthor{\bsnm{Csorgo},~\bfnm{Miklos}\binits{M.}} \AND
  \bauthor{\bsnm{Nasari},~\bfnm{Masoud~M}\binits{M.~M.}}
(\byear{2013}).
\btitle{Asymptotics of Randomly Weighted u-and v-statistics: Application to
  Bootstrap}.
\bjournal{Journal of Multivariate Analysis}
\bvolume{121}
\bpages{176--192}.
\end{barticle}
\endbibitem

\bibitem[\protect\citeauthoryear{D{\'{i}}az
  et~al.}{2021}]{diaz2021nonparametric}
\begin{barticle}[author]
\bauthor{\bsnm{D{\'{i}}az},~\bfnm{Iv{\'{a}}n}\binits{I.}},
  \bauthor{\bsnm{Williams},~\bfnm{Nicholas}\binits{N.}},
  \bauthor{\bsnm{Hoffman},~\bfnm{Katherine~L}\binits{K.~L.}} \AND
  \bauthor{\bsnm{Schenck},~\bfnm{Edward~J}\binits{E.~J.}}
(\byear{2021}).
\btitle{Nonparametric causal effects based on longitudinal modified treatment
  policies}.
\bjournal{Journal of the American Statistical Association}
\bpages{1--16}.
\end{barticle}
\endbibitem

\bibitem[\protect\citeauthoryear{Garreau, Jitkrittum and
  Kanagawa}{2017}]{garreau2017large}
\begin{barticle}[author]
\bauthor{\bsnm{Garreau},~\bfnm{Damien}\binits{D.}},
  \bauthor{\bsnm{Jitkrittum},~\bfnm{Wittawat}\binits{W.}} \AND
  \bauthor{\bsnm{Kanagawa},~\bfnm{Motonobu}\binits{M.}}
(\byear{2017}).
\btitle{Large sample analysis of the median heuristic}.
\bjournal{arXiv preprint arXiv:1707.07269}.
\end{barticle}
\endbibitem

\bibitem[\protect\citeauthoryear{Huling, Greifer and
  Chen}{2023}]{huling2023independence}
\begin{barticle}[author]
\bauthor{\bsnm{Huling},~\bfnm{Jared~D}\binits{J.~D.}},
  \bauthor{\bsnm{Greifer},~\bfnm{Noah}\binits{N.}} \AND
  \bauthor{\bsnm{Chen},~\bfnm{Guanhua}\binits{G.}}
(\byear{2023}).
\btitle{Independence weights for causal inference with continuous treatments}.
\bjournal{Journal of the American Statistical Association}
\bvolume{in press}
\bpages{1--25}.
\end{barticle}
\endbibitem

\bibitem[\protect\citeauthoryear{Huling and Mak}{2024}]{huling2020energy}
\begin{barticle}[author]
\bauthor{\bsnm{Huling},~\bfnm{Jared~D.}\binits{J.~D.}} \AND
  \bauthor{\bsnm{Mak},~\bfnm{Simon}\binits{S.}}
(\byear{2024}).
\btitle{Energy Balancing of Covariate Distributions}.
\bjournal{Journal of Causal Inference}
\bvolume{12}
\bpages{20220029}.
\bdoi{doi:10.1515/jci-2022-0029}
\end{barticle}
\endbibitem

\bibitem[\protect\citeauthoryear{Kennedy}{2019}]{kennedy2019nonparametric}
\begin{barticle}[author]
\bauthor{\bsnm{Kennedy},~\bfnm{Edward~H}\binits{E.~H.}}
(\byear{2019}).
\btitle{Nonparametric causal effects based on incremental propensity score
  interventions}.
\bjournal{Journal of the American Statistical Association}
\bvolume{114}
\bpages{645--656}.
\end{barticle}
\endbibitem

\bibitem[\protect\citeauthoryear{Korolyuk and
  Borovskich}{1994}]{korolyuk1989theory}
\begin{bbook}[author]
\bauthor{\bsnm{Korolyuk},~\bfnm{Vladimir~S}\binits{V.~S.}} \AND
  \bauthor{\bsnm{Borovskich},~\bfnm{Yu~V}\binits{Y.~V.}}
(\byear{1994}).
\btitle{Theory of U-statistics, Mathematics and its Applications}
\bvolume{273}.
\bpublisher{Kluwer Academic Publishers Group, Dordrecht}.
\end{bbook}
\endbibitem

\bibitem[\protect\citeauthoryear{Mak and Joseph}{2018}]{mak2018support}
\begin{barticle}[author]
\bauthor{\bsnm{Mak},~\bfnm{Simon}\binits{S.}} \AND
  \bauthor{\bsnm{Joseph},~\bfnm{V~Roshan}\binits{V.~R.}}
(\byear{2018}).
\btitle{Support points}.
\bjournal{The Annals of Statistics}
\bvolume{46}
\bpages{2562--2592}.
\end{barticle}
\endbibitem

\bibitem[\protect\citeauthoryear{Mu{\~n}oz and Van
  Der~Laan}{2012}]{munoz2012population}
\begin{barticle}[author]
\bauthor{\bsnm{Mu{\~n}oz},~\bfnm{Iv{\'a}n~D{\'i}az}\binits{I.~D.}} \AND
  \bauthor{\bsnm{Van Der~Laan},~\bfnm{Mark}\binits{M.}}
(\byear{2012}).
\btitle{Population intervention causal effects based on stochastic
  interventions}.
\bjournal{Biometrics}
\bvolume{68}
\bpages{541--549}.
\end{barticle}
\endbibitem

\bibitem[\protect\citeauthoryear{Patterson}{1989}]{patterson1989strong}
\begin{barticle}[author]
\bauthor{\bsnm{Patterson},~\bfnm{Ronald~Frank}\binits{R.~F.}}
(\byear{1989}).
\btitle{Strong convergence'for U-statistics in arrays of row-wise exchangeable
  random variables}.
\bjournal{Stochastic Analysis and Applications}
\bvolume{7}
\bpages{89--102}.
\end{barticle}
\endbibitem

\bibitem[\protect\citeauthoryear{Rizzo}{2002}]{rizzo2002test}
\begin{barticle}[author]
\bauthor{\bsnm{Rizzo},~\bfnm{Maria~L}\binits{M.~L.}}
(\byear{2002}).
\btitle{A test of homogeneity for two multivariate populations}.
\bjournal{Proceedings of the American Statistical Association, Physical and
  Engineering Sciences Section}.
\end{barticle}
\endbibitem

\bibitem[\protect\citeauthoryear{Sejdinovic
  et~al.}{2013}]{sejdinovic2013equivalence}
\begin{barticle}[author]
\bauthor{\bsnm{Sejdinovic},~\bfnm{Dino}\binits{D.}},
  \bauthor{\bsnm{Sriperumbudur},~\bfnm{Bharath}\binits{B.}},
  \bauthor{\bsnm{Gretton},~\bfnm{Arthur}\binits{A.}} \AND
  \bauthor{\bsnm{Fukumizu},~\bfnm{Kenji}\binits{K.}}
(\byear{2013}).
\btitle{Equivalence of distance-based and {RKHS}-based statistics in hypothesis
  testing}.
\bjournal{The Annals of Statistics}
\bvolume{41}
\bpages{2263--2291}.
\bdoi{10.1214/13-AOS1140}
\end{barticle}
\endbibitem

\bibitem[\protect\citeauthoryear{Serfling}{1980}]{serfling1980approximation}
\begin{bbook}[author]
\bauthor{\bsnm{Serfling},~\bfnm{Robert~J}\binits{R.~J.}}
(\byear{1980}).
\btitle{Approximation Theorems of Mathematical Statistics}.
\bpublisher{John Wiley \& Sons}.
\end{bbook}
\endbibitem

\bibitem[\protect\citeauthoryear{Stellato et~al.}{2020}]{osqp}
\begin{barticle}[author]
\bauthor{\bsnm{Stellato},~\bfnm{B.}\binits{B.}},
  \bauthor{\bsnm{Banjac},~\bfnm{G.}\binits{G.}},
  \bauthor{\bsnm{Goulart},~\bfnm{P.}\binits{P.}},
  \bauthor{\bsnm{Bemporad},~\bfnm{A.}\binits{A.}} \AND
  \bauthor{\bsnm{Boyd},~\bfnm{S.}\binits{S.}}
(\byear{2020}).
\btitle{{OSQP}: an operator splitting solver for quadratic programs}.
\bjournal{Mathematical Programming Computation}
\bvolume{12}
\bpages{637--672}.
\bdoi{10.1007/s12532-020-00179-2}
\end{barticle}
\endbibitem

\bibitem[\protect\citeauthoryear{Sz{\'e}kely, Rizzo and
  Bakirov}{2007}]{szekely2007measuring}
\begin{barticle}[author]
\bauthor{\bsnm{Sz{\'e}kely},~\bfnm{G{\'a}bor~J}\binits{G.~J.}},
  \bauthor{\bsnm{Rizzo},~\bfnm{Maria~L}\binits{M.~L.}} \AND
  \bauthor{\bsnm{Bakirov},~\bfnm{Nail~K}\binits{N.~K.}}
(\byear{2007}).
\btitle{Measuring and testing dependence by correlation of distances}.
\bjournal{The Annals of Statistics}
\bvolume{35}
\bpages{2769--2794}.
\end{barticle}
\endbibitem

\bibitem[\protect\citeauthoryear{Sz{\'e}kely and
  Rizzo}{2013}]{szekely2013energy}
\begin{barticle}[author]
\bauthor{\bsnm{Sz{\'e}kely},~\bfnm{G{\'a}bor~J}\binits{G.~J.}} \AND
  \bauthor{\bsnm{Rizzo},~\bfnm{Maria~L}\binits{M.~L.}}
(\byear{2013}).
\btitle{Energy statistics: A class of statistics based on distances}.
\bjournal{Journal of Statistical Planning and Inference}
\bvolume{143}
\bpages{1249--1272}.
\end{barticle}
\endbibitem

\bibitem[\protect\citeauthoryear{Wong and Chan}{2017}]{wong2017kernel}
\begin{barticle}[author]
\bauthor{\bsnm{Wong},~\bfnm{Raymond~KW}\binits{R.~K.}} \AND
  \bauthor{\bsnm{Chan},~\bfnm{Kwun Chuen~Gary}\binits{K.~C.~G.}}
(\byear{2017}).
\btitle{Kernel-based covariate functional balancing for observational studies}.
\bjournal{Biometrika}
\bvolume{105}
\bpages{199--213}.
\end{barticle}
\endbibitem

\end{thebibliography}
